\documentclass[11pt]{article}

\usepackage{fullpage}

\usepackage[utf8]{inputenc}
\usepackage{amsmath}
\usepackage{amsthm}
\usepackage{bbm}
\usepackage{amsfonts}
\usepackage{amssymb}
\usepackage{units}
\usepackage{xcolor}
\usepackage{graphicx}
\usepackage[ruled,linesnumbered]{algorithm2e}
\usepackage{comment}
\usepackage{xfrac}
\usepackage{tikz}
\usetikzlibrary{arrows.meta,positioning,shapes,calc}
\usepackage{subcaption}
\let\oldnl\nl
\newcommand{\nonl}{\renewcommand{\nl}{\let\nl\oldnl}}

\usepackage{tikz}
\usetikzlibrary{arrows.meta, positioning}
\usepackage{enumitem}
\usepackage[hidelinks]{hyperref}

\newtheorem{theorem}{Theorem}[section]
\newtheorem{claim}{Claim}[theorem]
\newtheorem{lemma}[theorem]{Lemma}

\newtheorem{corollary}[theorem]{Corollary}
\newtheorem{definition}[theorem]{Definition}

\newcommand{\dist}{\mathrm{dist}}
\newcommand{\e}{\mathbb{E}}

\newcommand{\lra}{\leftrightarrows}
\newcommand{\B}{EB}
\newcommand{\C}{S}
\newcommand{\s}{ES}

\newcommand{\T}{ET}

\newcommand{\eps}{\varepsilon}

\newcommand{\fc}{\widehat{C}}
\newcommand{\ft}{\widehat{T}}
\newcommand{\fb}{\widehat{B}}

\title{Tree Covers for Directed Graphs: \\ Distance Oracles, Routing Schemes and Spanners}

\newcommand{\email}[1]{\href{mailto:#1}{\texttt{#1}}}

\author{
Shiri Chechik $^*$
\and
Gur Lifshitz
\thanks{Tel Aviv University, Israel
(\email{shiri.chechik@gmail.com},
\email{gur.lifshitz@gmail.com}).}
}

\begin{document}

\maketitle

\begin{abstract}

In a seminal work [STOC'01; JACM'05], Thorup and Zwick showed that every weighted undirected graph admits a tree cover where each vertex belongs to $O(kn^{1/k})$ trees in expectation, and every pair has a path of stretch at most $2k-1$ in some tree. 

In this paper, we extend this structural property to directed graphs with nonnegative edge weights. Let $\dist_G(u \leftrightarrows v) = \dist_G(u \to v) + \dist_G(v \to u)$ denote the roundtrip distance. For every integer $k$, we construct a collection $\mathcal{T}$ of tree-pairs, each consisting of an in-tree and an out-tree rooted at the same vertex. Each vertex belongs to $O(kn^{1/k})$ tree-pairs in expectation, and for every ordered pair $(u, v)$ some tree-pair $T_{in}\cup T_{out} \in \mathcal{T}$ satisfies $\dist_{T_{in}\cup T_{out}}(u \to v) \le \dist_G(u \to v) + (k-1)\dist_G(u \leftrightarrows v).$

We call this a \emph{$k$-strong (roundtrip) tree cover}. While prior techniques partition distances into geometric scales and build separate structures for each scale, our construction handles all distances simultaneously within a single, unified hierarchy. This yields the first directed tree cover for general $k$ matching the classic Thorup-Zwick stretch-size trade-off. Our cover immediately implies improved constructions of roundtrip spanners, path-reporting distance oracles, and compact routing schemes, as well as their source-wise and prioritized variants. We also provide faster algorithms for these structures: for general $k \ge 4$, all the above objects can be constructed in $\widetilde{O}(\min\{mn, n^{2.5}\})$ expected time, breaking the $mn$ barrier in dense directed graphs.

\end{abstract}

\pagebreak

\section{Introduction}

In 1990, motivated by the growing size of communication networks, Awerbuch and Peleg~\cite{AP90} studied how to reduce the resources required by distributed algorithms. A central difficulty was that traditional algorithms often required vertices to maintain global information about the network. As networks grew, storing this information required more space, while maintaining and updating it imposed an increasing workload on individual vertices. One of the representations they proposed was a \emph{tree cover}: a collection of trees in the original graph satisfying two requirements: (1) each vertex belongs to only a few trees, and this number is called the \emph{overlap} of the vertex. (2) For every pair of vertices, some tree contains a path between them whose length is at most a small multiple of their distance in the graph. The smallest such factor that suffices for all pairs is called the \emph{stretch} of the cover.

Following a line of work on sparse covers~\cite{AP90,AP92,ABCP93,Cohen93}, Thorup and Zwick (TZ)~\cite{TZ05} showed how to construct a tree cover of stretch $2k-1$ for any weighted \emph{undirected} graph with $n$ vertices and $m$ edges. Their construction takes $O(kmn^{1/k})$ expected time, and each vertex belongs to $O(kn^{1/k})$ trees in expectation.

This construction also yielded faster algorithms for constructing \emph{spanners}: subgraphs that preserve all pairwise distances up to a prescribed stretch. In particular, the union of the trees forms a $(2k-1)$-spanner with $O(kn^{1+1/k})$ edges in expectation. For every fixed $k$, this stretch--size tradeoff is optimal under Erd\H{o}s's girth conjecture. The same techniques yielded \emph{approximate distance oracles}, data structures that answer distance queries without storing all pairwise distances explicitly. Their oracle uses $O(kn^{1+1/k})$ space and answers queries in $O(k)$ time with stretch $2k-1$. Their oracle also supports \emph{path reporting}: it can return a path achieving the same stretch, in additional time linear in the number of edges reported.

Their tree covers also led to improved \emph{compact routing schemes}~\cite{TZ01routing}, which forward packets through the network using small routing tables stored at the vertices. For every integer $k\ge2$, their scheme uses $\widetilde{O}(kn^{1/k})$ bits per vertex and routes each packet along a path whose length is at most $4k-5$ times the shortest-path distance.

The situation changes substantially in \emph{directed graphs}. Ordinary one-way distances can encode arbitrary reachability information. Let $L$ and $R$ be two sets of $\Theta(n)$ vertices, with an arbitrary set of directed edges from $L$ to $R$. Then $\dist_G(l\to r)=1$ if $(l,r)\in E$, and $\dist_G(l\to r)=\infty$ otherwise. Thus, any finite multiplicative approximation must encode an arbitrary Boolean matrix, requiring $\Omega(n^2)$ bits in general. The same obstruction rules out tree covers with low overlap and sparse spanners: in the complete bipartite case, no edge can be omitted. Since each tree has fewer edges than vertices, covering all $\Theta(n^2)$ edges requires overlap $\Omega(n)$.

A natural way around this one-way reachability barrier is the \emph{roundtrip distance}, introduced by Cowen and Wagner~\cite{CowenWagner00}, defined by $\dist_G(u\lra v)=\dist_G(u\to v)+\dist_G(v\to u)$\footnote{$\dist_G(u\to v)$ denotes the shortest-path distance
from $u$ to $v$ in $G$.}. It is the length of a shortest directed closed walk containing $u$ and $v$. Roundtrip distance is symmetric and satisfies the triangle inequality, and hence induces a metric on every strongly connected component.

Roundtrip distance is a natural measure in many applications that involve travel or communication in both directions. For example,
sending a message from $u$ to $v$ often requires a response from $v$ to $u$, even just to let $u$ know that $v$ received the message. Another example is planning trips in road networks, where one-way streets can make the outward and return routes very different.
Applications that repeatedly compare such trips motivate compact data structures for estimating roundtrip distances and reporting the corresponding routes.

Within each strongly connected component, one can represent this metric by the complete undirected graph $G^{\rm rt}$, where $\{u,v\}$ has weight $\dist_G(u\lra v)$, and apply undirected tree-cover constructions such as TZ to $G^{\rm rt}$. This yields a tree cover of the roundtrip metric, whose virtual edges represent pairs of directed paths rather than edges of $G$. Passing to $G^{\rm rt}$ preserves roundtrip distances but loses the original edges of $G$: This construction \emph{does not directly yield a tree cover},
spanner, path-reporting distance oracle, or compact routing scheme for $G$. Moreover, $G^{\rm rt}$ is dense and implicit, and constructing it explicitly requires all-pairs roundtrip distances, which take $O(mn)$ time to compute.

Roditty, Thorup, and Zwick~\cite{RTZ08} extended tree covers to directed graphs using \emph{double trees}, each consisting of an ingoing tree $T_{\rm in}$ and an outgoing tree $T_{\rm out}$ on the same vertex set, rooted at a common center. They constructed separate covers for geometrically increasing distance scales and took their union. For edge weights in $[1,W]$, this yields a $(2k+\varepsilon)$-roundtrip double-tree cover: for every pair $u,v$ with finite roundtrip distance, some double tree satisfies \( \dist_{T_{\rm in}\cup T_{\rm out}}(u\lra v) \le (2k+\varepsilon)\dist_G(u\lra v). \)
The expected overlap of each vertex is $O((k^2/\varepsilon)n^{1/k}\log(nW))$.

The union of these double trees forms a $(2k+\varepsilon)$-roundtrip spanner with $O((k^2/\varepsilon)n^{1+1/k}\log(nW))$ expected edges. The cover also yields a distance oracle with the same stretch and expected space bound, and $O((k^2/\varepsilon)n^{1/k}\log(nW))$ expected query time. They also obtained a compact roundtrip routing scheme with
stretch $4k+\varepsilon$ and $\widetilde{O}((k/\varepsilon)n^{1/k}\log(nW))$ bits of local storage per vertex.

For $k=2$, they also constructed an optimal $3$-roundtrip spanner with $O(n^{3/2})$ edges for arbitrary positive edge weights, matching the undirected stretch--size tradeoff. Following a line of work on faster roundtrip spanner constructions~\cite{Pachocki18,Chechik20,dalirrooyfard20,CL21},
Harbuzova et al.~\cite{HarbuzovaJinVWXu24} achieved the same stretch and size in $\widetilde{O}(m\sqrt n)$ time, matching the running time of the TZ undirected construction for $k=2$, up to logarithmic factors. For $k=3$, Kadria and Roditty~\cite{KR25} matched the TZ undirected routing guarantee, obtaining roundtrip stretch $7$ with $\widetilde{O}(n^{1/3})$ bits of local storage per vertex.

Cen, Duan, and Gu~\cite{CenDuanGu20} obtained \((2k-1)\)-roundtrip spanners with \(O(kn^{1+1/k}\log n)\) edges. However, a vertex in their construction can belong to many more than \(kn^{1/k}\) tree--pairs, so the construction does not yield the corresponding tree-cover or distance-oracle guarantees. Even for spanners, an additional $k\log n$ factor remains compared with the undirected bound of $O(n^{1+1/k})$, which is optimal under Erd\H{o}s's girth conjecture~\cite{Althofer93}.

\paragraph{Main problem}
For general $k$, the central question is whether the TZ framework extends to directed graphs: Can one construct tree covers directly in directed graphs with the TZ stretch--overlap tradeoff, and use them to match the undirected stretch and space bounds for spanners and approximate distance oracles?

At the structural level, previous roundtrip constructions for general $k$~\cite{RTZ08,Pachocki18,Chechik20,CenDuanGu20} handle distance scales separately, and then solve for geometrically increasing distance scales. These results build collections of ingoing and outgoing trees that approximate roundtrip distances within a prescribed range, and then combine the resulting subgraphs across scales.

\subsection{Our Results}

Our goal is to extend the TZ tree-cover framework to directed graphs: low-overlap tree covers contained in $G$, from which
spanners, path-reporting distance oracles, and routing schemes can be derived.
For $k\ge4$, all the all-pairs structures below can be constructed in expected time $\widetilde O(\min\{mn,\;n^{2.5}\})$.

\paragraph{Roundtrip tree covers.}
We first obtain a direct directed analogue of the TZ tree cover. A \emph{tree--pair} is the union of an ingoing and an outgoing shortest-path tree rooted at the same center.

\begin{theorem}[\(2k\)-Roundtrip Tree Cover]
\label{thm:intro-2k-tree-cover}
For every integer \(k\ge1\), every directed graph \(G=(V,E)\) with
nonnegative edge weights admits a collection \(\mathcal T\) of tree--pairs
such that, for every \(u,v\in V\), some \(T\in\mathcal T\) satisfies
\[
\dist_T(u\lra v)\le 2k\,\dist_G(u\lra v).
\]
Every vertex belongs in expectation to \(O(k n^{1/k})\) tree--pairs.
\end{theorem}

\paragraph{Strong tree covers.}
Our main structural result goes further. A roundtrip tree cover requires one tree--pair to approximate both directions between \(u\) and \(v\). Our key idea is to relax this requirement by allowing the two directions to use different tree--pairs, and use this flexibility to obtain a stronger guarantee for each direction separately. We say that a tree cover has \emph{\(k\)-strong (roundtrip) stretch} if, for every ordered pair \(u,v\), some tree--pair \(T\) satisfies
\[
\dist_T(u\to v)
\le
\dist_G(u\to v)+(k-1)\dist_G(u\lra v).
\]
Applying the guarantee in both directions shows that the union of the tree--pairs has \((2k-1)\)-roundtrip stretch.
\begin{theorem}[\(k\)-Strong Tree Cover]
For every integer \(k\ge1\), every directed graph \(G=(V,E)\) with
nonnegative edge weights admits a \(k\)-strong tree cover in which every
vertex belongs in expectation to \(O(k n^{1/k})\) tree--pairs.
\end{theorem}

Our results improve on the double-tree cover of Roditty, Thorup, and Zwick~\cite{RTZ08} in stretch, expected overlap and running time.
Their construction gives roundtrip stretch \(2k+\varepsilon\) and expected overlap
\(O((k^2/\varepsilon)n^{1/k}\log(nW))\) for edge weights in \([1,W]\) in $\widetilde{O}(mn)$.
Our first cover reduces the roundtrip stretch to \(2k\), while reducing the expected overlap to \(O(kn^{1/k})\).
Our strong cover achieves the same overlap bound and provides a roundtrip stretch \(2k-1\).

Previous constructions partition distances into geometric scales and build a separate structure for each scale. Our construction handles all distances simultaneously, without this partition, using a single hierarchy. It applies to arbitrary nonnegative edge weights, with no dependence on \(\varepsilon\) or the edge-weight range. For fixed \(k\ge4\), our covers can be constructed in \(\widetilde O(\min\{mn,n^{5/2}\})\) expected time, giving a polynomial improvement over the \(mn\) bound when \(m=n^{3/2+\Omega(1)}\).

The strong cover also matches the stretch--overlap tradeoff obtained abstractly by applying TZ to \(G^{\rm rt}\).
Our contribution is to realize this guarantee using tree--pairs contained in \(G\), while preserving the overlap bound and avoiding the explicit construction of the roundtrip metric.

Next, we present the applications of our tree covers.
\paragraph{Spanners and path-reporting distance oracles.}
Taking the union of the tree--pairs immediately gives a \(k\)-strong spanner with \(O(k n^{1+1/k})\) expected edges.

\begin{corollary}
For every integer \(k\ge1\), every directed graph with \(n\) vertices
and nonnegative edge weights admits a \(k\)-strong spanner with
\(O(k n^{1+1/k})\) expected edges.
\end{corollary}

This gives a \((2k-1)\)-roundtrip spanner, removing the \(\log n\) factor from the \(O(kn^{1+1/k}\log n)\) edge bound of
Cen, Duan, and Gu~\cite{CenDuanGu20}, while also providing the stronger guarantee for each direction separately. Under Erd\H{o}s's girth conjecture, this edge bound is optimal up to a factor of \(k\).

The same cover yields a path-reporting \(k\)-strong distance oracle of
expected size \(O(k n^{1+1/k})\) and expected query time \(O(k n^{1/k})\). Here and below, query time excludes the time required to output the reported path.

\begin{corollary}[\(k\)-Strong Path-Reporting Distance Oracle]
For every integer \(k\ge1\), every directed graph with \(n\) vertices
and nonnegative edge weights admits a path-reporting distance oracle
with \(k\)-strong stretch, expected size \(O(k n^{1+1/k})\),
and expected query time \(O(k n^{1/k})\).
\end{corollary}

We reduce the query time to \(O(k)\) at the cost of increasing the stretch from \(k\) to \(2k-2\).

\begin{theorem}[\((2k-2)\)-Strong Path-Reporting Distance Oracle]
For every integer \(k\ge2\), there is a path-reporting distance oracle of
expected size \(O(k n^{1+1/k})\), query time \(O(k)\), and
\((2k-2)\)-strong stretch.
\end{theorem}

Thus, we match the \(O(k)\) query time and \(O(kn^{1+1/k})\) expected space of the classical TZ oracle. Our roundtrip stretch is \(4k-5\), whereas the path-reporting oracle obtained from the routing construction of Roditty, Thorup, and Zwick~\cite{RTZ08} has roundtrip stretch \(4k+\eps\) and query time \(O((k/\eps)\log(nW))\) for edge weights in \([1,W]\).

\paragraph{Compact routing.}
The same hierarchy also admits a distributed implementation.

\begin{theorem}[\((2k-2)\)-Strong Compact Routing]
For every integer \(k\ge2\), there is a compact routing scheme with
\((2k-2)\)-strong stretch, local tables of size \(\widetilde O(kn^{1/k})\) words, labels of \(\widetilde O(k)\) words and packet headers of \(O(\log^2 n/\log\log n)\) bits. The initial routing decision takes \(O(k)\) time, and each subsequent decision takes \(O(1)\) time.
\end{theorem}

This gives roundtrip stretch \(4k-5\), matching the stretch and local-space bounds of the TZ routing scheme for undirected graphs~\cite{TZ01routing}, and improving the \(4k+\varepsilon\) scheme of Roditty, Thorup, and Zwick~\cite{RTZ08}.  Our scheme also removes their dependence on \(\eps\) and the edge-weight range, applies to arbitrary nonnegative weights, and provides the stronger guarantee for each direction separately.

\paragraph{Improving PRDO query time.}
Our directed construction supports not only known improvements
to the undirected Thorup--Zwick construction, but also new ones.
We demonstrate this by improving PRDO query time in both settings.
For undirected graphs, Wulff--Nilsen~\cite{WulffNilsen13}
reduced the query time to \(O(\log k)\), while preserving
the \(2k-1\) stretch and \(O(kn^{1+1/k})\) space bounds.
We further reduce the query time to \(O(\log\log k)\)
without increasing either bound.
We then adapt the same query algorithm to our directed
construction, obtaining \(O(\log\log k)\) query time
with \((2k-2)\)-strong stretch.

\begin{theorem}
For every integer \(k\ge2\), every \(n\)-vertex graph
with nonnegative edge weights admits a path-reporting
distance oracle with expected size \(O(kn^{1+1/k})\)
and query time \(O(\log\log k)\), achieving
\begin{enumerate}
    \item \(2k-1\) stretch for undirected graphs;
    \item \((2k-2)\)-strong stretch for directed graphs.
\end{enumerate}
\end{theorem}

\paragraph{Source-wise and prioritized structures.}
Our framework also extends to source-wise and prioritized tree covers. When approximation guarantees are required only for pairs in \(S\times V\), the dependence on \(n^{1/k}\) in the overlap bound can be replaced by \(|S|^{1/k}\).

\begin{theorem}[Source-wise Strong Tree Cover]
For every integer \(k\ge1\), every directed graph \(G=(V,E)\) with
nonnegative edge weights and every nonempty source set \(S\subseteq V\)
admit a collection \(\mathcal T\) of tree--pairs such that, for every \((u,v)\in S\times V\) with finite roundtrip distance, some \(T\in\mathcal T\) satisfies
\[
\dist_T(u\to v)
\le \dist_G(u\to v)+(k-1)\dist_G(u\lra v).
\]
Every vertex belongs in expectation to \(O(k|S|^{1/k})\) tree--pairs.
\end{theorem}

Applying the theorem to both \(G\) and its reverse graph yields a source-wise \((2k-1)\)-roundtrip spanner with \(O(kn|S|^{1/k})\) expected edges. This improves the \(2k+\eps\) stretch and \(O((k^2/\eps)n|S|^{1/k}\log(nW))\) edge bound of Zhu and Lam~\cite{ZhuLam17}. Our construction also provides a strong tree cover with a per-vertex overlap bound, yielding source-wise path-reporting distance oracles.

We further obtain prioritized guarantees in the sense of Elkin, Filtser, and Neiman~\cite{EFN15}. Given a priority ordering \(x_1,\ldots,x_n\), we provide better stretch for pairs involving higher-priority vertices.

\begin{theorem}[Prioritized Strong Tree Cover]
For every integer \(k\ge4\), every directed graph \(G=(V,E)\) with
\(n\ge2\) vertices, nonnegative edge weights, and a priority ordering
\(x_1,\ldots,x_n\) admits a collection \(\mathcal T\) of tree--pairs
with the following guarantee.
For every ordered pair \((u,v)\) with finite roundtrip distance,
let \(x_j\) be its higher-priority endpoint and let
\(q_j=\max\{1,\lceil k\log j/\log n\rceil\}\).
Then some \(T\in\mathcal T\) satisfies
\[
\dist_T(u\to v)
\le \dist_G(u\to v)+(q_j-1)\dist_G(u\lra v).
\]
Every vertex belongs in expectation to \(O(k^2 n^{1/k})\) tree--pairs.
\end{theorem}

The resulting spanners, path-reporting distance oracles, and compact
routing schemes are presented in Section~\ref{sec:further-app}.

\subsection{Related Work}
\label{sec:related}

The work most directly related to our directed roundtrip constructions was discussed in the Introduction. Here we briefly review several adjacent lines of work relevant to the applications and extensions of our framework.

\paragraph{Distance oracles and path reporting.}
For undirected graphs, Wulff-Nilsen~\cite{WulffNilsen13} reduced the query time of the Thorup--Zwick oracle to \(O(\log k)\), while preserving its \(2k-1\) stretch, \(O(kn^{1+1/k})\) space, and path-reporting capability. For distance estimation, subsequent improvements by Chechik~\cite{Chechik14,Chechik15} achieved \(2k-1\) stretch, \(O(n^{1+1/k})\) space, and \(O(1)\) query time, without path reporting. A separate line of work developed smaller path-reporting
oracles, trading stretch and query time for space~\cite{ElkinPettie15,ElkinNeimanWulffNilsen16,ElkinShabat23,NeimanShabat24}. Chechik and Zhang~\cite{ChechikZhang24} obtained a path-reporting oracle with \(O(n^{1+1/k})\) space, \(O(k)\) stretch, and \(O(\log\log k)\) query time. All query bounds exclude the time needed to report the path.

\paragraph{Compact routing.}
For undirected graphs, Chechik~\cite{ChechikRouting13} obtained stretch approximately \(3.68k\) with a worst-case local-storage bound. Kadria and Roditty~\cite{KR25} reduced the stretch to approximately \(2.64k\) under an average local-storage bound: individual vertices may store more, while the total storage is \(\widetilde O(n^{1+1/k})\).

\paragraph{Roundtrip spanners and emulators.}
Zhu and Lam~\cite{ZhuLam18} derandomized the roundtrip spanner construction of Roditty, Thorup, and Zwick and improved its size to \(O((k/\varepsilon)n^{1+1/k}\log(nW))\) edges, retaining
stretch \(2k+\varepsilon\) for edge weights in \([1,W]\). A separate line of work focused on faster constructions. For integer edge weights, Pachocki et al.~\cite{Pachocki18} constructed roundtrip spanners with stretch \(O(k\log n)\) and \(\widetilde O(n^{1+1/k})\) edges in \(\widetilde O(mn^{1/k})\) time.
Chechik et al.~\cite{Chechik20} obtained deterministic constructions with stretch \(O(k\log\log n)\) and randomized constructions with stretch \(O(k\log k)\), both with \(\widetilde O(n^{1+1/k})\) edges and \(\widetilde O(m^{1+1/k})\) construction time. Chechik and Lifshitz~\cite{CL21} constructed a \(4\)-roundtrip spanner with \(O(n^{3/2})\) edges in \(O(n^2\log n)\) time. Allowing virtual edges outside the original graph leads to \emph{roundtrip emulators}. For every integer \(k\ge3\), Harbuzova et al.~\cite{HarbuzovaJinVWXu24} constructed a \((2k-1)\)-roundtrip emulator with \(O(kn^{1+1/k})\) expected edges in \(O(kn^2\log n)\) time.

\paragraph{Source-wise and prioritized structures.}
Nonuniform variants of the Thorup--Zwick framework have been studied extensively in undirected graphs. Roditty, Thorup, and
Zwick~\cite{RodittyTZ05} considered source-restricted distance oracles, while source-wise spanners were studied, among others, by Parter~\cite{Parter14}. Prioritized metric structures were introduced by
Elkin, Filtser, and Neiman~\cite{EFN15}, with subsequent work improving
source-wise and prioritized spanners and path-reporting distance
oracles~\cite{NeimanShabat24Pairwise}.

For directed graphs and a source set \(S\subseteq V\), Zhu and Lam~\cite{ZhuLam17} constructed source-wise roundtrip spanners with stretch \(2k+\varepsilon\) and \(O((k^2/\varepsilon)n|S|^{1/k}\log(nW))\) edges for edge weights in \([1,W]\). Zhu, Han, and Lam~\cite{ZhuHanLam21} obtained \(O(k\log n)\) stretch with \(O(n|S|^{1/k}\log^2 n)\) edges and \(O(m|S|^{1/k}\log^5 n)\) construction time. Stafford and Zhu~\cite{StaffordZhu23} later obtained a \(3\)-roundtrip source-wise spanner with \(\widetilde O(n\sqrt{|S|})\) expected edges.


\section{Preliminaries}
\label{sec:prelim}

Let \(G=(V,E)\) be a directed graph with nonnegative edge weights, where
\(n=|V|\) and \(m=|E|\). For a directed graph \(H\) and \(u,v\in V(H)\),
let \(\dist_H(u\to v)\) denote the shortest-path distance from \(u\) to
\(v\), with value \(\infty\) if \(v\) is unreachable from \(u\).
The \emph{roundtrip distance} is
\[
\dist_H(u\lra v)
=
\dist_H(u\to v)+\dist_H(v\to u).
\]
Equivalently, it is the length of a shortest directed closed walk containing \(u\) and \(v\). We refer to such a walk as the cycle \(u\lra v\).

For \(V'\subseteq V\), let \(G[V']\) denote the induced subgraph of \(G\). A \emph{strongly connected component} is a maximal set of mutually reachable vertices. We may assume that \(G\) is strongly connected, since every pair with finite roundtrip distance lies in one strongly connected component, and our constructions can be applied separately to each component.

\paragraph{Tree--pairs and tree covers.}
A \emph{tree--pair} \(T\) rooted at \(w\) is the union of an ingoing and an outgoing shortest-path tree rooted at \(w\). The two trees may have different vertex sets.

\begin{definition}[\(k\)-Roundtrip Tree Cover]
A collection \(\mathcal T\) of tree--pairs is a \(k\)-roundtrip tree cover
if, for every \(u,v\in V\), some \(T\in\mathcal T\) containing \(u,v\)
satisfies
\(
\dist_T(u\lra v)\le k\,\dist_G(u\lra v).
\)
The \textbf{overlap} of a vertex is the number of tree--pairs containing it.
\end{definition}

\paragraph{Strong stretch.}
A subgraph \(H\subseteq G\) provides \(k\)-strong stretch for an ordered pair
\((u,v)\) if
\[
\dist_H(u\to v)
\le
\dist_G(u\to v)+(k-1)\dist_G(u\lra v)= k\cdot\dist_G(u \to v) + (k-1)\dist_G(v \to u).
\]
A collection of tree--pairs is a \emph{\(k\)-strong tree cover} if every
ordered pair is \(k\)-strongly approximated by some tree--pair.
Applying the guarantee in both directions shows that \(k\)-strong stretch implies \((2k-1)\)-roundtrip stretch.

\paragraph{Spanners.}
A subgraph \(H\subseteq G\) is a \emph{\(k\)-roundtrip spanner} if
\(\dist_H(u\lra v)\le k\,\dist_G(u\lra v)\) for every \(u,v\in V\).
It is a \emph{\(k\)-strong spanner} if it provides \(k\)-strong stretch for
every ordered pair.

\paragraph{Path-reporting distance oracles.}
A \emph{path-reporting distance oracle} returns, for every ordered query
\((u,v)\), an explicit directed \(u\to v\) path in \(G\).
It has \(t\)-strong stretch if the reported path satisfies the strong-stretch
bound above. Unless stated otherwise, query time excludes the time required
to output the path.

\subsection{The Thorup--Zwick Blueprint}

Thorup and Zwick (TZ)~\cite{TZ05} developed a hierarchical sampling framework underlying several fundamental constructions for undirected weighted graphs. For our purposes, it is most useful to view this framework as a low-overlap tree cover.
The construction samples a nested hierarchy
\[
A_0=V\supseteq A_1\supseteq\cdots\supseteq A_k=\emptyset,
\]
where every vertex of \(A_{i-1}\) is included independently in \(A_i\) with probability \(n^{-1/k}\).
For every \(v\in V\), let
\(
p_i(v)=\arg\min_{w\in A_i}\dist_G(v,w), \,
R_i(v)=\dist_G(v,p_i(v)),
\)
and set \(R_k(v)=\infty\).
Each center \(w\in A_i\setminus A_{i+1}\) defines the cluster
\[
C(w,R_{i+1})
=
\{v\in V:\dist_G(w,v)<R_{i+1}(v)\}.
\]
Thus, \(v\in C(w,R_{i+1})\) precisely when \(w\) is closer to \(v\) than its nearest center in \(A_{i+1}\).
For each such center, TZ constructs a shortest-path tree \(T(w,i+1)\) spanning its cluster.

\begin{theorem}[Thorup--Zwick tree cover~\cite{TZ05}]
\label{thm:tz-tree-cover}
The trees \(T(w,i+1)\), over all
\(0\le i\le k-1\) and \(w\in A_i\setminus A_{i+1}\), form a
\((2k-1)\)-approximate tree cover: for every \(u,v\in V\), some tree \(T\) containing both vertices satisfies
\[
\dist_T(u,v)\le(2k-1)\dist_G(u,v).
\]
Every vertex belongs in expectation to \(O(k n^{1/k})\) trees, and the cover can be constructed in \(O(kmn^{1/k})\) expected time.
\end{theorem}

Two properties of the construction will be crucial for us.
First, every cluster \(C(w,R_{i+1})\) is closed under shortest paths from \(w\): if \(v\in C(w,R_{i+1})\), then a shortest \(w\)-to-\(v\) path lies entirely in the cluster. Hence one shortest-path tree rooted at \(w\) simultaneously preserves its distances to all vertices in the cluster.

Second, the random sampling implies that every vertex belongs in
expectation to only \(O(n^{1/k})\) clusters per level. Since each tree spans its cluster without adding vertices, the same overlap bound holds for the trees.

\section{Technical Overview}
\label{tech-overview}

\subsection{Why the Naive Thorup-Zwick Adaptation Fails}

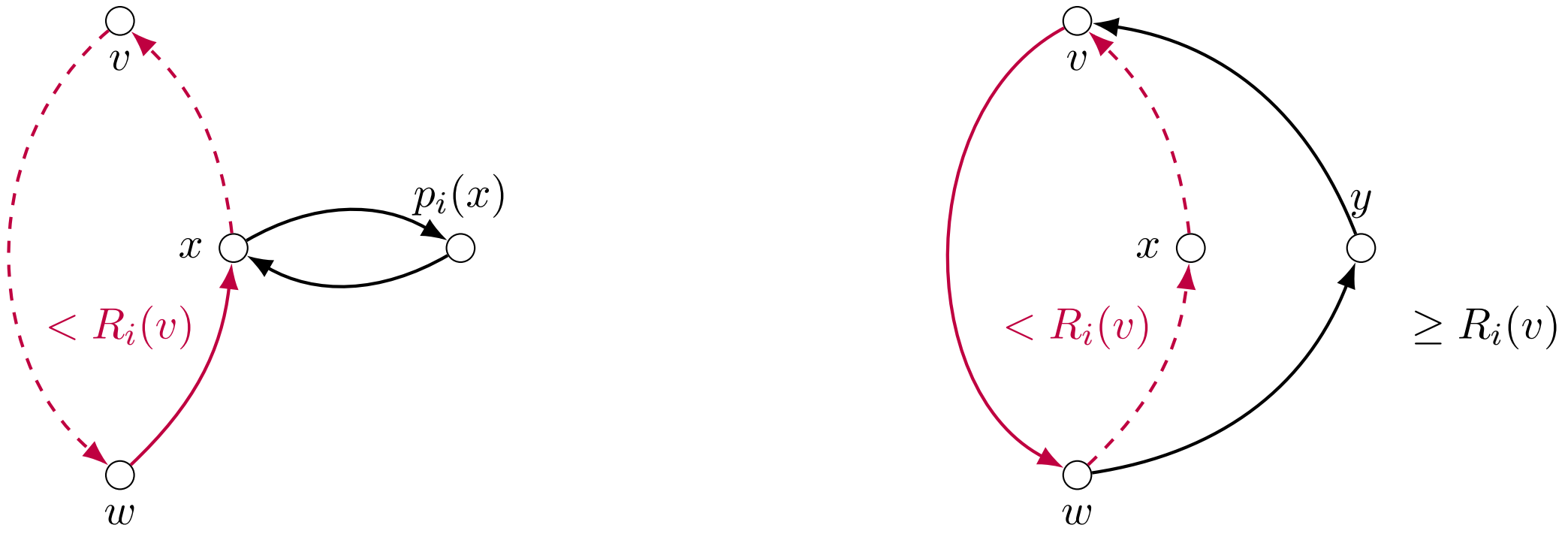
\begin{figure}[t]
\centering

\begin{subfigure}[t]{0.47\textwidth}
\centering
\begin{tikzpicture}[
 vertex/.style={circle, draw, minimum size=1mm, inner sep=0pt},
 directed/.style={->, >=Latex, thick}
]

\node[circle,draw,inner sep=2.5pt,label=below:$w$] (w) at (0,-2) {};
\node[circle,draw,inner sep=2.5pt,label=left:$x$] (x) at (1,0) {};
\node[circle,draw,inner sep=2.5pt,label=below:$v$] (v) at (0,2) {};
\node[circle,draw,inner sep=2.5pt,label=above:$p_i(x)$] (p) at (3,0) {};

\node[purple] at (0,-0.7) {\( < R_i(v) \)};

\draw[directed, purple, bend right=20] (w) to (x);
\draw[directed, dashed, purple, bend right=20] (x) to (v);
\draw[directed, dashed, purple, bend right=50] (v) to (w);
\draw[directed, bend left=30] (x) to (p);
\draw[directed, bend left=30] (p) to (x);

\end{tikzpicture}
\caption{$C(w,R_i)$ may contain $v$ even though $v$ is unreachable from $w$: 
$R_i(v) > \dist_G(w\!\lra\!v)$, but the shortest $w\!\to\!v$ path passes through a vertex $x$ for which $R_i(x) \le \dist_G(w\!\lra\!x)$, thus $x\notin C(w,R_i)$.}
\label{fig:cluster-fail}
\end{subfigure}
\hfill
\begin{subfigure}[t]{0.47\textwidth}
\centering
\begin{tikzpicture}[
 vertex/.style={circle, draw, minimum size=1mm, inner sep=0pt},
 directed/.style={->, >=Latex, thick}
]

\node[circle,draw,inner sep=2.5pt,label=below:$w$] (w) at (0,-2) {};
\node[circle,draw,inner sep=2.5pt,label=left:$x$] (x) at (1,0) {};
\node[circle,draw,inner sep=2.5pt,label=below:$v$] (v) at (0,2) {};
\node[circle,draw,inner sep=2.5pt,label=above:$y$] (y) at (2.5,0) {};

\node[purple] at (0,-0.7) {\( < R_i(v) \)};
\node at (3.6,-0.7) {\( \ge R_i(v) \)};

\draw[directed, dashed, purple, bend right=20] (w) to (x);
\draw[directed, dashed, purple, bend right=20] (x) to (v);
\draw[directed, bend right=30] (w) to (y);
\draw[directed, bend right=30] (y) to (v);
\draw[directed, purple, bend right=62] (v) to (w);

\end{tikzpicture}
\caption{While $R_i(v) > \dist_G(w\!\lra\!v)$, the vertex $x \in w\!\to\!v$ has $R_i(x)\le \dist_G(w\!\lra\!x)$ and is not in $C(w,R_i)$. 
Inside the remaining cluster, the shortest roundtrip $w\!\lra\!v$ (via $y$) is longer than $R_i(v)$.}
\label{fig:SCC-fail}
\end{subfigure}

\caption{Failures of naïve cluster definition.}
\label{fig:cluster-and-scc-failures}
\end{figure}

A natural attempt to extend the Thorup-Zwick hierarchy to directed graphs is to substitute undirected distances with roundtrip distances. For every vertex $v \in V$ and level $i$, define the pivot and radius by$$ p_i(v) = \arg\min_{w \in A_i} \dist_G(v \lra w) \quad \text{and} \quad R_i(v) = \dist_G(v \lra p_i(v)), $$and for each center $w \in A_i \setminus A_{i+1}$, define the cluster$$C(w, R_{i+1}) = \{v \in V : \dist_G(w \lra v) < R_{i+1}(v)\}.$$At first glance, this mirrors the undirected Thorup-Zwick construction. The fundamental obstacle, however, is that these clusters are not closed under the shortest paths that witness their roundtrip distances. Specifically, even if $v \in C(w, R_{i+1})$, a directed shortest path from $w$ to $v$ (or from $v$ to $w$) may traverse an intermediate vertex $x$ satisfying$$R_{i+1}(x) \le \dist_G(w \lra x).$$Such a vertex $x$ falls outside $C(w, R_{i+1})$, even though it lies directly on a shortest roundtrip path between the center and a cluster member. As a result, the roundtrip distance between $w$ and $v$ is not necessarily preserved in the induced subgraph $G[C(w, R_{i+1})]$. The internal distance $\dist_{G[C(w, R_{i+1})]}(w \lra v)$ can strictly exceed $\dist_G(w \lra v)$, and may even be infinite (see Figure~\ref{fig:cluster-and-scc-failures}). Therefore, this naive clustering fails to directly yield valid trees. Conversely, naively taking the closure by adding all shortest path vertices to the clusters breaks the expected overlap bound.

\subsection{Strict Clusters and the $2k$-Roundtrip Tree Cover}

To repair the issue above, we iteratively remove vertices whose radius
constraint cannot be satisfied using paths that remain inside the cluster.
Starting from \(C(w,R_{i+1})\), we compute the roundtrip distances from \(w\)
inside the induced subgraph. Any vertex \(v\) for which this distance is at
least \(R_{i+1}(v)\) is removed, and the distances are recomputed.
We continue until no such vertex remains. The resulting set is the \emph{strict cluster} \(\C(w,R_{i+1})\), and the precise procedure appears in Section~\ref{sec:2k-tree-cover}.

Equivalently, \(\C(w,R_{i+1})\) is the largest subset
\(S\subseteq C(w,R_{i+1})\) such that every \(v\in S\) satisfies
\[
\dist_{G[S]}(w\lra v)<R_{i+1}(v).
\]
For every center \(w\in A_i\setminus A_{i+1}\), we construct an ingoing and an outgoing shortest-path tree rooted at \(w\) in
\(G[\C(w,R_{i+1})]\). Their union is the tree-pair \(T(w,i+1)\),
and our cover consists of all of them.

Crucially, every surviving vertex has a roundtrip distance to the center smaller than its radius at the next level. We update the radii using the roundtrip distances in the tree-pairs we
construct: \(R_i(v)\) is the smallest roundtrip distance \emph{found} between \(v\) and a center \(w\in A_j\setminus A_{j+1}\), over all \(i\le j\), measured inside \(T(w,j+1)\).

This roughly explains the low overlap: if there were many more than \(n^{1/k}\) centers in \(A_i\) whose strict clusters contain \(v\), we would expect one of them to have been sampled into \(A_{i+1}\). Call this center \(w\). Since \(R_{i+1}\le R_{i+2}\), we have \(\C(w,R_{i+1})\subseteq\C(w,R_{i+2})\).
Thus, the tree-pair constructed for \(w\) at the higher level would give \(R_{i+1}(v)\) at most the roundtrip distance between \(v\) and \(w\) inside \(\C(w,R_{i+1})\). But membership in \(\C(w,R_{i+1})\) requires this distance to be strictly smaller than \(R_{i+1}(v)\), a contradiction.

The stretch analysis follows the same general idea as the undirected TZ argument, but instead of tracking only the radii of \(u\) and \(v\), we track the radii of all vertices on the cycle \(u\lra v\).

Let \(r\) be the length of \(u\lra v\). We find a level \(i\) and a vertex \(x\in u\lra v\) such that \(R_i(x)\le ir\), while \(R_{i+1}(y)>(i+1)r\) for every \(y\in u\lra v\). The pivot \(w=p_i(x)\) is connected to \(x\) by a roundtrip of length at most \(ir\) inside its strict cluster. Together with \(u\lra v\), this gives a roundtrip from \(w\) to every \(y\in u\lra v\) of length at most \((i+1)r<R_{i+1}(y)\), which means that every $y$ survives the pruning of the strict cluster. Thus $y \in \C(w,R_{i+1})$.

\begin{figure}[t]
\centering
\begin{tikzpicture}[directed/.style={->, >=Latex, thick}]

\definecolor{cyclecolor}{RGB}{35,105,170}
\definecolor{pivotcolor}{RGB}{155,65,140}

\node[circle,draw,fill=white,inner sep=2.5pt,label=below:$u$] (u) at (-2,-1) {};
\node[circle,draw,fill=white,inner sep=2.5pt,label=below:$x$] (x) at (0,0) {};
\node[circle,draw,fill=white,inner sep=2.5pt,label=below:$v$] (v) at (2,-1) {};
\node[circle,draw,fill=white,inner sep=2.5pt,label=above:$p_i(x)$] (p) at (0,2) {};
\node[circle,draw,fill=white,inner sep=2.5pt,label=above:$y$] (y) at (0.6,-2) {};

\draw[directed, cyclecolor, bend left=20] (u) to (x);
\draw[directed, cyclecolor, bend left=20] (x) to (v);
\draw[directed, cyclecolor, bend left=15] (v) to (y);
\draw[directed, cyclecolor, bend left=25] (y) to (u);

\draw[directed, pivotcolor, bend left=20] (x) to (p);
\draw[directed, pivotcolor, bend left=20] (p) to (x);

\node[pivotcolor] at (-1.3,1.0) {\(R_i(x) \le ir\)};
\node[cyclecolor] at (0,-1.0) {\(\dist_G(u \lra v) \le r\)};

\node[anchor=west,align=center] at (2.1,-0.6) {
  Roundtrip from \(p_i(x)\) through \(y\):\\[5pt]
  \(\textcolor{pivotcolor}{R_i(x)}
    +\textcolor{cyclecolor}{r}\)\\[3pt]
  \(\le (i+1)r<R_{i+1}(y)\)
};

\end{tikzpicture}
\caption{
The tree paths between \(p_i(x)\) and \(x\), together with the cycle \(u\lra v\), have total length at most \((i+1)r<R_{i+1}(y)\) for every cycle vertex \(y\). Thus the entire cycle survives in \(\C(p_i(x),R_{i+1})\).
}
\label{fig:strict-cluster-cycle}
\end{figure}
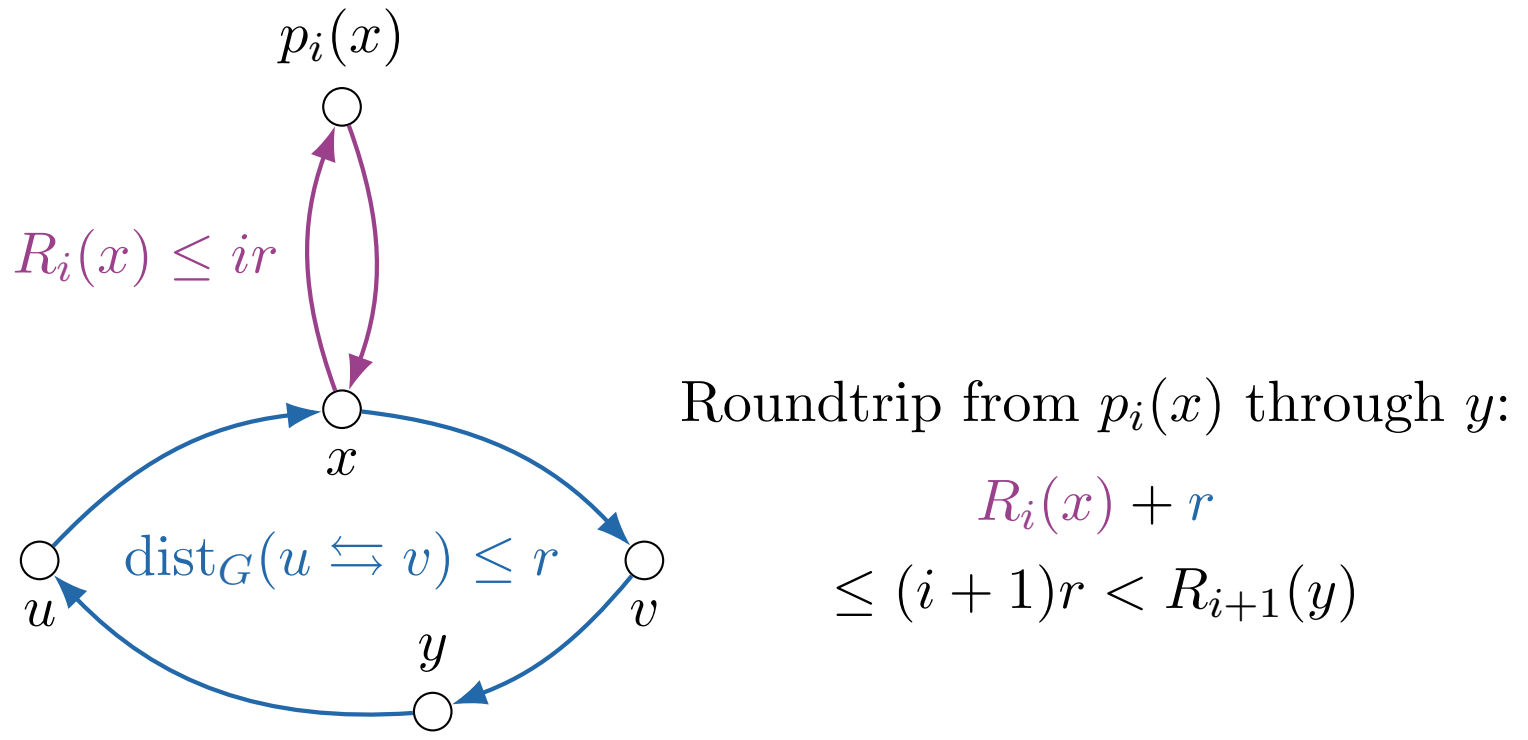

In the resulting tree--pair, both \(u\) and \(v\) have roundtrip distance at most \((i+1)r\le kr\) to \(w\), giving a roundtrip
between them of length at most \(2kr\). See Figure~\ref{fig:strict-cluster-cycle}.

\subsection{From Roundtrip to Strong Tree Covers}

The \(2k\)-roundtrip construction approximates \(u\) and \(v\) by capturing an entire shortest cycle containing them inside one strict cluster.
For strong stretch we need less: for an ordered pair \((u,v)\), we only need to preserve the directed path \(u\to v\); the return path \(v\to u\) may be handled by a different tree--pair.

To see why this distinction helps, let \(r=\dist_G(u\lra v)\), and suppose that some \(x\in u\to v\) satisfies \(R_i(x)\le ir\).
If every vertex on the entire cycle has level-\((i+1)\) radius larger than \((i+1)r\), the strict-cluster argument already captures the whole
cycle. The new case is when all vertices on \(u\to v\) have large radius at level \(i+1\), but a small-radius vertex appears on the return path \(v\to u\). Such a vertex may prevent the cycle from surviving in \(\C(p_i(x),R_{i+1})\), even though it should not prevent us from preserving the path \(u\to v\).

This motivates using the connectivity of a larger strict cluster while retaining the smaller radius threshold for membership.
For \(i\le k-2\) and a center \(w\in A_i\setminus A_{i+1}\), let
\(T(w,i+2)\) be the tree--pair of \(\C(w,R_{i+2})\), and define the \emph{expanded strict cluster}
\[
\s(w,R_{i+1})
=
\left\{
v \in\C(w,R_{i+2})
\;\middle|\;
\dist_{T(w,i+2)}(w\lra v)<R_{i+1}(v)
\right\}.
\]
We let \(\T(w,i+1)\) be the tree--pair obtained from ingoing and outgoing
shortest-path trees rooted at \(w\) in \(G[\s(w,R_{i+1})]\).
No further pruning is performed, even if some vertices lose one direction of their connection to the center.

The expanded cluster lies between two consecutive strict clusters:
\[
\C(w,R_{i+1})
\subseteq
\s(w,R_{i+1})
\subseteq
\C(w,R_{i+2}).
\]
Thus, \(\C(w,R_{i+2})\) supplies additional connectivity, whereas
\(R_{i+1}\) still determines which vertices are included in the tree--pair
added at level \(i\).
The radii are updated using the roundtrip distances in these expanded
tree--pairs.

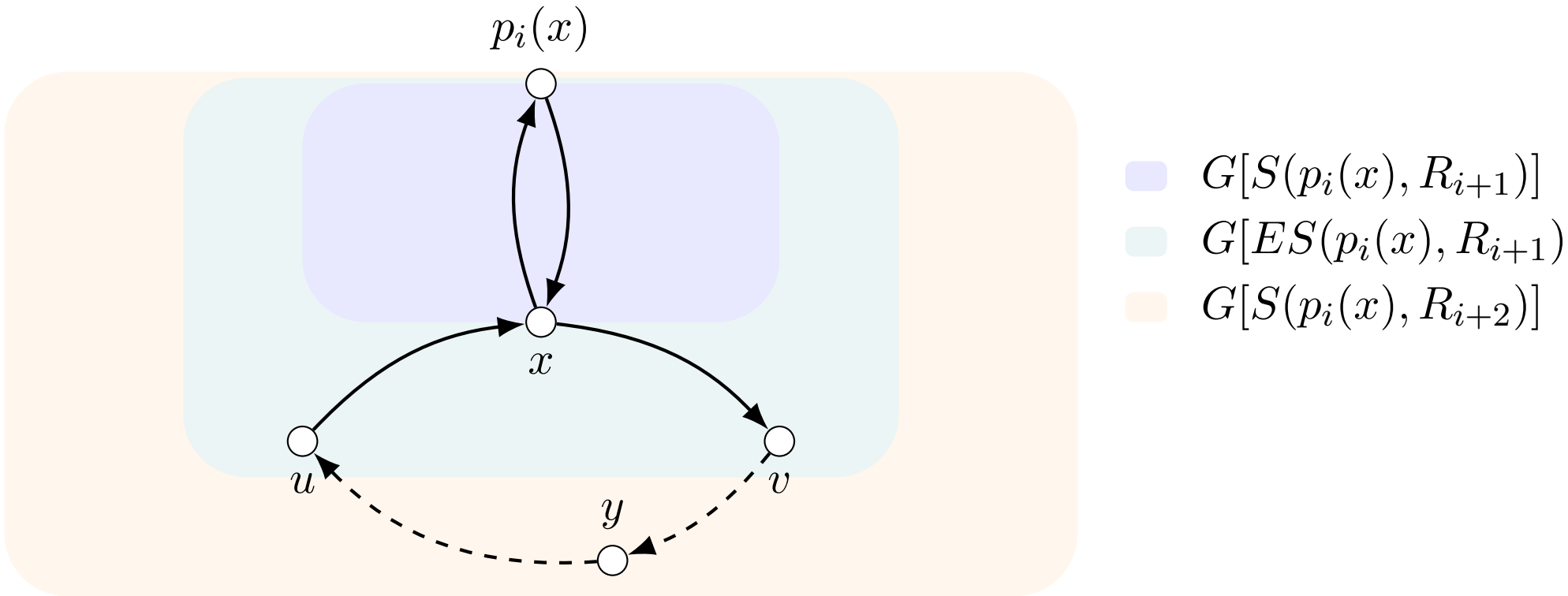
\begin{figure}[h]
\centering
\begin{tikzpicture}[
 directed/.style={->, >=Latex, thick}
]

\fill[orange!7, rounded corners=15pt]
 (-4.5,-2.3) rectangle (4.5,2.1);

\fill[teal!8, rounded corners=15pt]
 (-3,-1.3) rectangle (3,2.05);

\fill[blue!9, rounded corners=15pt]
 (-2,0) rectangle (2,2);

\node[circle,draw,fill=white,inner sep=2.5pt,label=below:$u$] (u) at (-2,-1) {};
\node[circle,draw,fill=white,inner sep=2.5pt,label=below:$x$] (x) at (0,0) {};
\node[circle,draw,fill=white,inner sep=2.5pt,label=below:$v$] (v) at (2,-1) {};
\node[circle,draw,fill=white,inner sep=2.5pt,label=above:$p_i(x)$] (p) at (0,2) {};
\node[circle,draw,fill=white,inner sep=2.5pt,label=above:$y$] (y) at (0.6,-2) {};

\draw[directed, bend left=20] (u) to (x);
\draw[directed, bend left=20] (x) to (v);
\draw[directed, dashed, bend left=15] (v) to (y);
\draw[directed, dashed, bend left=25] (y) to (u);

\draw[directed, bend left=20] (x) to (p);
\draw[directed, bend left=20] (p) to (x);

\fill[blue!9, rounded corners=2pt]
 (4.9,1.1) rectangle (5.25,1.35);
\node[anchor=west] at (5.4,1.225)
 {$G[\C(p_i(x),R_{i+1})]$};

\fill[teal!8, rounded corners=2pt]
 (4.9,0.55) rectangle (5.25,0.8);
\node[anchor=west] at (5.4,0.675)
 {$G[\s(p_i(x),R_{i+1})]$};

\fill[orange!7, rounded corners=2pt]
 (4.9,0) rectangle (5.25,0.25);
\node[anchor=west] at (5.4,0.125)
 {$G[\C(p_i(x),R_{i+2})]$};

\end{tikzpicture}
\caption{
Expanded strict clusters separate the two roles of the radii.
A small-radius vertex \(y\) on the return path \(v\to u\) may prevent the
whole cycle from lying in \(\C(p_i(x),R_{i+1})\), while the larger strict
cluster \(\C(p_i(x),R_{i+2})\) still supplies the connectivity needed to
include \(u\to v\) in \(\s(p_i(x),R_{i+1})\).
}
\label{fig:expanded-strict-technical-overview}
\end{figure}

\paragraph{Strong stretch.}
As in the \(2k\)-roundtrip analysis, we track a vertex \(x\) on the cycle with \(R_i(x)\le ir\). If \(x\) lies on \(u\to v\), we only need this directed path to survive in the expanded cluster, rather than the entire cycle. We can then follow the path from \(u\) to \(v\), taking a detour from \(x\) to its pivot and back. This detour costs at most \(ir\), giving a walk of length at most \(\dist_G(u\to v)+ir\).

If \(x\) lies on the return path, visiting \(x\) before finishing at \(v\) requires an extra traversal of the cycle, of length \(r\).
We therefore require the smaller bound \(R_i(x)\le(i-1)r\) in this case: the extra traversal and the detour together still cost at most \(ir\).

We show that, at some level \(i<k\), an expanded cluster contains one of these walks. Its ingoing and outgoing shortest-path trees then give a route through the pivot of no greater length, yielding the \(k\)-strong guarantee.

\paragraph{Expected overlap.}
The overlap analysis is very similar to that of the \(2k\)-roundtrip
cover. If many more than \(n^{1/k}\) centers in \(A_i\) had expanded
clusters containing \(v\), we would expect one of them, say \(w\),
to have been sampled into \(A_{i+1}\).
Since \(\C(w,R_{i+2})\subseteq\s(w,R_{i+2})\), the tree-pair constructed for \(w\) at the higher level provides a roundtrip to \(v\) no longer than the one used to test membership in \(\s(w,R_{i+1})\).
This would make \(R_{i+1}(v)\) at most that distance, contradicting
the strict inequality required for membership.
Thus, the same sampling intuition gives \(O(n^{1/k})\) expected overlap per level, and \(O(kn^{1/k})\) overall.

\subsection{Making the Construction Fast}

\paragraph{Fast clusters.} The bottleneck in the constructions above is computing strict clusters.
Naively, after removing even a single violating vertex, we may need to recompute the ingoing and outgoing shortest-path trees. Thus, one strict cluster may require \(O(n)\) Dijkstra computations and
\(\widetilde O(mn)\) time, leading to \(\widetilde O(mn^2)\) time overall.

The overlap argument relies on two properties: (1) the clusters are
nested across radius levels, and (2) every retained vertex satisfies
its radius constraint when distances are measured in the cluster
one level higher. Crucially, it does not require this constraint to
remain satisfied after pruning.

We call our new clusters \emph{fast clusters}.
To ensure (1), we construct them recursively, starting with
\(\fc(w,R_k)=V\). To ensure (2), we remove vertices that violate their radius constraints, but perform only \emph{one pruning round} at each radius level:
\[
\fc(w,R_{i+1})
=
\left\{
v\in\fc(w,R_{i+2}) :
\dist_{\ft(w,i+2)}(w\lra v)<R_{i+1}(v)
\right\},
\]
where \(\ft(w,i+2)\) consists of ingoing and outgoing shortest-path
trees rooted at \(w\) in \(G[\fc(w,R_{i+2})]\).
We then recompute the trees in the remaining induced subgraph,
without repeating the pruning at the same radius level.

The stretch argument is similar to that for the \(k\)-strong tree cover.
For the overlap analysis, the resulting distances may exceed the radius constraints, but the argument uses the distances before pruning. Indeed, if \(w\) is chosen as a center at level \(i+1\), its tree-pair is exactly \(\ft(w,i+2)\), so the same distances used in the membership test give an upper bound on \(R_{i+1}(v)\).

Each center requires only \(O(k)\) pairs of shortest-path searches,
reducing the total construction time to \(\widetilde O(kmn)\).

\paragraph{Sparsification.}
We further accelerate the construction using the sparsification technique of Chechik and Lifshitz, applied at levels \(i>k/2\).
If an edge of a cycle is deleted, the center responsible for its deletion allows us to reroute through it with at most a factor-two increase in the cycle length. Since this center belongs to a high level, the stretch analysis can restart there, and the smaller number of remaining levels compensates for this increase.
The resulting sparsity, together with a separate treatment of level \(0\), gives \(\widetilde O(n^{2.5})\) construction time for \(k\ge4\). Combining this with the preceding construction yields \(\widetilde O(\min\{mn,n^{2.5}\})\) preprocessing time.

\subsection{Applications of the Tree Cover}

Taking the union of the tree--pairs gives a \(k\)-strong spanner, while searching the tree--pairs shared by the query vertices gives a \(k\)-strong path-reporting distance oracle.
To obtain \(O(k)\) query time, we adapt the TZ pivot search:
we examine the destination's pivots and handle level \(0\) separately, obtaining \((2k-2)\)-strong stretch.

To reduce the query time further, we first obtain an
\(O(k)\)-approximate numerical distance estimate in constant time.
Using this estimate, we restrict the search to a sequence of \(O(\log k)\) levels that suffices to find a path with the required stretch.
Binary search over this sequence, with constant-time access to each level, gives \(O(\log\log k)\) query time.

For compact routing, we must also control the space used at each vertex.
We adapt the iterative sampling technique of Thorup and
Zwick~\cite{TZ05}, also used for directed routing by Roditty, Thorup, and Zwick~\cite{RTZ08}. Additional sampling rounds shrink oversized routing dictionaries while providing alternative routes through new tree--pairs.

Following the source-restricted framework of Roditty, Thorup, and
Zwick~\cite{RodittyTZ05}, we obtain source-wise covers by starting
the hierarchy from \(S\) and sampling at rate \(|S|^{-1/k}\).
For prioritized guarantees, following the framework of Elkin,
Filtser, and Neiman~\cite{EFN15}, we combine source-wise covers for
prefixes of the priority ordering, using different stretch parameters and applying the constructions in both orientations of the graph.

\section{Strict Clusters and the \(2k\)-Roundtrip Tree Cover}
\label{sec:2k-tree-cover}

In this section, we give the precise construction underlying Theorem~\ref{thm:intro-2k-tree-cover} and prove its \(2k\)-roundtrip stretch and \(O(k n^{1/k})\) expected-overlap guarantees.

\subsection{Construction}

As in TZ, we sample a nested hierarchy
\[
A_0=V\supseteq A_1\supseteq\cdots\supseteq A_k=\emptyset,
\]
where every vertex of \(A_{i-1}\) is included independently in \(A_i\) with probability \(n^{-1/k}\). We initialize \(R_k(v)=\infty\) for every \(v\in V\), and process the levels from \(k-1\) down to \(0\).

Suppose that level \(i\) is being processed, so that the radius function \(R_{i+1}\) is already defined. For every center \(w\in A_i\setminus A_{i+1}\), define its \emph{cluster}
\[
C(w,R_{i+1})
=
\left\{
v\in V :
\dist_G(w\lra v)<R_{i+1}(v)
\right\}.
\]
As discussed in Section~\ref{tech-overview}, this set might not preserve the paths witnessing these roundtrip distances. We therefore replace it by a strict cluster.

\paragraph{Strict clusters.}
Starting from \(C(w,R_{i+1})\), repeatedly compute the ingoing and outgoing shortest-path trees rooted at \(w\) in the current induced subgraph and remove every vertex \(v\) whose internal roundtrip distance to \(w\) is at least \(R_{i+1}(v)\). The process terminates when every remaining vertex satisfies its radius constraint. We denote the resulting set by \(\C(w,R_{i+1})\). Algorithm~\ref{alg:strictC} gives the precise procedure.

\begin{algorithm}[H] \label{alg:strictC}
\SetAlgoLined
\KwIn{Graph \(G(V,E)\), radius values \(R_{i+1}(\cdot)\), vertex \(w\), and initial cluster \(C(w,R_{i+1})\).}
\KwOut{The strict cluster of \(w\) with respect to \(R_{i+1}\).}

\(X\gets C(w,R_{i+1})\)\;

Run both ingoing and outgoing Dijkstra searches from \(w\)
within \(G[X]\), and let \(T\) be the union of the resulting trees.\\[1mm]

\While{\(\exists v \in X\) such that
\(\dist_T(v \lra w) \ge R_{i+1}(v)\)}{
    Remove \(v\) from \(X\).\\
    Recompute \(T\) in the new \(G[X]\).
}
\Return{\(X\)}\;

\caption{\(\textsc{StrictC}(G, R_{i+1}, w, C(w,R_{i+1}))\):
Computation of the strict cluster.}
\end{algorithm}

\vspace{1pt}

For every \(w\in A_i\setminus A_{i+1}\), after computing
\(\C(w,R_{i+1})\), we compute an ingoing and an outgoing shortest-path tree
rooted at \(w\) in \(G[\C(w,R_{i+1})]\). Their union is denoted by
\(T(w,i+1)\).

After all level-\(i\) tree--pairs have been constructed, define
\[
R_i(v)
=
\min\left\{
R_{i+1}(v),
\min_{w\in A_i\setminus A_{i+1}}
\dist_{T(w,i+1)}(v\lra w)
\right\},
\]
where the second distance is infinite when
\(v\notin\C(w,R_{i+1})\).
If the second term determines \(R_i(v)\), we denote a minimizing center by
\(p_i(v)\) and call it the \(i\)-pivot of \(v\).

The output is the collection
\(
\mathcal T
=
\left\{
T(w,i+1):
0\le i\le k-1,\;
w\in A_i\setminus A_{i+1}
\right\}.
\)
Algorithm~\ref{alg:2k-tree-cover} summarizes the construction.

\begin{algorithm}[H]
\SetAlgoLined
\SetAlgoNoEnd
\KwIn{Directed graph \(G=(V,E)\), integer \(k\ge 1\).}
\KwOut{A collection \(\mathcal T\) of tree--pairs.}

\(A_0 \gets V,\quad A_k \gets \emptyset,\quad \mathcal T \gets \emptyset\)\;
\(R_k(v) \gets \infty\) for every \(v\in V\)\;

\For{\(i\gets 1\) \KwTo \(k-1\)}{
    Include each vertex of \(A_{i-1}\) in \(A_i\) independently
    with probability \(n^{-1/k}\)\;
}

\For{\(i\gets k-1\) \KwTo \(0\)}{
    \ForEach{\(w\in A_i\setminus A_{i+1}\)}{
        Compute the preliminary cluster \(C(w,R_{i+1})\)\;
        Compute
        \(\C(w,R_{i+1})
          \gets
          \textsc{StrictC}(G,R_{i+1},w,C(w,R_{i+1}))\)\;
        Compute ingoing and outgoing shortest-path trees rooted at \(w\)
        in \(G[\C(w,R_{i+1})]\)\;
        Let \(T(w,i+1)\) be their union and add it to \(\mathcal T\)\;
    }

    \ForEach{\(v\in V\)}{
        \(R_i(v)\gets
        \min\left\{
            R_{i+1}(v),
            \displaystyle
            \min_{w\in A_i\setminus A_{i+1}}
            \dist_{T(w,i+1)}(v\lra w)
        \right\}\)\;
    }
}

\Return \(\mathcal T\)\;

\caption{\(\textsc{2k-TreeCover}(G,k)\): construction of the
\(2k\)-roundtrip tree cover.}
\label{alg:2k-tree-cover}
\end{algorithm}

\subsection{Stretch}

\begin{lemma}
\label{lem:2k-stretch}
The collection \(\mathcal T\) is a \(2k\)-roundtrip tree cover.
Namely, for every \(u,v\in V\), there exists a tree--pair \(T\in\mathcal T\) containing \(u\) and \(v\) such that
\(
\dist_T(u\lra v)\le 2k\,\dist_G(u\lra v).
\)
\end{lemma}

\begin{proof}
Fix \(u,v\in V\), let \(r=\dist_G(u\lra v)\), and fix a shortest cycle \(u\lra v\) of length \(r\).
For every level \(i\), let \(m_i=\min_{y\in u\lra v}R_i(y)\).

Every vertex is itself a center at some level, and from that level downward its radius remains zero. Hence \(m_0=0\), while \(m_k=\infty\).
Let \(i\) be the largest index such that \(m_i\le ir\).
Then \(i\le k-1\) and \(m_{i+1}>(i+1)r\).

Choose \(x\in u\lra v\) with \(R_i(x)=m_i\).
Thus \(R_i(x)\le ir\), whereas \(R_{i+1}(y)>(i+1)r\) for every \(y\in u\lra v\). 
In particular, \(R_i(x)<R_{i+1}(x)\), so \(x\) has an \(i\)-pivot \(w=p_i(x)\), and
\[
\dist_{T(w,i+1)}(w\lra x)=R_i(x)\le ir.
\]

We claim that \(u\lra v\subseteq\C(w,R_{i+1})\).
First, for every \(y\in u\lra v\), the cycle \(u\lra v\) gives
\(\dist_G(x\lra y)\le r\), and therefore
\[
\dist_G(w\lra y)
\le
\dist_{T(w,i+1)}(w\lra x)+\dist_G(x\lra y)
\le
(i+1)r
<
R_{i+1}(y).
\]
Hence \(u\lra v\subseteq C(w,R_{i+1})\).

Suppose, for contradiction, that some vertex of \(u\lra v\) is removed
while constructing the strict cluster, and let \(y\) be the first
such vertex.
At the moment \(y\) is removed, every vertex of \(u\lra v\) is still
present. Moreover, the paths between \(w\) and \(x\) in \(T(w,i+1)\)
lie in the final strict cluster, and therefore are also still present.
Together with the appropriate portions of \(u\lra v\), these paths
give an internal roundtrip from \(w\) to \(y\) of length at most
\[
\dist_{T(w,i+1)}(w\lra x)+r
\le (i+1)r<R_{i+1}(y),
\]
contradicting the pruning rule.
Thus \(u\lra v\subseteq\C(w,R_{i+1})\).

It remains to bound the distance in the resulting tree--pair.
For every \(y\in u\lra v\), the induced subgraph
\(G[\C(w,R_{i+1})]\) contains a roundtrip from \(w\) to \(y\) of
length at most \((i+1)r\).
Since \(T(w,i+1)\) consists of ingoing and outgoing shortest-path
trees in this induced subgraph,
\(\dist_{T(w,i+1)}(w\lra y)\le(i+1)r\).
In particular,
\[
\dist_{T(w,i+1)}(u\lra v)
\le
\dist_{T(w,i+1)}(u\lra w)
+
\dist_{T(w,i+1)}(w\lra v)
\le
2(i+1)r
\le
2kr.
\]
Thus \(T(w,i+1)\) contains \(u\) and \(v\) and satisfies the claimed stretch bound. See Figure~\ref{fig:strict-cluster-cycle}.
\end{proof}

\subsection{Overlap}

For every vertex \(v\) and level \(i\), define its level-\(i\)
\emph{strict bunch} by
\[
B_i(v)
=
\{\,w\in A_i\setminus A_{i+1}
    : v\in\C(w,R_{i+1})\,\},
\]
and let \(B(v)=\bigcup_{i=0}^{k-1}B_i(v)\).
By construction, \(w\in B_i(v)\) exactly when the tree--pair
\(T(w,i+1)\) contains \(v\). Thus, the overlap of \(v\) is \(|B(v)|\).

We will use the following monotonicity property of strict clusters.

\begin{claim}
\label{clm:strict-monotonicity}
For every \(w\in V\) and \(i<k\),
\(
\C(w,R_i)\subseteq\C(w,R_{i+1}).
\)
\end{claim}

\begin{proof}
By construction, \(R_i(z)\le R_{i+1}(z)\) for every \(z\in V\), and hence
\(C(w,R_i)\subseteq C(w,R_{i+1})\).

Suppose that some vertex of \(\C(w,R_i)\) is removed while constructing
\(\C(w,R_{i+1})\), and let \(z\) be the first such vertex. At the moment
\(z\) is removed, every vertex of \(\C(w,R_i)\) is still present.
Therefore, if \(C'\) is the current cluster,
\[
\dist_{G[C']}(w\lra z)
\le
\dist_{G[\C(w,R_i)]}(w\lra z)
<
R_i(z)
\le
R_{i+1}(z),
\]
contradicting the pruning rule.
\end{proof}

\begin{lemma}
\label{lem:strict-bunch-size}
For every \(v\in V\) and \(0\le i\le k-1\),
\(
\e[|B_i(v)|]=O(n^{1/k}).
\)
So
\(\e[|B(v)|]=O(k n^{1/k})\).
\end{lemma}

\begin{proof}
Extend the definitions of strict clusters and their tree--pairs to every root \(w\in V\).

Fix \(v\in V\) and \(i<k-1\).
Fix a vertex \(w\) and the levels of all other vertices.
Suppose that placing \(w\) at level \(i\) makes \(w\in B_i(v)\).
By Claim~\ref{clm:strict-monotonicity} and the strict-cluster condition,
\[
\dist_{T(w,i+2)}(w\lra v)
\le
\dist_{T(w,i+1)}(w\lra v)
<
R_{i+1}(v).
\]
Now move \(w\) to level \(i+1\).
This leaves \(R_{i+2}\) and the tree--pairs of all other level-\((i+1)\)
centers unchanged.
The displayed inequality therefore makes \(w\) the unique minimizing
center in the update of \(R_{i+1}(v)\), so \(p_{i+1}(v)=w\).

If placing \(w\) in \(A_i\setminus A_{i+1}\) makes \(w\in B_i(v)\), then placing it in \(A_{i+1}\setminus A_{i+2}\) makes \(p_{i+1}(v)=w\).
The second choice is at least \(n^{-1/k}\) times as likely as the first.
Thus the following inequality holds for every fixed choice of the other vertex levels:
\begin{align*}
\Pr\!\left[
w\in B_i(v)
\,\middle|\,
\text{the levels of }V\setminus\{w\}
\right]
&\le
n^{1/k}\Pr\!\left[
p_{i+1}(v)=w
\,\middle|\,
\text{the levels of }V\setminus\{w\}
\right]
\\
\Pr[w\in B_i(v)]
&\le
n^{1/k}\Pr[p_{i+1}(v)=w].
\end{align*}
Since \(v\) has at most one pivot at each level, summing over \(w\) gives
\[
\e[|B_i(v)|]
\le
n^{1/k}\sum_{w\in V}\Pr[p_{i+1}(v)=w]
\le n^{1/k}.
\]

At the top level,
\(\e[|B_{k-1}(v)|]\le\e[|A_{k-1}|]=n^{1/k}\).
Summing yields
\(\e[|B(v)|]\le k n^{1/k}\).
\end{proof}

\section{$k$-Strong Tree Cover}
\label{sec:k-strong}

We now refine the previous construction to obtain the \(k\)-strong tree cover. The main difference is that a strong tree cover does not require the same tree--pair to approximate both directions between \(u\) and \(v\). For an ordered pair \((u,v)\), we only need one tree--pair that provides a short \(u\to v\) path; the reverse direction may use a different tree--pair. Accordingly, the stretch argument will try to preserve only a shortest \(u\to v\) path rather than the entire shortest cycle.

The strict clusters of the previous section are not quite sufficient for this purpose. A vertex on the return path \(v\to u\) that fails its radius condition may be removed during the pruning process and thereby destroy paths needed for vertices on \(u\to v\), even though that vertex itself is irrelevant to the direction being approximated.

We make two changes to overcome this issue. First, we strengthen the radius update by requiring the radii to decrease at a controlled rate between consecutive levels. This strengthening could already have been imposed in the previous construction without worsening its stretch or overlap guarantees. Second, we construct each tree--pair using an \emph{expanded strict cluster}, which lies between the strict cluster of the current level and that of the next level. Unlike a strict cluster, however, the expanded cluster need not preserve both directions between the center and every vertex: for some \(v\), only the \(w\to v\) path or only the \(v\to w\) path may remain.

\subsection{Construction}

The sampling of the sets \(A_i\) is the same as in
Algorithm~\ref{alg:2k-tree-cover}. Namely, the algorithm samples a nested hierarchy
\[
A_0=V\supseteq A_1\supseteq\cdots\supseteq A_k=\emptyset,
\]
where every vertex of \(A_{i-1}\) is included independently in \(A_i\)
with probability \(n^{-1/k}\).
The algorithm initializes \(R_k(v)=R_{k+1}(v)=\infty\) for every \(v\in V\), and
processes the levels from \(k-1\) down to \(0\).

\paragraph{Expanded strict clusters.}
Fix a level \(i\) and a center \(w\in A_i\setminus A_{i+1}\). The algorithm first computes the larger strict cluster
\(\C(w,R_{i+2})\) together with the tree--pair \(T(w,i+2)\).

We then define the \emph{expanded strict cluster}
\[
\s(w,R_{i+1})
=
\left\{
v\in\C(w,R_{i+2})
:
\dist_{T(w,i+2)}(w\lra v)<R_{i+1}(v)
\right\}.
\]
Thus,
\[
\C(w,R_{i+1})
\subseteq
\s(w,R_{i+1})
\subseteq
\C(w,R_{i+2}).
\]

The tree--pair added to the cover at level \(i\), denoted
\(\T(w,i+1)\), is obtained by running ingoing and outgoing shortest-path
searches from \(w\) in \(G[\s(w,R_{i+1})]\).
Unlike a strict cluster, \(\s(w,R_{i+1})\) need not preserve a short
roundtrip between \(w\) and every vertex; in particular, for some vertex
only one of the two directions may remain reachable inside the expanded
cluster.

The algorithm does not compute \(\s(w,R_{i+1})\) explicitly.
Instead, after computing \(T(w,i+2)\), it directly constructs the ingoing and outgoing trees \(\T(w,i+1)\), restricting the searches to vertices \(v\in\C(w,R_{i+2})\) satisfying
\( \dist_{T(w,i+2)}(w\lra v)<R_{i+1}(v). \)

\paragraph{Radius update.}
After constructing all level-\(i\) tree--pairs, the algorithm sets
\[
R_i(v)
=
\min\left\{
\frac{i}{i+1}R_{i+1}(v),
\min_{w\in A_i\setminus A_{i+1}}
\dist_{\T(w,i+1)}(v\lra w)
\right\},
\]
where the second distance is infinite if one of the two directions between
\(v\) and \(w\) is missing from \(\T(w,i+1)\).

The output is the collection
\(
\mathcal T
=
\left\{
\T(w,i+1)
:
0\le i\le k-1,\;
w\in A_i\setminus A_{i+1}
\right\}.
\)
Algorithm~\ref{alg:k-strong-tree-cover} summarizes the construction.

\begin{algorithm}[H]
\SetAlgoLined
\SetAlgoNoEnd
\KwIn{Directed graph \(G=(V,E)\), integer \(k\ge 1\).}
\KwOut{A collection \(\mathcal T\) of tree--pairs.}

\(A_0 \gets V,\quad A_k \gets \emptyset,\quad \mathcal T \gets \emptyset\)\;
\(R_k(v),R_{k+1}(v) \gets \infty\) for every \(v\in V\)\;

\For{\(i\gets 1\) \KwTo \(k-1\)}{
    Include each vertex of \(A_{i-1}\) in \(A_i\) independently
    with probability \(n^{-1/k}\)\;
}

\For{\(i\gets k-1\) \KwTo \(0\)}{
    \ForEach{\(w\in A_i\setminus A_{i+1}\)}{
        Compute \(\C(w,R_{i+2})\) and its tree--pair \(T(w,i+2)\)\;

        Compute ingoing and outgoing shortest-path trees rooted at $w$, restricted to vertices \(v\in\C(w,R_{i+2})\) satisfying
        \(
        \dist_{T(w,i+2)}(w\lra v)<R_{i+1}(v);
        \)
        
        Let \(\T(w,i+1)\) be their union and add it to \(\mathcal T\)\;
    }

    \ForEach{\(v\in V\)}{
        \(R_i(v)\gets
        \min\left\{
            \dfrac{i}{i+1}R_{i+1}(v),
            \displaystyle
            \min_{w\in A_i\setminus A_{i+1}}
            \dist_{\T(w,i+1)}(v\lra w)
        \right\}\)\;
    }
}

\Return \(\mathcal T\)\;

\caption{\(\textsc{StrongTreeCover}(G,k)\): construction of the
\(k\)-strong tree cover.}
\label{alg:k-strong-tree-cover}
\end{algorithm}

\subsection{Strong Stretch}

\begin{lemma}
\label{lem:k-strong-stretch}
The collection \(\mathcal T\) is a \(k\)-strong tree cover. Namely, for
every ordered pair \(u,v\in V\), there exists a tree--pair
\(\T\in\mathcal T\) such that
\[
\dist_{\T}(u\to v)
\le
\dist(u\to v)+(k-1)\dist(u\lra v).
\]
\end{lemma}

\begin{proof}
Fix \(u,v\in V\), and let \(r=\dist_G(u\lra v)\). For every \(y\in u\lra v\), define
\[
\delta(y)=
\begin{cases}
0, & y\in u\to v,\\
-1, & y\in (v\to u) \setminus (u \to v).
\end{cases}
\]

Let \(i\) be the largest level for which there exists \(x\in u\lra v \) satisfying \(R_i(x)\le(i+\delta(x))r\). Such a level exists since \(R_0(u)=0\), and \(i\le k-1\) since \(R_k=\infty\). By maximality, for every \(y\in u\lra v \),
\[
R_{i+1}(y)>(i+1 +\delta(y))r
\qquad\text{and}\qquad
R_{i+2}(y)>(i+2+\delta(y))r.
\]

We first show that $p_i(x)$ exists. If \(i>0\),
\[
\frac{i}{i+1}R_{i+1}(x) >  \frac{(i+1 +\delta(x))r \cdot i}{i+1} \ge (i + \delta(x))r \ge R_i(x),
\]
so $R_i(x)$ cannot come from the shrinking term. If \(i=0\), we choose \(x=u\) and \(w=u\).

We set $w := p_i(x)$. For every \(y\in u\lra v\),
\[
R_i(x)+\dist_G(x\lra y)
\le
(i+1+\delta(x))r
\le
(i+2+\delta(y))r
<
R_{i+2}(y).
\]
The paths between \(w\) and \(x\) lie in
\(\s(w,R_{i+1})\subseteq\C(w,R_{i+2})\).
Therefore, by the same first-removed-vertex argument as in
Lemma~\ref{lem:2k-stretch}, the entire cycle survives in
\(\C(w,R_{i+2})\). Consequently, for every \(y\in u\lra v \),
\[
\dist_{T(w,i+2)}(w\lra y)
\le
R_i(x)+\dist_G(u \lra v).
\]

Now consider the directed \(u\)-to-\(v\) walk along \(u\lra v\) that passes through \(x\). Every vertex \(y\) on this walk satisfies \(\delta(y)\ge\delta(x)\). Hence
\[
\dist_{T(w,i+2)}(w\lra y)
\le
R_i(x)+r
\le
(i+1+\delta(x))r
\le
(i+1+\delta(y))r
<
R_{i+1}(y).
\]
Thus the entire walk is contained in \(\s(w,R_{i+1})\).

The length of this walk is \(\dist_G(u\to v)-\delta(x)r\). Therefore, routing through \(w\) in \(\T(w,i+1)\) gives
\[
\begin{aligned}
\dist_{\T(w,i+1)}(u\to v)
\le
\dist_G(u\to v)-\delta(x)r+R_i(x)
\le
\dist_G(u\to v)+ir.\\
\end{aligned}
\]
\end{proof}

We note here that this construction also has $2k$-roundtrip stretch, like Algorithm~\ref{alg:2k-tree-cover}. Lemma~\ref{lem:2k-stretch} applies here also.

\subsection{Overlap}

For every vertex \(v\) and level \(i\), define its level-\(i\)
\emph{expanded bunch} by
\[
\B_i(v)
=
\{\,w\in A_i\setminus A_{i+1}
    : v\in\s(w,R_{i+1})\,\},
\]
and let \(\B(v)=\bigcup_{i=0}^{k-1}\B_i(v)\).
The overlap of \(v\) is at most \(|\B(v)|\).

\begin{lemma}
\label{lem:expanded-bunch-size}
For every \(v\in V\),
\(
\e[|\B(v)|]=O(k n^{1/k}).
\)
\end{lemma}

\begin{proof}
We show that
\(\e[|\B_i(v)|]=O(n^{1/k})\) for every level \(i\).

Fix \(i<k-1\), and condition on
\(A_i,A_{i+2},\ldots,A_k\).
The radius function \(R_{i+2}\) is now fixed.
Consider all \(w\in A_i\setminus A_{i+2}\) such that
\(v\in\C(w,R_{i+2})\), ordered by increasing
\(
\dist_{T(w,i+2)}(w\lra v).
\)
Let \(w'\) be the first vertex in this ordering that belongs to
\(A_{i+1}\).

Since
\(\C(w',R_{i+2})\subseteq\s(w',R_{i+2})\), the level-\((i+1)\)
tree--pair rooted at \(w'\) gives
\[
R_{i+1}(v)
\le
\dist_{T(w',i+2)}(w'\lra v).
\]
On the other hand, if \(w\in\B_i(v)\), then by the definition of the
expanded strict cluster,
\[
\dist_{T(w,i+2)}(w\lra v)<R_{i+1}(v).
\]
Thus every \(w\in\B_i(v)\) appears before \(w'\) in the ordering.

Conditioned as above, each vertex of
\(A_i\setminus A_{i+2}\) independently belongs to \(A_{i+1}\) with
probability \(\Theta(n^{-1/k})\). Hence the expected number of vertices
before the first such vertex is \(O(n^{1/k})\), and therefore
\(
\e[|\B_i(v)|]=O(n^{1/k}).
\)

For \(i=k-1\), trivially
\(|\B_{k-1}(v)|\le |A_{k-1}|\), whose expectation is \(n^{1/k}\).
Summing over the \(k\) levels gives
\(\e[|\B(v)|]=O(k n^{1/k})\).
\end{proof}

\section{Fast Constructions}
\label{sec:fast-constructions}
\subsection{Fast Clusters in $\widetilde O(km)$ Time}

The main bottleneck in both constructions is computing the strict
clusters. The original procedure may recompute the ingoing and
outgoing trees after each vertex deletion, requiring
$\widetilde O(mn)$ time per center. We introduce \emph{fast clusters},
which require only one pruning round per radius level, giving
$\widetilde O(km)$ time per center. In the resulting construction,
a single cluster definition replaces both the strict and expanded
clusters.

To motivate the definition, consider a center
$w\in A_i\setminus A_{i+1}$, whose expanded cluster is formed
using $R_{i+1}$. The proof of
Lemma~\ref{lem:expanded-bunch-size} uses the following two properties:
\begin{enumerate}
    \item The clusters satisfy
    \[
    \C(w,R_{i+1})\subseteq \s(w,R_{i+1}) \subseteq\C(w,R_{i+2}).
    \]
    \item If $v\in\s(w,R_{i+1})$, then
    \[
    \dist_{T(w,i+2)}(w\lra v)<R_{i+1}(v).
    \]
\end{enumerate}
The first property ensures that if a candidate center $w'$
is sampled at level $i+1$, then
\[
R_{i+1}(v)
\le
\dist_{\T(w',i+2)}(w'\lra v)
\le
\dist_{T(w',i+2)}(w'\lra v).
\]
Together with the second property, this gives the comparison
used in the overlap argument. No radius constraint is required
on the distances in $\T(w,i+1)$ after the filtering step.

We enforce the comparison used in the overlap argument directly. Denote the fast
clusters by $\fc(w,R_{i+1})$ and their associated tree--pairs
by $\ft(w,i+1)$. Set $\fc(w,R_k)=V$. For $i<k-1$, define
\[
\fc(w,R_{i+1})
=
\left\{
v\in\fc(w,R_{i+2}) :
\dist_{\ft(w,i+2)}(w\lra v)<R_{i+1}(v)
\right\},
\]
where $\ft(w,i+2)$ consists of the ingoing and outgoing
shortest-path trees rooted at $w$ in
$G[\fc(w,R_{i+2})]$.

Thus, we first compute the cluster at the higher radius level,
and then delete all vertices violating the new radius constraint
in a single round.
We recompute the trees in the remaining induced subgraph,
but do not repeat the pruning at the same radius.
The resulting distances need not satisfy the radius constraints:
the overlap argument uses the distances before this pruning round.

This is also exactly the filtering step used in the definition
of an expanded cluster. Hence no separate expanded-cluster
definition is needed for the fast construction.

\begin{algorithm}[H]
\SetAlgoLined
\KwIn{Graph $G=(V,E)$, a level-$i$ center $w$, and radius
functions $R_{i+1}\le\cdots\le R_k=\infty$.}
\KwOut{The fast cluster $\fc(w,R_{i+1})$.}

$\fc(w,R_k)\gets V$\;
Compute $\ft(w,k)$ by running Dijkstra searches from $w$ in $G$\;

\For{$j=k-2,k-3,\ldots,i$}{
    $\displaystyle
    \fc(w,R_{j+1})\gets
    \left\{
    v\in\fc(w,R_{j+2}) :
    \dist_{\ft(w,j+2)}(w\lra v)<R_{j+1}(v)
    \right\}$\;

    Compute $\ft(w,j+1)$ by running Dijkstra searches from $w$ in $G[\fc(w,R_{j+1})]$\;
}
\Return{$\fc(w,R_{i+1})$}\;

\caption{$\textsc{FastC}(G,w,i)$:
Computation of a fast cluster of a level-$i$ center.}
\label{alg:fastC}
\end{algorithm}

For a center $w\in A_i\setminus A_{i+1}$, we execute the
procedure only when processing level $i$, using the radius
functions already computed at higher levels. We add
$\ft(w,i+1)$ to the cover and use its distances to update
$R_i$, retaining the shrinking rule of the $k$-strong
construction.

There are $O(k)$ pairs of Dijkstra searches per center,
and each vertex is processed as a center at most once.
The total running time is therefore $\widetilde O(kmn)$.

The proof of Lemma~\ref{lem:k-strong-stretch} carries over:
the first-removed-vertex argument still applies to every
pruning round, and the final filtering step is unchanged.
That proof does not require all retained vertices to satisfy
their radius constraints after the trees are recomputed.
The construction thus retains the $k$-strong and
$2k$-roundtrip stretch guarantees.

For every vertex $v$ and level $i$, define its level-$i$
\emph{fast bunch} by
\[
\fb_i(v)
=
\left\{
w\in A_i\setminus A_{i+1} :
v\in\fc(w,R_{i+1})
\right\}, \quad \text{ and } \quad
\fb(v)=\bigcup_{i=0}^{k-1}\fb_i(v).
\]
Since $\ft(w,i+1)$ is contained in
$G[\fc(w,R_{i+1})]$, the overlap of $v$ is at most
$|\fb(v)|$.

\begin{lemma}
\label{lem:fast-bunch-size}
For every $v\in V$ and $0\le i\le k-1$,
\(
\e[|\fb_i(v)|]=O(n^{1/k}).
\)
Consequently,
\(
\e[|\fb(v)|]=O(k n^{1/k}).
\)
\end{lemma}

\begin{proof}
Fix $v$ and $i<k-1$, and condition on
$A_i,A_{i+2},\ldots,A_k$.
The radius functions $R_{i+2},\ldots,R_k$ are now fixed.
For the analysis, define the corresponding fast clusters
and trees for every possible center, whether or not it
is sampled.

Order all vertices $w\in A_i\setminus A_{i+2}$ such that
$v\in\fc(w,R_{i+2})$ by increasing
\[
\dist_{\ft(w,i+2)}(w\lra v),
\]
breaking ties deterministically.
This ordering is fixed under the conditioning.
Let $w'$ be the first vertex in the ordering that belongs
to $A_{i+1}$.

If $w'$ exists, it is a level-$(i+1)$ center, and its
tree--pair is $\ft(w',i+2)$. The radius update therefore gives
\[
R_{i+1}(v)
\le
\dist_{\ft(w',i+2)}(w'\lra v).
\]
On the other hand, every $w\in\fb_i(v)$ belongs to the
ordering and, by the single pruning round defining
$\fc(w,R_{i+1})$, satisfies
\[
\dist_{\ft(w,i+2)}(w\lra v)<R_{i+1}(v).
\]
Thus every such $w$ precedes $w'$.
If $w'$ does not exist, we bound $|\fb_i(v)|$ by the
length of the entire ordering.

Conditioned as above, each vertex of
$A_i\setminus A_{i+2}$ independently belongs to $A_{i+1}$
with probability $\Theta(n^{-1/k})$.
Indeed, writing $q=n^{-1/k}$, this probability is
$q/(1+q)$ for $i<k-2$, and $q$ for $i=k-2$, since
$A_k=\emptyset$.
The number of vertices before the first sampled vertex,
or the entire ordering if none is sampled, therefore has
expectation $O(n^{1/k})$.

For $i=k-1$, trivially
\[
|\fb_{k-1}(v)|\le |A_{k-1}|,
\]
whose expectation is $n^{1/k}$. Summing over the levels
completes the proof.
\end{proof}

\subsection{Constructing a Tree cover in $\widetilde O(n^{2.5})$ time}

Here, we show how to reduce the running time from $O(n^3)$ for dense
graphs to $\widetilde O(n^{2.5})$. We use the Chechik--Lifshitz sparsification, with the main idea being to sparsify the graph on every level \(i > k/2\). In particular, after processing level \(\lfloor k/2  + 1 \rfloor\), the remaining graph has \(O(n^{3/2+1/k})\) edges.

The construction produces a nested sequence of
subgraphs \(G:=G_k\supseteq G_{k- 1}\supseteq\cdots\supseteq G_1\). For \(i>k/2\), \(G_i\) is obtained from \(G_{i+1}\) after the sparsification step at level \(i\); otherwise, \(G_i=G_{i+1}\).

Throughout this section, we write \(\dist_i\) for distances in \(G_i\), and let $N_i^+(v)$ denote the outgoing neighbors of $v$ in $G_i$.

\paragraph{Chechik--Lifshitz sparsification.}

Fix a level \(i\), and let \(G_{i+1}\) denote the current graph before the sparsification step at level \(i\). We use the centers \(w\in A_i\setminus A_{i+1}\) as \textit{eliminators}. For an edge \(e=(a,b)\in E(G_{i+1})\), we say that \textit{\(w\) eliminates \(e\)} if
\[
2\dist_{i+1}(a\to w)+\dist_{i+1}(w\to b)
\le
2\omega(e)+\dist_{i+1}(b\to w).
\tag{CL}
\label{eq:CL-elimination}
\]

Note that \(\omega(e)\ge\dist_{i+1}(a\to b)\), so there is a symmetry property: if \((a,w)\in E(G_{i+1})\) also, then either \(w\) eliminates \((a,b)\), or \(b\) eliminates \((a,w)\). This means that if we pick a uniformly random out-neighbor of \(a\), it will eliminate at least half of \(a\)'s outgoing edges in expectation.

The relevant consequence of~\eqref{eq:CL-elimination} is the following.

\begin{lemma}[CL]
\label{lem:CL-replacement}
Let \(C\) be the cycle \(u\lra v\) in \(G_{i+1}\), of length \(r\). Suppose that an edge \(e=(a,b)\) on the \(u\to v\) portion of \(C\) is eliminated by \(w\). Then
\[
\dist_{i+1}(u\to w)+\dist_{i+1}(w\to v)
+\dist_{i+1}(v\to u)\le 2r.
\]

\begin{figure}[h]
 \centering
 \begin{tikzpicture}[directed/.style={->, >=Latex, thick}]

 \node[circle,draw,inner sep=2.5pt,label=below left:$a$] (a) at (0,-1) {};
 \node[circle,draw,inner sep=2.5pt,label=below:$b$] (b) at (0.75,0) {};
 \node[circle,draw,inner sep=2.5pt,label=left:$w$] (w) at (-0.75,0) {};

 \node[circle,draw,inner sep=2.5pt,label=below:$u$] (u) at (2,-2) {};
 \node[circle,draw,inner sep=2.5pt,label=right:$v$] (v) at (2,0.3) {};

 \draw[directed] (a) to (w);
 \draw[line width=4pt, green, opacity=0.3] (a) to (w);

 \draw[directed, purple, dashed] (a) to node[below]{$e$} (b);

 \draw[directed, bend left=30] (w) to (b);
 \draw[bend left=30,line width=4pt, green, opacity=0.3] (w) to (b);

 \draw[directed, bend left=30] (b) to (w);

 \draw[directed, purple, bend left=20] (b) to (v);
 \draw[line width=4pt, green, opacity=0.3, bend left=20] (b) to (v);

 \draw[directed, purple, bend left=50] (v) to (u);
 \draw[line width=4pt, green, opacity=0.3, bend left=50] (v) to (u);

 \draw[directed, purple, bend left=40] (u) to (a);
 \draw[line width=4pt, green, opacity=0.3, bend left=40] (u) to (a);

 \end{tikzpicture}
 \caption{The purple cycle is a shortest roundtrip through \(u\) and \(v\).
 If \(w\) eliminates the edge \(e=(a,b)\), then replacing \(e\) by
 \(a\to w\to b\) gives the green roundtrip
 \(u\to a\to w\to b\to v\to u\), whose length is at most twice the
 length of the purple cycle.}
 \label{fig:CL-replacement}
\end{figure}
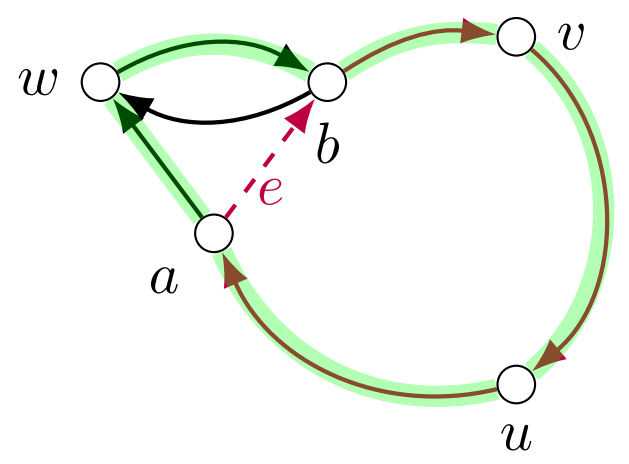

\end{lemma}

\paragraph{How we use that.}

The main idea is that after each round \(i>k/2\), we sparsify \(G_{i+1}\) to obtain \(G_i\), whose maximum out-degree is \(\widetilde O(n^{i/k})\). Roughly, suppose that \(v\) has out-degree \(\deg_{i+1}(v)\gg n^{i/k}\) in \(G_{i+1}\). Since every vertex belongs to \(A_i\) with probability \(n^{-i/k}\), we expect roughly \(\deg_{i+1}(v)\cdot n^{-i/k}\gg1\) sampled out-neighbors. Thus, we expect some of the out-neighbors of \(v\) to be centers in \(A_i\setminus A_{i+1}\). For each such center \(w\), we scan the edges eliminated by \(w\) and delete them. As long as the out-degree of \(v\) remains much larger than \(n^{i/k}\), we expect to find another sampled center among its remaining out-neighbors, so this process continues until the desired degree bound is reached.

For the stretch, let \(u,v\in V\), and let \(r\) be the length of \(u\lra v\) in \(G\). If none of its edges is deleted, we use the original stretch argument. Otherwise, let \(i>k/2\) be the first level for which \(u\lra v\nsubseteq G_i\). The deleted edge belonged to \(G_{i+1}\), and by Lemma~\ref{lem:CL-replacement}, if it was eliminated by a center \(w\), then \(G_{i+1}\) contains a roundtrip through \(u,v,w\) of length at most \(2r\). We restart the stretch argument using this new roundtrip and \(R_i(w)=0\). Since \(i>k/2\), fewer than \(k/2\) levels at or above \(i\) need to be considered. Thus, the factor \(2\) increase in the roundtrip length is compensated by the smaller number of levels in the stretch argument, and the same \(k\)-strong stretch bound follows.

\paragraph{The Construction.}
We assume that \(k\ge4\). The construction follows the
\(k\)-strong tree-cover construction with fast clusters, using
progressively sparser graphs at the upper levels and a direct
implementation of the fast clusters at level \(0\).

The algorithm samples a nested hierarchy
\[
A_0=V\supseteq A_1\supseteq\cdots\supseteq A_{k-1}
\supseteq A_k=\emptyset,
\]
where, for \(1\le i\le k-1\), every vertex of \(A_{i-1}\)
is included independently in \(A_i\) with probability \(n^{-1/k}\).
It initializes \(G_k:=G\) and \(R_k(v):=\infty\) for every
\(v\in V\), and processes the levels from \(k-1\) down to \(1\).

At level \(i\), for every center \(w\in A_i\setminus A_{i+1}\),
the algorithm computes \(\fc(w,R_{i+1})\) and its corresponding
tree-pair \(\ft(w,i+1)\) on \(G_{i+1}\), using the radius functions
already computed at higher levels. It then updates
\[
R_i(v)
=
\min\left\{
\frac{i}{i+1}R_{i+1}(v),
\min_{w\in A_i\setminus A_{i+1}}
\dist_{\ft(w,i+1)}(v\lra w)
\right\}
\]
for every \(v\in V\), where the second distance is infinite
if either direction between \(v\) and \(w\) is missing from
\(\ft(w,i+1)\).

\paragraph{Sparsifying \(G_{i+1}\).}
If \(i>k/2\), the algorithm applies
Algorithm~\ref{alg:sparsify} with centers \(A_i\setminus A_{i+1}\)
to obtain \(G_i\).
The procedure first computes the distances to and from these centers
in \(G_{i+1}\). For each vertex \(a\), it maintains the set \(X\)
of remaining out-neighbors of \(a\). It repeatedly chooses a center
\(w\in X\cap(A_i\setminus A_{i+1})\) uniformly at random, records it
in \(\mathcal E_i(a)\), and removes every edge \(e=(a,b)\) with
\(b\in X\) satisfying
\[
2\dist_{i+1}(a\to w)+\dist_{i+1}(w\to b)
\le
2\omega(e)+\dist_{i+1}(b\to w).
\]
All comparisons use distances in the unchanged input graph
\(G_{i+1}\). If \(i\le k/2\), the algorithm sets \(G_i:=G_{i+1}\).

\paragraph{Level \(i=0\).}
At level \(0\), the fast cluster satisfies
\[
\fc(w,R_1)
=
\left\{
v\in V:
\dist_{G_1}(w\lra v)<R_1(v)
\right\}.
\]
We compute this set directly, without computing the intermediate
fast clusters rooted at \(w\).

This step uses the distances to and from \(A_{k-1}\) computed
at the top level.
For each \(w\in V\setminus A_1\), run an outgoing truncated
Dijkstra search from \(w\) in \(G_1\).
Once a vertex \(x\) is extracted with tentative distance \(d\),
do not expand it if some \(p\in A_{k-1}\) satisfies
\[
2\dist_k(w\to p)+\dist_k(p\to x)
\le
2d+\dist_k(x\to p).
\]

Let \(X\) consist of \(w\) and the vertices expanded by the search.
Construct \(G_1[X]\) by scanning the outgoing edges of the
vertices in \(X\), and compute distances to and from \(w\)
in this subgraph using Dijkstra searches.
Retain every \(v\in X\) satisfying
\(\dist_{G_1[X]}(w\lra v)<R_1(v)\).
The retained vertices form \(\fc(w,R_1)\).

Finally, compute \(\ft(w,1)\) by running incoming and outgoing
shortest-path searches from \(w\) in \(G_1[\fc(w,R_1)]\).
The output is the collection
\[
\mathcal T
=
\left\{
\ft(w,i+1):
0\le i\le k-1,\;
w\in A_i\setminus A_{i+1}
\right\}.
\]

\begin{algorithm}[H]
\SetAlgoLined
\SetAlgoNoEnd
\KwIn{A directed weighted graph \(G=(V,E)\) and an integer \(k\ge4\).}
\KwOut{A tree cover \(\mathcal T\).}

Sample \(A_0=V\supseteq A_1\supseteq\cdots
\supseteq A_{k-1}\supseteq A_k=\emptyset\) as in the original construction\;
\(G_k\gets G\), \(\mathcal T\gets\emptyset\), Set \(R_k(v)\gets\infty\) for every \(v\in V\)\;

\For{\(i=k-1,k-2,\ldots,1\)}{
    \ForEach{\(w\in A_i\setminus A_{i+1}\)}{
        $\fc(w, R_{i+1}) \gets \textsc{FastC}(G_{i+1},w,i)$\;
        Compute $\ft(w, i+1)$ by running Dijkstra searches from $w$ in $G_{i+1}[\fc(w, R_{i+1})]$\;
        \(\mathcal T\gets\mathcal T\cup\{ \ft(w, i+1) \}\)\;
}
\ForEach{\(v\in V\)}{
\(
R_i(v)\gets
\min\left\{
    \frac{i}{i+1}R_{i+1}(v), \,
    \min_{w\in A_i\setminus A_{i+1}}
    \dist_{\ft(w,i+1)}(v\lra w)
\right\}
\)\;
}
\If{\(i>k/2\)}{
        \((G_i,\cdot) \gets \textsc{Sparsify}(G_{i+1},A_i\setminus A_{i+1})\)\;
        }
        \Else{
        $G_i \gets G_{i+1}$
        }
}

\(\mathcal T \gets \mathcal T \cup
\textsc{LevelZero}(G_1,A_1,A_{k-1},R_1,\dist_k)\)\;

\Return{\(\mathcal T\)}\;

\caption{\(\textsc{FastTreeCover}(G,k)\).}
\label{alg:fast-tree-cover}
\end{algorithm}

\begin{algorithm}[H]
\SetAlgoLined
\SetAlgoNoEnd
\KwIn{The graph \(G_{i+1}\), and a set of centers
\(A_i\setminus A_{i+1}\).}
\KwOut{The graph \(G_i\) and eliminator sets
\(\mathcal E_i(a)\), for \(a\in V\).}

\(E_i\gets\emptyset\)\;

Compute $\dist_{i+1}(w \to a), \dist_{i+1}(a \to w)$ for every $w\in A_i \setminus A_{i+1}$ and $a \in V$;

\ForEach{\(a\in V\)}{
    \(X\gets N_{i+1}^+(a)\), \(\mathcal E_i(a)\gets\emptyset\)\;

    \While{\(X\cap(A_i\setminus A_{i+1})\ne\emptyset\)}{
        Choose \(w\) uniformly at random from
        \(X\cap(A_i\setminus A_{i+1})\)\;
        \(\mathcal E_i(a)\gets\mathcal E_i(a)\cup\{w\}\)\;

        \ForEach{\(e=(a,b)\in E(G_{i+1})\) with \(b\in X\)}{
            \If{\(2\dist_{i+1}(a\to w)+\dist_{i+1}(w\to b)
            \le 2\omega(e)+\dist_{i+1}(b\to w)\)}{
                \(X\gets X\setminus\{b\}\)\;
            }
        }
    }

    \(E_i\gets E_i\cup\{(a,b)\in E(G_{i+1}):b\in X\}\)\;
}
\Return{\(\left((V,E_i),\mathcal E_i\right)\)}\;

\caption{\(\textsc{Sparsify}(G_{i+1},
A_i\setminus A_{i+1})\).}
\label{alg:sparsify}
\end{algorithm}

\begin{algorithm}[H]
\SetAlgoLined
\SetAlgoNoEnd
\KwIn{The graph \(G_1\), the sets \(A_1,A_{k-1}\), the radius
function \(R_1\), and the distances \(\dist_k\) to and from
\(A_{k-1}\).}
\KwOut{The level-\(0\) tree-pairs \(\mathcal T_0\).}

\(\mathcal T_0\gets\emptyset\)\;

\ForEach{\(w\in V\setminus A_1\)}{
    Run an outgoing truncated Dijkstra search from \(w\) in \(G_1\).
    Once \(x\) is extracted with tentative distance \(d\), do not expand it if some \(p\in A_{k-1}\) satisfies
    \[
        2\dist_k(w\to p)+\dist_k(p\to x)
        \le 2d+\dist_k(x\to p).
    \]

    \(X\gets\{w\}\cup
    \{\text{vertices expanded by the search}\}\)\;

    Compute distances to and from \(w\) in \(G_1[X]\) using Dijkstra searches\;
    \(
        \fc(w,R_1)\gets
        \left\{
            v\in X:
            \dist_{G_1[X]}(w\lra v)<R_1(v)
        \right\};
    \)

    Compute \(\ft(w,1)\) in \(G_1[\fc(w,R_1)]\)\;
    \(\mathcal T_0\gets\mathcal T_0\cup\{\ft(w,1)\}\)\;
}
\Return{\(\mathcal T_0\)}\;

\caption{\(\textsc{LevelZero}(G_1,A_1,A_{k-1},R_1,\dist_k)\).}
\label{alg:level-zero}
\end{algorithm}

\subsubsection{Running Time}

We process strongly connected components separately and assume \(4\le k\le\lceil\log n\rceil\). Larger values of \(k\) can be replaced by \(\lceil\log n\rceil\) while preserving the stated guarantees.

\begin{lemma}
\label{lem:sparsification-size}
With high probability, after every sparsification step \(i>k/2\), every vertex has out-degree \(\widetilde O(n^{i/k})\) in \(G_i\), and \(|\mathcal E_i(a)|=O(\log n)\) for every \(a\in V\). Given the required distances, the sparsification takes \(\widetilde O(n+|E(G_{i+1})|)\) time. Consequently, \(G_1\) has maximum out-degree \(\widetilde O(n^{1/2+1/k})\).
\end{lemma}

\begin{proof}
We sketch the adaptation of the sampling and elimination argument of Chechik and Lifshitz~\cite{CL21}. Fix a vertex \(a\), and condition on \(A_{i+1},\ldots,A_k\) and all earlier sparsification choices. This fixes \(G_{i+1}\). Since earlier steps removed all edges entering \(A_{i+1}\), each remaining out-neighbor belongs to \(A_i\setminus A_{i+1}\) independently with probability \(\Theta(n^{-i/k})\).

The elimination inequality guarantees that, for every pair of remaining neighbors, at least one eliminates the other. In particular, a uniformly chosen remaining sampled neighbor eliminates at least half of the remaining sampled neighbors in expectation. Thus only \(O(\log n)\) eliminators are chosen before the sample is exhausted, with high probability.

To bound the total number of remaining neighbors, fix a possible sequence of these eliminators. The remaining neighborhood is then fixed, and the probability that it contains no sampled vertex decreases exponentially with its size times \(n^{-i/k}\). A union bound over the \(n^{O(\log n)}\) possible sequences shows that at most \(O(n^{i/k}\log^2 n)\) neighbors remain, with high probability.

Each eliminator requires one scan of the outgoing edges of \(a\), giving the running-time bound. The bound for \(G_1\) follows because the last sparsification occurs at level \(\lfloor k/2\rfloor+1\).
\end{proof}

\begin{lemma}
\label{lem:level-zero-cluster}
For every \(w\in V\setminus A_1\),
\(\fc(w,R_1)=\{v\in V:\dist_{G_1}(w\lra v)<R_1(v)\}\).
Algorithm~\ref{alg:level-zero} computes this set without constructing the intermediate fast clusters. Moreover, with high probability, its set \(X\) has size \(\widetilde O(n^{1-1/k})\) for every root.
\end{lemma}

\begin{proof}
Every vertex in the recursively defined cluster satisfies the stated distance inequality. For the reverse inclusion, fix \(v\) with \(r=\dist_{G_1}(w\lra v)<R_1(v)\). We show that a shortest roundtrip \(w\lra v\) survives all filtering rounds before the final one. The key point is that a sufficiently small radius anywhere on this roundtrip would also force \(R_1(v)\le r\).

More precisely, we claim that \(R_j(y)>(j-1)r\) for every vertex \(y\) on the roundtrip and every \(2\le j\le k\). Otherwise, choose the largest violating level \(j<k\), and a vertex \(x\) with \(R_j(x)\le(j-1)r\). By maximality, \(R_{j+1}(y)>jr\) throughout the roundtrip. The shrinking term therefore exceeds \((j-1)r\), so \(R_j(x)\) is realized by a pivot \(p_j(x)\).

The paths between the pivot and \(x\), together with the roundtrip, give a roundtrip to each vertex \(y\) of length at most \(R_j(x)+r\le jr<R_{j+1}(y)\). Since \(G_1\subseteq G_{j+1}\), the first-removed-vertex argument from Lemma~\ref{lem:k-strong-stretch} places the entire roundtrip in the pivot's fast cluster. Hence \(R_j(v)\le jr\), and the shrinking rule gives \(R_1(v)\le R_j(v)/j\le r\), a contradiction.

Thus every filtering threshold \(R_j\), for \(j\ge2\), exceeds \(r\) throughout the roundtrip. While the roundtrip is present, the internal roundtrip distance from \(w\) to each of its vertices is at most \(r\), so all these filtering rounds retain it. The final round retains \(v\), proving the identity.

We next verify that the truncated search preserves this roundtrip. Suppose that \(x\) is its first vertex, starting from \(w\), that is not expanded, and let \(p\in A_{k-1}\) reject it. Its extracted distance is at most the length of the prefix from \(w\) to \(x\). As in Lemma~\ref{lem:CL-replacement}, the elimination inequality allows us to replace this prefix by paths through \(p\), obtaining a closed walk through \(w,p\) of length at most \(2r\). Hence \(\dist_G(w\lra p)\le2r\), and consequently \(\dist_G(v\lra p)\le3r\). The top-level radius and the shrinking rule now give \(R_1(v)\le R_{k-1}(v)/(k-1)\le3r/(k-1)\le r\), again a contradiction. Therefore \(X\) contains the required roundtrip for every such \(v\), and computing distances in \(G_1[X]\) gives exactly the stated cluster.

Finally, every expanded vertex passes all elimination tests also when its extracted distance is replaced by its true distance from \(w\) in \(G\). These comparisons are fixed before sampling \(A_{k-1}\) and have the same pairwise symmetry used in Lemma~\ref{lem:sparsification-size}. Since each vertex is sampled with probability \(n^{-1+1/k}\), the same sampling argument bounds the number of vertices surviving all comparisons by \(\widetilde O(n^{1-1/k})\), with high probability. This also bounds \(|X|\).
\end{proof}

\begin{lemma}
\label{lem:fast-tree-cover-time}
The construction runs in time \(\widetilde O(n^{5/2})\) with high probability.
\end{lemma}

\begin{proof}
At each level \(i>k/2\), there are \(\widetilde O(n^{1-i/k})\) centers and \(\widetilde O(n^{1+(i+1)/k})\) edges in the input graph, with high probability; at the top level, the latter follows from \(m\le n^2\). Since each fast-cluster computation uses \(O(k)\) shortest-path searches and \(k=O(\log n)\), the total work at this level, including sparsification, is \(\widetilde O(n^{2+1/k})\). At the remaining positive levels, all searches use \(G_1\), which has \(\widetilde O(n^{3/2+1/k})\) edges. There are at most \(|A_1|=\widetilde O(n^{1-1/k})\) centers across these levels, giving total time \(\widetilde O(n^{5/2})\).

At level \(0\), Lemmas~\ref{lem:sparsification-size} and~\ref{lem:level-zero-cluster} give \(|X|=\widetilde O(n^{1-1/k})\) and maximum out-degree \(\widetilde O(n^{1/2+1/k})\). Thus scanning outgoing edges of \(X\), constructing \(G_1[X]\), and performing the subsequent searches costs \(\widetilde O(n^{3/2})\) per root. Each vertex is extracted at most once, so the elimination tests cost at most \(O(n|A_{k-1}|)=\widetilde O(n^{1+1/k})\) per root. Summing over all roots gives \(\widetilde O(n^{5/2})\).
The failure probabilities can be made arbitrarily small inverse polynomials.
\end{proof}

\subsubsection{Strong Stretch}

\begin{lemma}
\label{lem:fast-k-strong-stretch}
The collection \(\mathcal T\) is a \(k\)-strong tree cover.
Namely, for every ordered pair \(u,v\in V\), there exists
a tree-pair \(\T\in\mathcal T\) such that
\[
\dist_{\T}(u\to v)
\le
\dist(u\to v)+(k-1)\dist(u\lra v).
\]
\end{lemma}

\begin{proof}
Fix \(u,v\in V\), and let \(r=\dist_G(u\lra v)\). Suppose first that no edge of \(u\lra v\) is deleted. Then \(u\lra v\subseteq G_1\), so this roundtrip is contained in every graph used by the construction. The proof of Lemma~\ref{lem:k-strong-stretch} therefore applies unchanged.

Otherwise, let \(i>k/2\) be the first level at which an edge of \(u\lra v\) is deleted, and let \(w\in A_i\setminus A_{i+1}\) eliminate such an edge. By Lemma~\ref{lem:CL-replacement}, replacing this edge by the paths through \(w\) gives a closed walk in \(G_{i+1}\) of length at most \(2r\), containing \(w\) and every vertex of the original roundtrip.

Choose the largest level \(j\ge i\) for which some vertex \(x\) on this walk satisfies \(R_j(x)\le2(j-i)r\). Such a level exists because \(R_i(w)=0\), and \(j\le k-1\) because \(R_k=\infty\). By maximality, \(R_{j+1}(y)>2(j-i+1)r\) for every vertex \(y\) on the walk. Since \(\frac{j}{j+1}2(j-i+1)\ge2(j-i)\), the shrinking term is greater than \(R_j(x)\). Thus \(R_j(x)\) is realized by a center \(p_j(x)\).

The modified walk lies in \(G_{j+1}\), since \(j\ge i\). The paths between \(p_j(x)\) and \(x\) in \(\ft(p_j(x),j+1)\) have total length \(R_j(x)\) and survive every pruning round. Together with the modified walk, they give a roundtrip to each vertex \(y\) of length at most \(R_j(x)+2r<R_{j+1}(y)\). All higher radius thresholds are at least \(R_{j+1}(y)\). Hence, by the first-removed-vertex argument of Lemma~\ref{lem:k-strong-stretch}, the entire walk survives in \(\fc(p_j(x),R_{j+1})\).

The original shortest \(u\)-to-\(v\) path also lies in the induced subgraph on this cluster: all its vertices survived, and all its edges are present in \(G_{j+1}\). Concatenating this path with one traversal of the modified closed walk starting at \(v\) gives a \(u\)-to-\(v\) walk through \(x\) of length at most \(\dist_G(u\to v)+2r\). Routing through \(p_j(x)\) adds at most \(R_j(x)\). The resulting additive error is therefore at most \(2r+R_j(x)\le2(j-i+1)r\le2(k-i)r\le(k-1)r\), where the last inequality follows from \(2i\ge k+1\).

\end{proof}

The construction also has \(2k\)-roundtrip stretch.

\subsubsection{Overlap}

Define \(\fb_i(v)\) and \(\fb(v)\) as before, using the fast clusters of the present construction.

\begin{lemma}
For every \(v\in V\) and \(0\le i\le k-1\), \(\e[|\fb_i(v)|]=O(n^{1/k})\). Consequently, the expected overlap of every vertex is \(O(kn^{1/k})\).
\end{lemma}

\begin{proof}
We adapt the proof of Lemma~\ref{lem:fast-bunch-size}. First, for fixed radius functions, deleting edges can only shrink the fast clusters and increase their internal distances. This follows by induction over the filtering rounds: both constructions start with \(V\), and containment of the current induced subgraphs implies that distances in the smaller graph are no shorter. The next filtering round therefore preserves the containment.

Fix \(v\) and \(i<k-1\). Use the conditioning of Lemma~\ref{lem:fast-bunch-size}, additionally fixing all sparsification choices at levels at least \(i+2\), and define its ordering using clusters and distances computed in \(G_{i+2}\). This graph and the required radius functions are fixed under the conditioning.

Since \(G_{i+1}\subseteq G_{i+2}\), every \(w\in\fb_i(v)\) belongs to this ordering. By the monotonicity above, its distance in the ordering is at most the distance used in its final filtering round in \(G_{i+1}\), and hence is less than \(R_{i+1}(v)\). The first vertex in the ordering sampled into \(A_{i+1}\), if one exists, has its actual tree-pair computed in \(G_{i+2}\). Thus the radius update bounds \(R_{i+1}(v)\) by its distance in the ordering, exactly as in the previous proof. Every \(w\in\fb_i(v)\) therefore precedes the first sampled vertex.

The conditional sampling probabilities are unchanged, so the geometric bound and the treatment of the top level in Lemma~\ref{lem:fast-bunch-size} apply unchanged.
\end{proof}

\paragraph{A tradeoff between stretch and construction time.}
The sparsification can be continued below level \(k/2\).
More generally, for an integer \(1\le t\le\lfloor (k-1)/2\rfloor\),
we apply Algorithm~\ref{alg:sparsify} after every level \(i>t\).
This reduces the construction time at the cost of a larger stretch,
while preserving the expected overlap.
The same analysis gives the following result.

\begin{theorem}
\label{thm:tree-cover-tradeoff}
For every integer \(k\ge4\) and every integer
\(1\le t\le\lfloor(k-1)/2\rfloor\), a directed graph with nonnegative
edge weights admits a
\((2k-2t-1)\)-strong tree cover with expected overlap
\(O(kn^{1/k})\).
The cover can be constructed in
\(\widetilde O(n^{2+t/k})\) time with high probability.
In particular, taking \(t=1\) gives a \((2k-3)\)-strong tree cover
in \(\widetilde O(n^{2+1/k})\) time.
\end{theorem}

The remaining \(n^{1/k}\) factor may be reduced by restricting the shortest-path searches at higher levels using the radius
bounds needed for the final fast clusters. The goal is to avoid exploring all vertices in these intermediate computations,
while preserving the paths needed to construct the final clusters.
Such a refinement could yield a \(2k\)-strong tree cover in \(\widetilde O(n^2)\) time. We do not develop this refinement here.

\section{Spanners, Distance Oracles and Routing Schemes}
\label{sec:applications}

In this section, we derive spanners, distance oracles, and routing schemes from our tree cover. We begin with two direct applications.

Taking the union of all tree-pairs gives a \(k\)-strong roundtrip spanner. Since every vertex belongs in expectation to \(O(kn^{1/k})\) tree-pairs, the union has \(O(kn^{1+1/k})\) edges in expectation.

\begin{corollary}
For every integer \(k\ge4\), a \(k\)-strong roundtrip spanner of any \(n\)-vertex directed graph with nonnegative edge weights can be constructed in \(\widetilde O(n^{5/2})\) time with high probability. The spanner has \(O(kn^{1+1/k})\) edges in expectation.
\end{corollary}

The tree cover also gives a simple distance oracle. For each vertex \(u\), store its fast bunch \(\fb(u)\) in a dictionary, together with the distances from \(u\) to each center and back in the corresponding tree-pair. To answer a query \((u,v)\), scan all centers \(w\in\fb(u)\). For each center, check whether \(w\in\fb(v)\) and, if so, compute the length of the route \(u\to w\to v\) within that tree-pair. Return the minimum length, or infinity if no finite route is found.

The stretch analysis guarantees that one of these routes satisfies the \(k\)-strong bound. Each candidate is checked in constant time, so the query takes \(O(|\fb(u)|)\) time, which is \(O(kn^{1/k})\) in expectation.

\begin{corollary}
For every integer \(k\ge4\), any \(n\)-vertex directed graph with nonnegative edge weights admits a path-reporting distance oracle with \(k\)-strong stretch, expected space \(O(kn^{1+1/k})\), and expected query time \(O(kn^{1/k})\). The oracle can be constructed in \(\widetilde O(n^{5/2})\) time with high probability.
\end{corollary}

\subsection{Distance Oracle with \(O(k)\) query time}
To reduce the query time to \(O(k)\), we consider routes through the destination's pivots, together with a level-\(0\) route. For every vertex \(u\), store \(\fc(u,R_1)\) in a dictionary. For each \(v\in \fc(u,R_1)\), store the exact distance \(\dist_{G_1}(u\to v)\) and the first edge of a shortest \(u\to v\) path in \(G_1\), breaking ties by the number of edges. These distances and first edges are obtained from the level-\(0\) searches before the final filtering step. For \(u\in A_1\), the dictionary is empty: if \(\dist_{G_1}(u\lra v)<R_1(v)\), the proof of Lemma~\ref{lem:level-zero-cluster} shows that this roundtrip
survives in the fast cluster constructed for \(u\).
The radius update and shrinking rule would then give \(R_1(v)\le\dist_{G_1}(u\lra v)\), a contradiction.

Given a query \((u,v)\), examine every defined pivot \(p_i(v)\), for \(0\le i<k\). Using the stored tree distances, compute the length of the route \(u\to p_i(v)\to v\) in \(\ft(p_i(v),i+1)\): the first part follows the incoming tree, and the second follows the outgoing tree. Discard candidates for which either path is missing. If \(v\in \fc(u,R_1)\), also consider the stored shortest-path distance from \(u\) to \(v\) in \(G_1\). Return the minimum candidate length. The corresponding path can be reported using the stored trees or the level-\(0\) next-hop entries.

\begin{lemma}
\label{lem:routing-linear-query}
The above procedure selects a route with \((2k-2)\)-strong stretch in \(O(k)\) time.
\end{lemma}

\begin{proof}
Fix \(u,v\), and let \(r=\dist_G(u\lra v)<\infty\).

The level-\(0\) construction provides exact distances in \(G_1\) for every \(v\in \fc(u,R_1)\), before the final filtering step. These distances and the stored next hops recover a shortest path in \(G_1\): every vertex \(x\) on such a path satisfies \(\dist_{G_1}(x\lra v)\le\dist_{G_1}(u\lra v)<R_1(v)\), and hence also stores an entry for \(v\). We break shortest-path ties by minimizing the number of edges.

We first show that either the level-\(0\) route is exact in \(G\), or some level \(i\ge1\) satisfies
\[
u\lra v\subseteq G_{i+1}
\qquad\text{and}\qquad
R_i(v)\le(2i-1)r.
\]
If \(u\lra v\subseteq G_1\), then \(v\in \fc(u,R_1)\) gives an exact route. Otherwise, the level-\(0\) characterization gives \(R_1(v)\le r\), so \(i=1\) suffices.

If an edge of \(u\lra v\) is deleted, let \(b\) be the first such level. The CL replacement lemma gives a roundtrip \(v\lra w\) in \(G_{b+1}\) of length at most \(2r\), where \(w\) is a level-\(b\) center and \(R_b(w)=0\). Choose the largest \(j\ge b\) for which some \(x\in v\lra w\) satisfies \(R_j(x)\le2(j-b)r\). At level \(j+1\), every vertex of this roundtrip has radius greater than \(2(j-b+1)r\). Thus the shrinking rule implies that \(p_j(x)\) exists, and the pruning argument shows that its cluster contains the roundtrip. Consequently,
\[
R_j(v)\le R_j(x)+2r
\le2(j-b+1)r
\le(2j-1)r,
\]
where the last inequality uses \(b\ge2\). Since \(j\ge b\), the original roundtrip \(u\lra v\) is contained in \(G_{j+1}\), establishing the required level.

Now choose the largest \(i\ge1\) such that \(u\lra v\subseteq G_{i+1}\) and at least one of the following holds:
\[
R_i(v)\le(2i-1)r,
\qquad\text{or}\qquad
R_i(x)\le(2i-2)r
\quad\text{for some }x\in u\lra v.
\]
Such a level exists by the preceding argument. Maximality gives
\[
R_{i+1}(v)>(2i+1)r,
\qquad
R_{i+1}(x)>2ir
\quad\text{for every }x\in u\lra v.
\]

We claim that \(R_i(v)\le(2i-1)r\). Otherwise, some \(x\in u\lra v\) satisfies \(R_i(x)\le(2i-2)r\). The shrinking rule implies that \(p_i(x)\) exists. Moreover, for every \(y\in u\lra v\),
\[
R_i(x)+r\le(2i-1)r<R_{i+1}(y).
\]
By the same first-removed-vertex argument used in the tree-cover proof, the entire roundtrip survives in \(\fc(p_i(x),R_{i+1})\). The radius update then gives \(R_i(v)\le R_i(x)+r\le(2i-1)r\), a contradiction.

Finally, \(R_{i+1}(v)>(2i+1)r\) ensures that the shrinking term cannot attain \(R_i(v)\), so \(p_i(v)\) exists. Since
\[
R_i(v)+r\le2ir<R_{i+1}(x)
\quad\text{for every }x\in u\lra v,
\]
the same pruning argument gives \(u\lra v\subseteq\fc(p_i(v),R_{i+1})\). Within this cluster, the shortest \(u\to v\) path followed by the tree path \(v\to p_i(v)\) bounds the incoming tree distance from \(u\) to \(p_i(v)\). Thus the candidate route through \(p_i(v)\) has length at most
\[
\dist_G(u\to v)+R_i(v)
\le \dist_G(u\to v)+(2k-3)r.
\]
The query returns a route of no greater length, proving the claimed stretch.
\end{proof}

\subsection{Routing schemes}
Compact routing seeks to route packets along short paths while storing only a small amount of information at each vertex. The first general space--stretch tradeoffs for undirected graphs were obtained by Peleg and Upfal~\cite{PU89}, and roundtrip routing in directed graphs was introduced by Cowen and Wagner~\cite{CowenWagner00}. We use the labeled routing model described in~\cite{TZ01routing,RTZ08}. The model has two phases.
During \textit{preprocessing}, the algorithm has access to the entire graph and assigns each vertex a label and a local routing table.
During \textit{routing}, a source \(u\) is given the label of its destination
\(v\) and uses this label and its own routing table to construct a
packet header. At each vertex, the packet is forwarded along an
outgoing edge, identified by its fixed port number, using only the
header and the local routing table. The header may also be updated.
We measure routing-table space in words, with each identifier, port
number, or stored distance occupying one word, and bound the label
length, header length, and routing-decision time separately.
For pairs with finite roundtrip distance, a scheme has \(t\)-strong
stretch if its route from \(u\) to \(v\) has length at most
\(\dist_G(u\to v)+(t-1)\dist_G(u\lra v)\). We prove the following theorem.
\begin{theorem}
For every integer \(k\ge4\), any \(n\)-vertex directed graph with
nonnegative edge weights admits a routing scheme with
\((2k-2)\)-strong stretch. With high probability, preprocessing takes \(\widetilde O(n^{5/2})\) time and every vertex has a routing table
of size \(\widetilde O(kn^{1/k})\) words.
The scheme uses labels of \(\widetilde O(k)\) words and packet headers of \(O(\log^2 n/\log\log n)\) bits.
The initial routing decision takes \(O(k)\) time, and each subsequent decision takes \(O(1)\) time.
\end{theorem}

Following the tree-cover approach of Thorup and Zwick~\cite{TZ01routing}, Roditty, Thorup, and Zwick~\cite{RTZ08} reduced directed roundtrip
routing to selecting a suitable tree-pair and routing within it. The tree-routing schemes of Fraigniaud and Gavoille~\cite{FG01} and Thorup and Zwick~\cite{TZ01routing} provide \(O(\log^2 n/\log\log n)\)-bit labels and constant-time forwarding. As explained in~\cite[Section~4]{RTZ08}, these schemes also support
tree-pairs: the packet first follows the incoming tree to the root and then the outgoing tree to its destination. 

\paragraph{Preprocessing} We apply this approach to our distance oracle by having each vertex store routing information for the tree-pairs in which it appears: every $u \in V$ stores $\fb(u)$.

\paragraph{The labels.} The label of a destination \(v\) contains, for every defined pivot \(p_i(v)\), the identifier of \(\ft(p_i(v),i+1)\), the distance from \(p_i(v)\) to \(v\), and the routing label of \(v\) in its outgoing
tree:
\[
\operatorname{label}(v)
=
\left(
v,\,
\left\{
\left(
\operatorname{ID}(\ft(p_i(v),i+1)),
\dist_{\ft(p_i(v),i+1)}(p_i(v)\to v),
\operatorname{label}^{\mathrm{out}}_{\ft(p_i(v),i+1)}(v)
\right)
:\ i \in [k] \right\}
\right),
\]
where \(\operatorname{label}^{\mathrm{out}}_T(v)\) denotes the routing label of \(v\) in the outgoing tree of the tree-pair \(T\).

\paragraph{The routing.} Using its local table and this label, for every $p_i(v) \in \fb(u)$, the source \(u\) can calculate the length of the route \(u\to p_i(v)\to v\) and select the shortest candidate.

This yields a \((2k-1)\)-strong routing scheme. Roughly, the same stretch argument shows that some pivot \(p_i(v)\) satisfies \(R_i(v)\le 2ir\), where \(r=\dist_G(u\lra v)\), and provides a route of length at most \(\dist_G(u\to v)+R_i(v)\). Since \(i\le k-1\), this gives the claimed bound. We omit the proof, as below we refine this argument to obtain \((2k-2)\)-strong stretch by handling level \(0\) separately.

\paragraph{Improving the stretch to \(2k-2\).}
In the distance oracle with \(O(k)\) query time, we obtained \((2k-2)\)-strong stretch by also considering
\(\dist_{G_1}(u\to v)\) whenever \(v\in\fc(u,R_1)\). Implementing these routes requires each vertex \(u\) to store distances and next-hop information for the destinations in \(\fc(u,R_1)\). However, these sets may be too large at individual vertices.

To address this, we adapt the recursive sampling technique of Thorup and Zwick~\cite{TZ01routing}, also used in the directed setting by Roditty, Thorup, and Zwick~\cite{RTZ08}. We perform additional sampling rounds that shrink the level-\(0\) dictionaries and provide alternative routes through new tree-pairs. Adapting this approach to our construction requires rebuilding the positive levels of the fast-cluster hierarchy on the fixed graph \(G_1\), while retaining the original construction. We show below that this gives the required space bound simultaneously at every vertex with high probability, while preserving the preprocessing bound, \(O(k)\) initial routing time, and \((2k-2)\)-strong stretch.

\paragraph{Additional preprocessing.}
We retain the original construction and denote its radii by \(R_i^{(0)}(v)\). All additional rounds operate on the fixed graph \(G_1\), without further sparsification. For every vertex \(u\), we initialize a dictionary containing the destinations in \(\fc(u,R_1)\), together with their distances and next-hop information in \(G_1\).

At the beginning of each round \(t\ge1\), let \(W\) consist of the vertices whose current dictionaries contain more than \(8n^{1/k}\) destinations. If \(W=\emptyset\), we stop. Otherwise, we sample
\(A_1^{(t)}\subseteq W\), including each vertex independently with probability
\(
\min\left\{1,\frac{n^{1-1/k}}{|W|}\right\}.
\)
For \(i=2,\ldots,k-1\), we obtain \(A_i^{(t)}\) by sampling each vertex of \(A_{i-1}^{(t)}\) independently with probability \(n^{-1/k}\), and set \(A_k^{(t)}=\emptyset\). Using this hierarchy, we run the fast-cluster construction on \(G_1\) for levels \(k-1,\ldots,1\), starting with \(R_k^{(t)}(v)=\infty\) for every \(v\).
This produces a separate collection of tree-pairs, radii, and pivots for round \(t\).

We then remove from the dictionary at each vertex \(u\) every destination \(v\) satisfying
\(
\dist_{G_1}(u\lra v)\ge R_1^{(t)}(v).
\)
Thus, after round \(t\), the dictionary at \(u\) contains
exactly the destinations in
\[
\left\{v\in V:
\dist_{G_1}(u\lra v)<
\min_{0\le j\le t}R_1^{(j)}(v)\right\}.
\]
These updates only filter the original dictionaries: we reuse their stored distances and next-hop information, without repeating the level-\(0\) searches.

Each vertex stores the routing information for all tree-pairs containing it, from both the original construction and the additional rounds. Tree-pair identifiers also specify their round. When the process terminates, each vertex stores its remaining dictionary, together with the corresponding distances and next-hop entries.
For each destination \(v\), we also record a round attaining the minimum of \(R_1^{(j)}(v)\); its pivots, together with the original pivots, will be used in the label of \(v\).

\paragraph{Additional labels.}
For each destination \(v\), choose a round attaining \(\min_j R_1^{(j)}(v)\). We append to the original label of \(v\) the same information for every defined pivot from this round: the tree-pair identifier, including the round, the distance from the pivot to \(v\) in its outgoing tree, and the routing label of \(v\) in that tree. Thus, the label still contains \(O(k)\) pivot entries.

\paragraph{Additional routing.}
The source \(u\) evaluates all available pivot routes specified in the augmented label of \(v\), using its local routing table as before. If \(v\) appears in the remaining dictionary at \(u\), it also considers the direct candidate \(\dist_{G_1}(u\to v)\). It selects the shortest candidate in \(O(k)\) time.

If a pivot route is selected, routing proceeds through the corresponding tree-pair as before.
Otherwise, the packet carries the identifier of \(v\) and follows the retained next-hop entries toward \(v\) in \(G_1\). Each subsequent routing decision takes \(O(1)\) time.

\begin{lemma}
\label{lem:routing-preprocessing}
With high probability, the additional preprocessing terminates
after \(O(\log n)\) rounds, the total preprocessing time is
\(\widetilde O(n^{5/2})\), and every vertex has a routing table
of size \(\widetilde O(kn^{1/k})\) words, simultaneously.
\end{lemma}

\begin{proof}
Fix the history before a round with \(W\ne\emptyset\), and
suppress the round superscript. Let \(q=n^{-1/k}\), and let
\(p=\min\{1,n^{1-1/k}/|W|\}\) be the sampling probability
for \(A_1\). Notice that \(p\ge q\).

Fix a destination \(v\), and condition on \(A_2,\ldots,A_k\).
The ordering used in the geometric argument of
Lemma~\ref{lem:fast-bunch-size} at level \(0\), restricted
to roots in \(W\setminus A_2\), is now fixed.
Each root in this ordering belongs to \(A_1\) independently
with probability
\[
\frac{p(1-q)}{1-pq}
\ge \frac{p}{1+p}
\ge \frac p2.
\]
By the level-\(0\) characterization and the radius-propagation
argument in Lemma~\ref{lem:level-zero-cluster}, every root
in \(W\) retaining \(v\) must precede the first sampled root
in this ordering. Thus the expected number of such roots
is at most \(2/p\). If \(p=1\), there are none.

Let \(W'\) be the set of vertices with oversized dictionaries
after the round. Dictionaries only shrink, so \(W'\subseteq W\).
For \(p<1\), summing over all destinations gives
\[
\mathbb E[|W'|\mid\text{history}]
\le
\frac{2n/p}{8n^{1/k}}
=
\frac{|W|}{4}.
\]
For \(p=1\), \(W'=\emptyset\).
Consequently, the probability that the process has not
terminated after \(t\) rounds is at most \(n4^{-t}\).
This proves the \(O(\log n)\) bound with high probability.

For the space bound, sampling above level \(1\) is unchanged.
The geometric tails from Lemma~\ref{lem:fast-bunch-size}
therefore bound the overlap at each positive level by
\(O(n^{1/k}\log n)\), with high probability.
At the top level, the same bound follows from a Chernoff
bound, since
\[
\mathbb E[|A_{k-1}|\mid\text{history}]
=
|W|p q^{k-2}
\le n^{1/k}.
\]
The same tail bounds hold for the original construction.
A union bound over all vertices, levels, and the first
\(O(\log n)\) rounds gives
\(\widetilde O(kn^{1/k})\) retained tree-pairs per vertex,
simultaneously.
Each contributes a polylogarithmic number of words, and
each final dictionary contains at most \(8n^{1/k}\) entries.
This gives the claimed routing-table size.

Finally, each round has
\(\widetilde O(n^{1-1/k})\) centers with high probability,
since \(\mathbb E[|A_1|\mid\text{history}]\le n^{1-1/k}\).
By Lemma~\ref{lem:sparsification-size},
\(G_1\) has \(\widetilde O(n^{3/2+1/k})\) edges.
As each center requires \(O(k)\) pairs of shortest-path
searches, and \(k=O(\log n)\) as in the earlier running-time
analysis, each round takes \(\widetilde O(n^{5/2})\) time.
Filtering the stored dictionaries adds at most \(O(n^2)\)
time per round. Constructing the tree-routing information
also fits within this bound.
Together with \(O(\log n)\) rounds and the original
preprocessing bound, this proves the claimed running time.
\end{proof}

\paragraph{Correctness and stretch.}
Fix \(u,v\) with \(r=\dist_G(u\lra v)<\infty\), and fix
shortest paths in both directions.
If either path is not contained in \(G_1\), the corresponding
case in the proof of Lemma~\ref{lem:routing-linear-query}
provides a route through an original pivot of \(v\) of length
at most \(\dist_G(u\to v)+(2k-3)r\).

Otherwise, both paths survive in \(G_1\).
If \(v\) belongs to the final dictionary at \(u\), the direct
candidate has length exactly \(\dist_G(u\to v)\).
If not, the round \(t\) chosen for the label of \(v\) satisfies
\[
R_1^{(t)}(v)=\min_j R_1^{(j)}(v)\le r.
\]
Starting from this inequality, the argument of
Lemma~\ref{lem:routing-linear-query} applies to that round
and provides a pivot route with the required stretch.
Its pivot is included in the label of \(v\).

Tree-pair routes are realized as described above.
For a selected direct route, every vertex \(x\) on a shortest
\(u\to v\) path in \(G_1\) satisfies
\[
\dist_{G_1}(x\lra v)
\le \dist_{G_1}(u\lra v)
< \min_j R_1^{(j)}(v).
\]
Hence every such vertex retains the entry for \(v\).
The next-hop entries therefore realize a shortest path in
\(G_1\), with minimum-hop tie-breaking ensuring termination
even with zero-weight edges.
Since the source selects the shortest candidate, the scheme
has \((2k-2)\)-strong stretch.

\section{Path Reporting Distance Oracles with $O(\log \log k)$ Query Time}
\label{sec:PRDO}

In this section, we present a new PRDO for \emph{undirected} graphs with stretch \(2k-1\), expected space \(O(kn^{1+1/k})\), and \(O(\log\log k)\) query time. This improves the \(O(\log k)\) query time of Wulff--Nilsen~\cite{WulffNilsen13} while preserving the stretch and space bounds, and provides a different
space--stretch tradeoff from Chechik and Zhang~\cite{ChechikZhang24},
who achieve \(O(n^{1+1/k})\) space, \(O(k)\) stretch, and \(O(\log\log k)\) query time.
We then show how, with minor modifications, the same query algorithm applies to our \emph{directed} construction, yielding a PRDO with \((2k-2)\)-strong stretch, expected space \(O(kn^{1+1/k})\), and \(O(\log\log k)\) query time.
In both cases, reporting the path takes additional time linear in its number of edges.

\subsection{Undirected Graphs}

Before describing the faster query algorithm, we recall the
two consecutive checks used by Wulff--Nilsen~\cite{WulffNilsen13}.
The following lemma summarizes their guarantees: success gives
a short path, while failure bounds the increase in the radius.
We include a proof for completeness.

We use the following consequence of the analysis of
Wulff--Nilsen~\cite{WulffNilsen13}.

\begin{lemma}[Consecutive Checks Lemma]
\label{lem:consecutive-checks}
Let \(r=\dist(u,v)\) and \(0\le i\le k-2\).
Check whether \(p_i(u)\in B(v)\) and, if not, whether
\(p_{i+1}(v)\in B(u)\).
If a check succeeds, let \(P\) be the concatenation of
the two shortest paths through the corresponding pivot. Then
\begin{equation}
\label{eq:consecutive-success}
    \operatorname{length}(P)\le 2R_i(u)+3r.
\end{equation}
If both checks fail, then
\begin{equation}
\label{eq:consecutive-failure}
    R_{i+2}(u)-R_i(u)\le 2r.
\end{equation}
At \(i=k-2\), at least one check succeeds.
\end{lemma}

\begin{proof}
If the first check succeeds, the path through \(p_i(u)\)
has length at most
\[
    \dist(u,p_i(u))+\dist(p_i(u),v)
    \le 2R_i(u)+r.
\]
Otherwise, \(p_i(u)\notin B(v)\), so the definition of the
TZ bunch implies
\[
    R_{i+1}(v)\le \dist(v,p_i(u))\le r+R_i(u).
\]
If the second check succeeds, the path through \(p_{i+1}(v)\)
therefore has length at most
\[
    \dist(u,p_{i+1}(v))+\dist(p_{i+1}(v),v)
    \le r+2R_{i+1}(v)\le 2R_i(u)+3r.
\]
If it also fails, the same bunch property gives
\[
    R_{i+2}(u)\le \dist(u,p_{i+1}(v))
    \le r+R_{i+1}(v)\le R_i(u)+2r.
\]
Finally, \(A_{k-1}\subseteq B(u)\), so the second check
succeeds at \(i=k-2\) whenever the first check fails.
\end{proof}

In particular, a successful check at a level \(i\) with
\(R_i(u)\le(k-2)r\) returns a path of stretch at most \(2k-1\); thus,
our goal is to find such a success quickly.

\paragraph{Overview of the idea.}
For simplicity, assume that \(k\ge4\) is even, \(r>0\), and the differences \(R_{i+2}(u)-R_i(u)\) are nondecreasing over the even indices \(i\). We consider the levels \(0,2,\ldots,k-2\), applying Lemma~\ref{lem:consecutive-checks}
at each tested level.

Suppose the checks fail at level \(i\) and succeed at level \(i+2\). Failure implies \(R_{i+2}(u)-R_i(u)\le2r\) by Lemma~\ref{lem:consecutive-checks}. By our monotonicity assumption, every preceding difference is also at most \(2r\), so
\[
R_{i+2}(u)\le \sum_{j=0}^{i/2}(R_{2j+2}(u) - R_{2j}(u)) \le (i+2)r\le(k-2)r.
\]
The successful checks therefore return a path of stretch at most \(2k-1\).
Thus, after testing level \(0\), we return its path if successful. Otherwise, we use \emph{binary search} between level \(0\) and the always-successful level \(k-2\).
We seek two consecutive selected levels where the checks fail at the first and succeed at the second. The search therefore takes \(O(\log k)\) checks. Note that the checks need not fail at all levels below some threshold and succeed at all levels above it.

To reduce the query time further, we preprocess pointers that skip levels. Write
\(b_i=(R_{i+2}(u)-R_i(u))/2\).
A failed check at \(i\) certifies \(b_i\le r\).
We can therefore jump from \(i\) to the largest even index \(j\) such that \(R_j(u)\le(k-2)b_i\):
if the checks succeed at \(j\), the returned path has the
required stretch.
Moreover, our monotonicity assumption gives \(R_{i+2}(u)\le(i+2)b_i\le(k-2)b_i\).

We first obtain a numerical estimate \(r\le\widehat r\le\alpha r\) in constant time,
where \(\alpha=O(k)\) (using~\cite{MendelNaor07}), and start at the largest even index \(s\) such that \(R_s(u)\le(k-2)\widehat r/\alpha\).
If the checks at level \(s\) succeed, the returned path has
stretch at most \(2k-1\), since
\[
R_s(u)\le(k-2)\widehat r/\alpha\le(k-2)r.
\]
Otherwise, \(s<k-2\), because the checks at level \(k-2\)
always succeed.
Moreover, \(b_s>\widehat r/\alpha\): if not, then
\(R_{s+2}(u)\le(s+2)b_s\le(k-2)\widehat r/\alpha\),
so \(s+2\) would also satisfy the condition defining \(s\),
contradicting the choice of \(s\) as the largest such index.

We prove below that within two pointer steps, either the last level is reached, the value of \(b_i\) doubles, or the remaining index range \(k-2-i\) halves.
Consequently, after \(O(\log k)\) steps we reach the last
level or a level with \(b_i>\widehat r\), which must succeed.
If the checks at \(s\) fail, we perform binary search on this pointer sequence until a failed level is followed immediately by a successful one.
By the definition of the pointers, that success returns a path of the required stretch.
The search takes \(O(\log\log k)\) checks, and preprocessing lets us access its midpoints in constant time.

Finally, to remove the monotonicity assumption, we use the lower convex hull of the points \((i,R_i(u))\). Its edge slopes replace the values \(b_i\):
they are increasing, failure still bounds the outgoing slope by \(r\), and the same pointer construction applies.

\subsubsection{Constructing the Data Structure}

We augment the standard Thorup--Zwick oracle~\cite{TZ05},
retaining its pivots, bunches, and reporting trees.
We assume \(G\) is connected and \(k\ge3\); disconnected graphs
are handled componentwise, and \(k\le2\) is handled by the
standard oracle.
We may also assume \(k=O(\log n)\), since larger values can
be replaced by \(\lceil\log_2 n\rceil\) without worsening
the claimed stretch or asymptotic space bound.
We ensure that \(A_{k-1}\ne\varnothing\) by resampling the
hierarchy if necessary, which takes a constant expected
number of attempts.

\paragraph{Selected levels.}
For even \(k\), we use the levels \(0,2,\ldots,k-2\).
For odd \(k\), we use \(0,1,3,\ldots,k-2\).
At each selected level \(i\), we use the checks of
Lemma~\ref{lem:consecutive-checks}, with one exception:
when \(k\) is odd, the check at level \(0\) tests only whether \(v\in B(u)\).
If so, it returns an exact shortest path. Otherwise,
\(R_1(u)\le r\). The checks at the last selected level \(k-2\) always succeed.

\paragraph{Lower convex hull.}
Fix a vertex \(u\). Among the selected levels, let \(i_0\)
be the largest index with \(R_{i_0}(u)=0\).
We compute the lower convex hull of the points
\((i,R_i(u))\) over the selected levels \(i\ge i_0\).
We retain only its corners and its endpoints,
\((i_0,0)\) and \((k-2,R_{k-2}(u))\).
We call these retained points \emph{hull points}.
For each hull point other than the last, let \(i^+\) denote the index of the next hull point, and define
\begin{equation}
\label{eq:hull-slope}
    b_i(u)=\frac{R_{i^+}(u)-R_i(u)}{i^+-i}.
\end{equation}
These slopes are increasing.
Since the hull starts at \((i_0,0)\), its preceding slopes are at most \(b_i(u)\), and hence
\begin{equation}
\label{eq:hull-radius}
    R_i(u)\le i\,b_i(u),
    \qquad
    R_{i^+}(u)\le i^+b_i(u).
\end{equation}

\begin{figure}[t]
\centering
\begin{tikzpicture}[
    x=0.6cm,y=0.5cm,
    >=Stealth,
    font=\small
]
\draw[->,black!60] (-0.4,0) -- (19.3,0)
    node[right] {Level};
\draw[->,black!60] (0,-3.2) -- (0,11.5)
    node[above] {Radius};

\draw[gray,dashed,thick]
    (0,-3) -- (18.5,8.1);

\draw[orange!80!black,dashed,thick]
    (0,0) -- (17.5,10.5);
\node[above left,text=orange!80!black]
    at (15.5,9.3) {$y=b_i(u)x$};

\draw[blue!65!black,thick]
    (2,0) -- (4,0.4) -- (8,1.8)
    -- (12,4.2) -- (16,7.4) -- (17.5,11.6);

\draw[blue!65!black,very thick]
    (8,1.8) -- (12,4.2);

\foreach \x/\y in {
    2/0, 4/0.4, 8/1.8, 12/4.2, 16/7.4, 17.5/11.6
}{
    \fill[blue!65!black] (\x,\y) circle (2.2pt);
}

\foreach \x/\y in {6/1.5, 10/3.9, 14/6.7}{
    \fill[gray] (\x,\y) circle (2.2pt);
}

\fill[gray] (0,0) circle (2.2pt);

\draw[black!25,dotted] (8,0) -- (8,4.8);
\draw[black!25,dotted] (12,0) -- (12,7.2);
\fill[orange!80!black] (8,4.8) circle (1.6pt);
\fill[orange!80!black] (12,7.2) circle (1.6pt);

\node[below left,text=gray] at (0,0) {$(0,0)$};
\node[below] at (2,0) {$(2,0)$};
\node[below] at (8,0) {$i$};
\node[below] at (12,0) {$i^+$};
\node[below right] at (8,1.8) {$(i,R_i(u))$};
\node[below right] at (12,4.2)
    {$(i^+,R_{i^+}(u))$};
\node[above left] at (17.5,11.6)
    {$(k-2,R_{k-2}(u))$};
\node[above left,text=orange!80!black] at (8,4.8)
    {$i\,b_i(u)$};
\node[above left,text=orange!80!black] at (12,7.2)
    {$i^+b_i(u)$};

\end{tikzpicture}
\caption{The lower convex hull (blue) of \(u\), starting
at the last point with radius zero, here \((2,0)\).
The point \((0,0)\) is discarded; the other gray points
lie strictly above the hull.
The dashed gray line extends the hull edge of slope \(b_i(u)\);
every selected point lies on or above this line.
The parallel orange line passes through the origin, illustrating
\(R_i(u)\le i\,b_i(u)\) and
\(R_{i^+}(u)\le i^+b_i(u)\).}
\label{fig:lower-hull}
\end{figure}
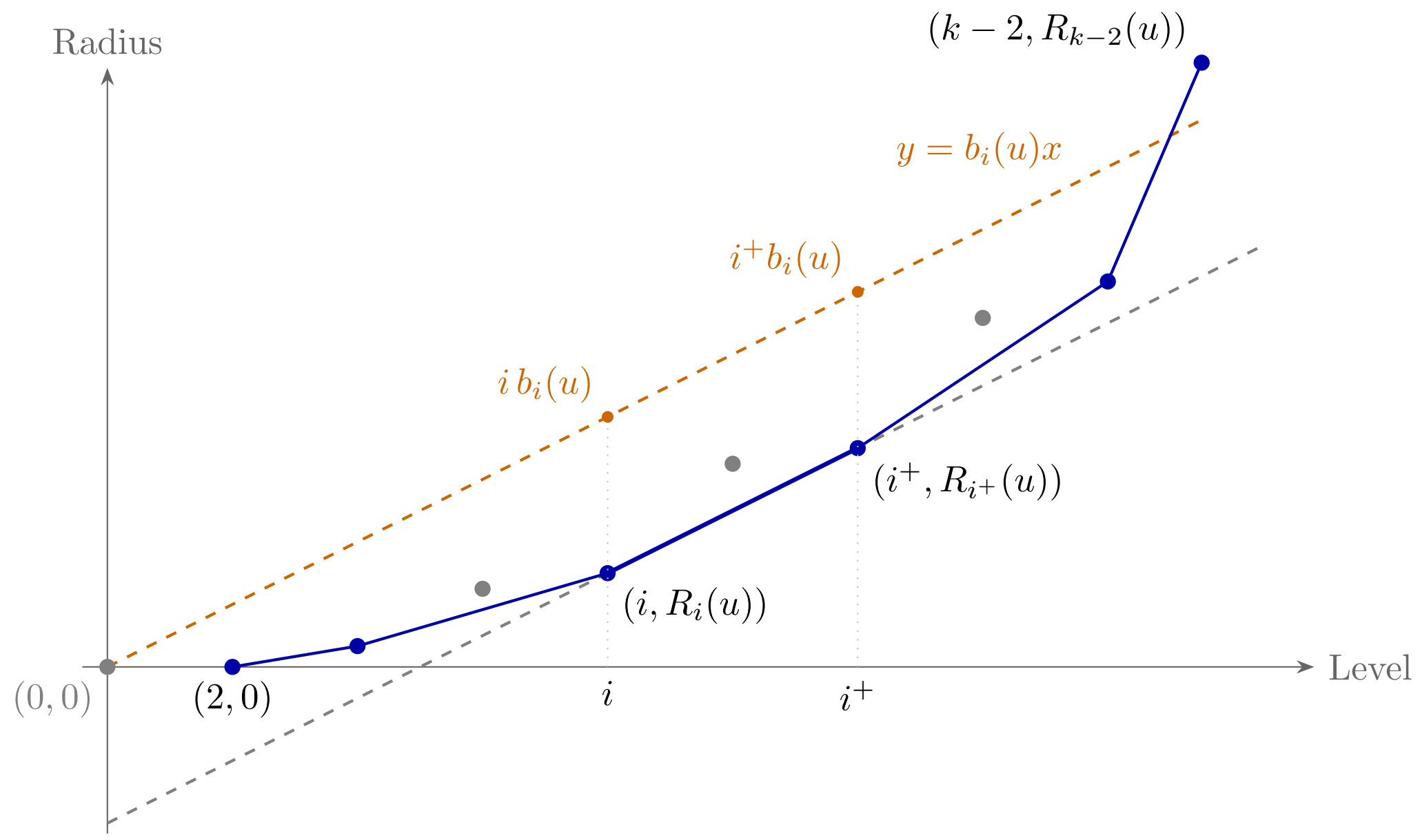

\paragraph{Jump pointers.}
For every hull point \(i<k-2\), define
\begin{equation}
\label{eq:hull-jump}
    j_u(i)=
    \max\bigl\{j:\ j\text{ is a hull index and }
    R_j(u)\le(k-2)b_i(u)\bigr\}.
\end{equation}
By \eqref{eq:hull-radius}, \(i^+\) satisfies this condition, so \(j_u(i)\ge i^+>i\).
We form a tree whose vertices are the hull indices:
for each \(i < k-2\), the parent of \(i\) is \(j_u(i)\).
Since \(j_u(i)>i\), repeatedly following parent pointers eventually reaches \(k-2\), which is the root.

We augment this tree with the level-ancestor structure of Bender and Farach-Colton~\cite{BenderFarachColton04}, using \(O(k)\) space and preprocessing time.
Given an index \(i\) of a hull point, and an integer \(t\), this structure returns its \(t\)-th ancestor in \(O(1)\) time.
Thus, we can access the index reached after \(t\) jump-pointer steps without traversing the intermediate indices.

\paragraph{Numerical estimates and starting levels.}
We also construct the numerical distance oracle of
Mendel and Naor~\cite{MendelNaor07}.
It uses \(O(n^{1+1/k})\) space and returns, in constant time, an estimate satisfying \(r\le\widehat r\le\alpha r\), where \(\alpha=128k\).
Given this estimate, we need to find, in constant time, the largest index \(s\) such that \((s,R_s(u))\) is a point of the lower convex hull for \(u\) and
\begin{equation}
\label{eq:query-start}
    R_s(u)\le(k-2)\widehat r/\alpha.
\end{equation}
This is a predecessor query among the radii of the hull points for \(u\). Storing these radii in a balanced search tree would support
this query in \(O(\log k)\) time, whereas we need to answer
it in constant time.

To support this query, we use the fact that every answer of the Mendel--Naor oracle is a stored tree-node label, and there are only \(O(n^{1+1/k})\) such labels~\cite{WulffNilsen13}.
During preprocessing, we collect the threshold \((k-2)\widehat r/\alpha\) corresponding to each label, together with all values \(R_i(u)\) for which \((i,R_i(u))\) is a hull point, over all \(u\in V\).
We sort these values and assign each distinct value its
rank in the sorted order. There are \(O(kn+n^{1+1/k})\) values, so each rank occupies \(O(\log n)\) bits. In other words, we form the set
\[
\begin{aligned}
&\bigcup_{u\in V}
\left\{
    R_i(u) \;\middle|\;
    (i,R_i(u)) \text{ is a point on the hull for } u
\right\}
\\
&\qquad{}\bigcup
\left\{
    \frac{(k-2)\widehat r}{\alpha}
    \;\middle|\;
    \widehat r \text{ is a possible oracle answer}
\right\}.
\end{aligned}
\]
We sort its distinct values and assign them consecutive ranks
starting from \(1\).
The set contains \(O(kn+n^{1+1/k})\) values, so each rank
occupies \(O(\log n)\) bits.

For each possible answer \(\widehat r\) of the numerical
oracle, we store the rank of the corresponding threshold
\((k-2)\widehat r/\alpha\) alongside the stored value
that produces this answer.
Thus, the oracle returns both \(\widehat r\) and the
threshold rank in constant time.

For each vertex \(u\), we build a \emph{static fusion tree}
of Fredman and Willard~\cite{FredmanWillard93} on \emph{the ranks} of the radii of its hull points.
Given the threshold rank returned by the numerical oracle, a predecessor query in this fusion tree returns exactly the index \(s\) in \eqref{eq:query-start}, since ranks preserve the order of the original values.
Each fusion tree contains \(O(k)\) keys of \(O(\log n)\)
bits and uses \(O(k)\) space. Since \(k=O(\log n)\), its query time is \(O(1)\).

The hull points, jump pointers, level-ancestor structures, and fusion trees use \(O(kn)\) additional space, while the threshold ranks stored with the numerical oracle use \(O(n^{1+1/k})\) additional space.
Together with the numerical oracle and the original Thorup--Zwick structures, the total expected space is \(O(kn^{1+1/k})\).

\subsubsection{Query}

Given a query \((u,v)\), we first obtain the estimate
\(\widehat r\) and use the fusion tree for \(u\) to find the starting index \(s\) in \eqref{eq:query-start}.

Let \(\textsc{Check}(u,v,i)\) perform the checks at selected level \(i\), including the special check at level \(0\) when \(k\) is odd.
It returns either failure or the pivot of the successful check. For the special check, this pivot is \(v\).
Together with \(u\) and \(v\), the returned pivot identifies the path in the stored reporting trees and serves as a constant-size path representation.
Each call takes \(O(1)\) time.
The actual path is reported only after the search terminates.

We first call \(\textsc{Check}(u,v,s)\).
If it succeeds, we return its path.
Otherwise, we search along the jump-pointer sequence
starting at \(s\).
Let \(h\) be the depth of \(s\) in the jump-pointer tree,
which we store during preprocessing.
For \(0\le t\le h\), let \(i_t\) denote the index of the
\(t\)-th ancestor of \(s\).
Thus, \(i_0=s\), \(i_h=k-2\), and
\(i_{t+1}=j_u(i_t)\).
The level-ancestor structure gives access to each \(i_t\)
in constant time.

We set
\[
    \ell=0,\qquad
    r=\min\bigl\{h,\;4\lceil\log_2\alpha\rceil\bigr\},
\]
and call \(\textsc{Check}(u,v,i_r)\).
As proved below, the choice of \(s\) using the initial
estimate \(\widehat r\) guarantees that this check succeeds.
Thus, the check at \(i_\ell=s\) fails and the check
at \(i_r\) succeeds.
We use binary search between these positions,
preserving this property, until \(r=\ell+1\).
The initial interval has length \(O(\log k)\), so
the binary search takes \(O(\log\log k)\) time.
We return the path found at \(i_r\).

Note that the checks need not be monotone along the sequence, meaning that a check may succeed at one position and fail at a later position. The search only maintains a failed check at its left endpoint and a successful check at its right endpoint.
At termination, these endpoints are consecutive in the jump-pointer sequence, so \(i_r=j_u(i_\ell)\).
As proved below, failure at \(i_\ell\) guarantees that the path returned at \(i_r\) has stretch at most \(2k-1\).

\begin{algorithm}[t]
\SetAlgoLined
\SetAlgoNoEnd
\KwIn{Query vertices \(u,v\).}
\KwOut{A constant-size representation of a \(u\)-to-\(v\) path.}
Obtain \(\widehat r\) and find \(s\) using
\eqref{eq:query-start}\;
\If{\(\textsc{Check}(u,v,s)\) succeeds}{
    \KwRet{the path representation returned by this check}\;
}
\(h\gets\) depth of \(s\) in the jump-pointer tree\;
\(\ell\gets0,\quad
  r\gets\min\bigl\{h,\;4\lceil\log_2\alpha\rceil\bigr\}\)\;
\While{\(r>\ell+1\)}{
    \(t\gets\lfloor(\ell+r)/2\rfloor\)\;
    \eIf{\(\textsc{Check}(u,v,i_t)\) fails}{
        \(\ell\gets t\)\;
    }{
        \(r\gets t\)\;
    }
}
\KwRet{the path representation returned by
\(\textsc{Check}(u,v,i_r)\)}\;
\caption{Answering a path-reporting distance query.
Here \(i_t\) is obtained by a level-ancestor query from \(s\).}
\label{alg:fast-prdo-query}
\end{algorithm}

\subsubsection{Analysis}

We first prove the stretch guarantee.
\begin{lemma}
\label{clm:fast-prdo-stretch}
The path returned by Algorithm~\ref{alg:fast-prdo-query}
has length at most \((2k-1)r\).
\end{lemma}

\begin{proof}
We first show that failure at a hull point \(i<k-2\)
implies
\begin{equation}
\label{eq:hull-failure}
    b_i(u)\le r.
\end{equation}
Except when \(i=0\) and \(k\) is odd, the point
\((i+2,R_{i+2}(u))\) lies on or above the line
extending the hull edge from \(i\) to \(i^+\).
Thus, failure and Lemma~\ref{lem:consecutive-checks} give
\[
    b_i(u)
    \le \frac{R_{i+2}(u)-R_i(u)}{2}
    \le r.
\]
When \(i=0\) and \(k\) is odd, we instead use the point
\((1,R_1(u))\).
Since \(R_0(u)=0\), the same geometric argument and
failure of the special check give
\(b_0(u)\le R_1(u)\le r\).

If the check at \(s\) succeeds, then
\[
    R_s(u)\le(k-2)\widehat r/\alpha\le(k-2)r,
\]
as $\hat{r}\le \alpha r$.

By Lemma~\ref{lem:consecutive-checks}, the returned path
has length at most
\(2R_s(u)+3r\le(2k-1)r\).
If \(s=0\) and \(k\) is odd, the special check returns
an exact shortest path, so the bound also holds.

Otherwise, the binary search maintains two positions
\(\ell<r\) such that the check at \(i_\ell\) fails and
the check at \(i_r\) succeeds.
At termination, \(r=\ell+1\), so
\(i_r=j_u(i_\ell)\).
By \eqref{eq:hull-failure} and the definition of the
jump pointers,
\[
    R_{i_r}(u)
    \le(k-2)b_{i_\ell}(u)
    \le(k-2)r.
\]
Since \(i_r>i_\ell\ge0\), the successful check at \(i_r\)
is governed by Lemma~\ref{lem:consecutive-checks}.
The returned path therefore has length at most
\(2R_{i_r}(u)+3r\le(2k-1)r\).
\end{proof}

We next bound the query time.
We first analyze two successive jumps.
\begin{claim}
\label{clm:two-jumps}
Let \(i<k-2\) be a hull index with \(b_i(u)>0\).
If two successive jumps from \(i\) reach an index
\(j<k-2\), then
\[
    b_j(u)\ge2b_i(u)
    \qquad\text{or}\qquad
    j > \frac{i}{2} + \frac{k-2}{2}.
\]
\end{claim}

\begin{proof}
Here \(j=j_u(j_u(i))>j_u(i)\).
Since \(j_u(i)\) is the largest hull index whose radius
is at most \((k-2)b_i(u)\), we have
\(R_j(u)>(k-2)b_i(u)\).
Every hull edge between \(i\) and \(j\) has slope
at most \(b_j(u)\).
Summing the radius increases along these edges and
using \eqref{eq:hull-radius}, we obtain
\[
\begin{aligned}
    (k-2)b_i(u)
    &< R_j(u)\\
    &\le R_i(u)+(j-i)b_j(u)\\
    &\le i\,b_i(u)+(j-i)b_j(u).
\end{aligned}
\]
If \(b_j(u)<2b_i(u)\), dividing by \(b_i(u)>0\) gives
\(k-2<i+2(j-i)=2j-i\), and hence
\[
    \frac{i}{2} + \frac{k-2}{2} < j.
\]
\end{proof}

We now bound the query time.
\begin{lemma}
\label{clm:fast-prdo-query-time}
Algorithm~\ref{alg:fast-prdo-query} returns a path
representation in \(O(\log\log k)\) time.
\end{lemma}

\begin{proof}
Obtaining \(\widehat r\), finding \(s\), and performing
each check or level-ancestor query take \(O(1)\) time.
If the check at \(s\) succeeds, the lemma follows.
Hence, assume that it fails, so \(s<k-2\).

Let \(h\) be the depth of \(s\) in the tree whose parent
pointers are \(j_u\), and let \(i_t\) be its \(t\)-th
ancestor for \(0\le t\le h\).
Thus, \(i_0=s\), \(i_h=k-2\), and
\(i_{t+1}=j_u(i_t)\) for \(0\le t<h\).

The algorithm initializes \(\ell=0\) and
\(r=\min\{h,4\lceil\log_2\alpha\rceil\}\).
We first prove that the check at \(i_r\) succeeds.
If \(r=h\), then \(i_r=k-2\), where the check always
succeeds.
We may therefore assume that \(r<h\), so
\(r=4\lceil\log_2\alpha\rceil\).

By the choice of \(s\) in \eqref{eq:query-start}
and by \eqref{eq:hull-radius},
\[
    (k-2)\widehat r/\alpha
    < R_{s^+}(u)
    \le s^+b_s(u)
    \le(k-2)b_s(u).
\]
Thus, \(b_s(u)>\widehat r/\alpha\), and in particular
\(b_s(u)>0\).

Consider the \(r/2=2\lceil\log_2\alpha\rceil\) pairs
\((i_0,i_2),(i_2,i_4),\ldots,(i_{r-2},i_r)\).
For each pair, Claim~\ref{clm:two-jumps} gives
\[
    b_{i_{t+2}}(u)\ge2b_{i_t}(u)
    \qquad\text{or}\qquad
    i_{t+2}>\frac{i_t}{2}+\frac{k-2}{2}.
\]
By the pigeonhole principle, at least
\(\lceil\log_2\alpha\rceil\) pairs satisfy the same
inequality.

If the first inequality held for \(\lceil\log_2\alpha\rceil\) pairs,
since the slopes never decrease,
\[
    b_{i_r}(u)\ge\alpha b_s(u)>\widehat r\ge r.
\]
By \eqref{eq:hull-failure}, the check at \(i_r\)
must succeed.

The second inequality reduces the difference from
\(k-2\) by more than half.
This difference is initially at most \(k-2\) and
never increases.
If the second inequality held for \(\lceil\log_2\alpha\rceil\) pairs, we would have
\[
    0<k-2-i_r<\frac{k-2}{\alpha}<1,
\]
where \(\alpha=128k\). This is impossible because \(k-2-i_r\) is an integer.

Thus, the binary search starts with a failed check
at \(i_\ell=s\) and a successful check at \(i_r\).
Its initial interval has length \(4\lceil\log_2\alpha\rceil=O(\log k)\),
so it performs \(O(\log\log k)\) iterations.
Each iteration uses one level-ancestor query and
one check, taking \(O(1)\) time.
The final check that returns the pivot also takes
\(O(1)\) time, giving a total query time of
\(O(\log\log k)\).
\end{proof}

\subsection{Extension to directed graphs}

We now adapt the construction to directed graphs with nonnegative edge weights, using our \(k\)-strong tree cover. Our goal is to obtain a \((2k-2)\)-strong approximation.

We use the fast-cluster construction (Algorithm~\ref{alg:fastC}), with all clusters
computed in the original graph \(G\).
The analysis also extends to the construction with edge sparsification, with minor modifications.

The main adaptation is that in the directed construction, a radius \(R_i(u)\) may be determined by the shrinking rule without a corresponding pivot. Thus, Lemma~\ref{lem:consecutive-checks} does not apply directly. For that, we use the fact that if \( R_i(u) < \frac{i}{i+1}R_{i+1}(u)\), then \(p_i(u)\) must be defined.

\subsubsection{The Construction}

We augment the fast-cluster tree cover with the same
auxiliary search structures as in the undirected construction.

For a query \((u,v)\), let \(r=\dist(u\lra v)\).
We apply the Mendel--Naor oracle to the roundtrip metric
\(G_{\mathrm{rt}}\), obtaining in constant time an estimate
\(r\le\widehat r\le\alpha r\), where \(\alpha=128k\). We use the same global ranking and static fusion trees
to find the starting hull index defined below\footnote{For $\widetilde{O}(\min\{n^{2.5},mn\})$ preprocessing time, we can first construct a $4$-roundtrip spanner in $\widetilde{O}(n^2)$ time~\cite{CL21} and apply the Mendel and Schwob numerical oracle~\cite{MendelSchwob09} to its roundtrip metric, increasing $\alpha$ by only a constant factor.}.

\paragraph{Level \(0\).}
We use the level-\(0\) dictionaries of the directed oracle.
For each stored entry \(u\) in the dictionary of \(v\),
we also store the incoming distance \(\dist(u\to v)\)
and the last edge of a shortest \(u\)-to-\(v\) path,
breaking shortest-path ties by the number of edges.
This information is obtained from the shortest-path
searches before the final filtering step.

Looking up \(u\) in the dictionary of \(v\) succeeds
exactly when \(\dist(u\lra v)<R_1(u)\).
A successful lookup gives a constant-size representation
of a shortest \(u\)-to-\(v\) path, which can be reported
using the stored incoming pointers.

\paragraph{Hull points.}
For each vertex \(u\), we compute the lower convex hull
of \((0,0)\) and the points \((2i-1,R_i(u))\), for
\(1\le i\le k-1\).
We retain its corners and endpoints, removing all points
of radius zero except the one with the largest index.
We call the retained points \emph{hull points} and
continue to index them by their original levels.
In particular, \((0,0)\), when retained, has index \(0\),
and the last hull point has index \(k-1\).

For each hull point \(i<k-1\), let \(i^+\) denote
the index of the next hull point.
We define \(b_i(u)\) to be the slope of the edge
between these points:
\[
    b_i(u)=
    \begin{cases}
        \displaystyle\frac{R_{i^+}(u)}{2i^+-1},
            & i=0,\\[6pt]
        \displaystyle\frac{R_{i^+}(u)-R_i(u)}
                              {2(i^+-i)},
            & i\ge1.
    \end{cases}
\]

For every hull point \(1\le i<k-1\), its point lies
on or below the segment joining \((0,0)\) to
\((2i+1,R_{i+1}(u))\). Moreover, \(R_{i+1}(u)>0\), since we removed all zero-radius points except the last. Hence
\[
    R_i(u)
    \le \frac{2i-1}{2i+1}R_{i+1}(u)
    < \frac{i}{i+1}R_{i+1}(u),
\]
so \(p_i(u)\) is defined.

For each hull point \(i<k-1\), define its parent by
\begin{equation}
\label{eq:directed-hull-jump}
    j_u(i)=
    \max\bigl\{j:\ j\text{ is a hull index and }
        R_j(u)\le(2k-3)b_i(u)\bigr\}.
\end{equation}
The same hull argument as before gives
\(R_{i^+}(u)\le(2i^+-1)b_i(u)\le(2k-3)b_i(u)\).
Thus \(j_u(i)\ge i^+>i\), and these parent pointers
form a tree rooted at \(k-1\).
We store the depths and augment the tree with
the same level-ancestor structure as before.

\paragraph{The starting index} \(s\) is the largest hull index satisfying
\begin{equation}
\label{eq:directed-query-start}
    R_s(u)\le(2k-3)\widehat r/\alpha.
\end{equation}
To find it in constant time, we apply the same ranking
construction to the hull radii and all possible thresholds
\((2k-3)\widehat r/\alpha\).
We store the threshold rank with each possible numerical
oracle answer and build a static fusion tree for each \(u\).

\paragraph{The subroutine \textsc{DiCheck}.}
The subroutine \(\textsc{DiCheck}(u,v,i)\) returns failure or a constant-size path representation.
For \(1\le i<k-1\), it first checks whether the outgoing tree of \(T(p_i(u),i+1)\) contains a path from \(p_i(u)\) to \(v\).
If so, it checks whether
\[
    \dist_{T(p_i(u),i+1)}(u\to p_i(u)\to v)
    \le R_i(u)+b_i(u).
\]
If either check fails, the subroutine returns failure.
Otherwise, it returns the pivot identifying the path
\(u\to p_i(u)\to v\).

At \(i=0\), it performs the dictionary lookup described above. At \(i=k-1\), it returns the path representation through \(p_{k-1}(u)\), without the length test.
Each call takes \(O(1)\) time.

\paragraph{The query.}
We follow Algorithm~\ref{alg:fast-prdo-query},
using the starting index in \eqref{eq:directed-query-start}
and replacing \(\textsc{Check}\) by \(\textsc{DiCheck}\).
If \(\textsc{DiCheck}(u,v,s)\) succeeds, we return
its path representation. If the check at \(s\) fails, let \(h\) be its depth in the parent-pointer tree, and let \(i_t\) be its \(t\)-th
ancestor, for \(0\le t\le h\).
Thus \(i_0=s\) and \(i_h=k-1\).
We initialize
\(\ell=0\) and
\(r=\min\{h,4\lceil\log_2\alpha\rceil\}\).
As proved below, the check at \(i_r\) succeeds.
We then run the same binary search on these ancestor
positions, maintaining failure at \(i_\ell\) and success
at \(i_r\), until \(r=\ell+1\).
Each midpoint index is obtained by one level-ancestor query.
We return the representation obtained by
\(\textsc{DiCheck}(u,v,i_r)\) and report the actual path
only after the search terminates.

The total expected space remains \(O(kn^{1+1/k})\).

\subsubsection{Analysis}

Fix a query \((u,v)\) with finite roundtrip distance
\(r=\dist(u\lra v)\).
We first prove the guarantees of \(\textsc{DiCheck}\),
then establish the stretch and query-time bounds.

For every positive hull index \(i<k-1\), convexity gives
\begin{equation}
\label{eq:directed-hull-bounds}
    R_i(u)\le(2i-1)b_i(u),
    \qquad
    R_j(u)\ge R_i(u)+2(j-i)b_i(u)
    \quad\text{for every }i<j\le k-1.
\end{equation}
The first inequality follows because all preceding hull
edges have slope at most \(b_i(u)\).
The second follows because the line containing the outgoing
hull edge at \(i\) lies on or below every input point.

\begin{lemma}
\label{lem:directed-checks}
Let \(i<k-1\) be a hull index.
If \(b_i(u)>r\), then \(\textsc{DiCheck}(u,v,i)\)
succeeds and returns a path \(P\) satisfying
\[
    \operatorname{length}(P)
    \le \dist(u\to v)+R_i(u).
\]
Consequently, if \(\textsc{DiCheck}(u,v,i)\) fails,
then \(b_i(u)\le r\).

At \(i=k-1\), the check always succeeds and returns
a path of length at most
\(\dist(u\to v)+R_{k-1}(u)\).
\end{lemma}

\begin{proof}
We use the pruning argument from the proof of
Lemma~\ref{lem:routing-linear-query}.
It shows that if \(p_j(x)\) is defined for
\(x\in u\lra v\), and
\[
    R_{j+1}(y)>R_j(x)+r
    \qquad\text{for every }y\in u\lra v,
\]
then all the vertices of $u \lra v$ are in $\ft(p_j(x),j+1)$.
As shown there, the radius update consequently gives
\(R_j(u)\le R_j(x)+r\).

At level \(0\), \(b_0(u)\le R_1(u)\), so \(b_0(u)>r\)
makes the dictionary lookup return a shortest path.
At level \(k-1\), the stored shortest-path trees give
the claimed length bound directly.

Now fix \(1\le i<k-1\) with \(b_i(u)>r\).
We show that
\[
    R_{i+1}(y)>R_i(u)+r
    \qquad\text{for every }y\in u\lra v.
\]
Suppose otherwise.
Choose the largest \(j>i\) for which some \(x\in u\lra v\)
satisfies
\[
    R_j(x)<R_i(u)+2(j-i)b_i(u)-r.
\]
Such a level exists at \(j=i+1\), since \(b_i(u)>r\).
It cannot be \(k-1\): the triangle inequality and
\eqref{eq:directed-hull-bounds} give
\[
    R_{k-1}(x)\ge R_{k-1}(u)-r
    \ge R_i(u)+2(k-1-i)b_i(u)-r.
\]
Thus \(j<k-1\).
By maximality, for every \(y\in u\lra v\),
\[
    R_{j+1}(y)
    \ge R_i(u)+2(j+1-i)b_i(u)-r
    >R_j(x)+2b_i(u).
\]

Moreover, \(R_i(u)\le(2i-1)b_i(u)\), so
\(R_j(x)<2j\,b_i(u)\).
Consequently,
\[
    \frac{j}{j+1}R_{j+1}(x)
    >
    \frac{j}{j+1}\bigl(R_j(x)+2b_i(u)\bigr)
    >R_j(x).
\]
Hence \(p_j(x)\) is defined.
Since \(2b_i(u)>r\), the pruning argument applies and gives
\[
    R_j(u)\le R_j(x)+r
    <R_i(u)+2(j-i)b_i(u)
    \le R_j(u),
\]
contradicting \eqref{eq:directed-hull-bounds}.

Therefore every \(y\in u\lra v\) satisfies
\(R_{i+1}(y)>R_i(u)+r\).
Since \(p_i(u)\) is defined, the same pruning argument
shows that \(u\lra v\) survives in its cluster.
The outgoing tree can follow the tree path from
\(p_i(u)\) to \(u\) and then a shortest \(u\)-to-\(v\) path.
Together with the incoming path from \(u\) to \(p_i(u)\),
this bounds the returned route by
\[
    R_i(u)+\dist(u\to v)
    \le R_i(u)+r
    <R_i(u)+b_i(u).
\]
Thus the check succeeds with the claimed length bound.
The failure bound follows by contraposition.
\end{proof}

Under sparsification, let \(b>k/2\) be the first deletion level on \(u\lra v\), and take \(j\ge b\) with \(R_j(u)\le2(j-b+1)r\), as in
Lemma~\ref{lem:routing-linear-query}.
Convexity and shrinking give \(b_i(u)\le r\) for every hull index \(i<b\) and \(R_1(u)\le r\). Hence the preceding proof applies whenever \(b_i(u)>r\), and the level-\(0\) lookup cannot succeed after a deletion.

\begin{lemma}
\label{lem:directed-query-stretch}
The query returns a path \(P\) satisfying
\[
    \operatorname{length}(P)
    \le \dist(u\to v)+(2k-3)r.
\]
Thus, the oracle has \((2k-2)\)-strong stretch.
\end{lemma}

\begin{proof}
We first show that every successful check at a hull
index \(i\) with \(R_i(u)\le(2k-3)r\) returns a path
satisfying the claimed bound.

At level \(0\), the returned path is exact.
At level \(k-1\), or when \(b_i(u)>r\), the bound follows
from Lemma~\ref{lem:directed-checks}.
Otherwise, \(1\le i\le k-2\) and \(b_i(u)\le r\).
The length test and \eqref{eq:directed-hull-bounds} give
\[
    \operatorname{length}(P)
    \le R_i(u)+b_i(u)
    \le 2i\,b_i(u)
    \le(2k-4)r,
\]
which also satisfies the claimed bound.

It remains to bound the radius at the returned level.
If the check at \(s\) succeeds, then
\eqref{eq:directed-query-start} gives
\[
    R_s(u)\le(2k-3)\widehat r/\alpha
    \le(2k-3)r.
\]
Otherwise, the binary search terminates with failure
at \(i_\ell\), success at \(i_r\), and \(r=\ell+1\).
The success of its initial right endpoint is established
in Lemma~\ref{lem:directed-query-time} below.
Since consecutive positions in the ancestor sequence
are connected by a parent pointer,
\(i_r=j_u(i_\ell)\).
Lemma~\ref{lem:directed-checks} and
\eqref{eq:directed-hull-jump} therefore imply
\[
    R_{i_r}(u)
    \le(2k-3)b_{i_\ell}(u)
    \le(2k-3)r.
\]
Thus, in both cases, the successful check returns
a path with the required bound.
\end{proof}

\begin{lemma}
\label{lem:directed-query-time}
Suppose \(\textsc{DiCheck}(u,v,s)\) fails.
Let \(h\) be the depth of \(s\) in the parent-pointer tree,
and let \(i_t\) be its \(t\)-th ancestor, for \(0\le t\le h\).
For
\[
    r=\min\{h,4\lceil\log_2\alpha\rceil\},
\]
the check \(\textsc{DiCheck}(u,v,i_r)\) succeeds.
The query takes \(O(\log\log k)\) time, excluding
the time required to report the path.
\end{lemma}

\begin{proof}
Since the check at \(s\) fails, \(s<k-1\).
The maximality of \(s\) and the hull geometry give
\[
    (2k-3)\widehat r/\alpha
    <R_{s^+}(u)
    \le(2k-3)b_s(u).
\]
Hence \(b_s(u)>\widehat r/\alpha\).
By Lemma~\ref{lem:directed-checks}, we also have
\(b_s(u)\le r\), so \(r>0\).

Write \(x_0=0\) and \(x_i=2i-1\) for \(i\ge1\).
The two-parent-pointer statement from the undirected
analysis takes the following form here:
for every hull index \(i<k-1\), if
\(j=j_u(i)<k-1\) and \(j_u(j)<k-1\), then
\[
    b_{j_u(j)}(u)\ge2b_i(u)
    \qquad\text{or}\qquad
    2k-3-x_{j_u(j)}
    <\frac{2k-3-x_i}{2}.
\]
Thus, after two parent-pointer edges, either the root
has been reached, the slope has at least doubled,
or the remaining horizontal distance to the root
has more than halved.

The proof is the same as in the undirected analysis,
using the coordinates \(x_i\) and replacing the root
coordinate \(k-2\) by \(2k-3\).
It uses only convexity, \(R_i(u)\le x_i b_i(u)\),
and the maximality of the parent in
\eqref{eq:directed-hull-jump}.

Let \(L=\lceil\log_2\alpha\rceil\).
The same counting argument from the undirected analysis
now gives the following conclusion at
\(r=\min\{h,4L\}\):
either \(i_r=k-1\), or \(i_r<k-1\) and
\(
    b_{i_r}(u)>\widehat r.
\)
Indeed, the conditions used by that argument remain valid:
the initial slope exceeds \(\widehat r/\alpha\),
the horizontal coordinates are integers, and their total
range is at most \(2k-3<\alpha\le2^L\).

The check at \(i_r\) therefore succeeds.
At the root this holds by definition; otherwise,
\(b_{i_r}(u)>\widehat r\ge r\), and
Lemma~\ref{lem:directed-checks} applies.

The binary search starts with failure at position \(0\)
and success at position \(r\), and uses
\(O(1+\log r)=O(\log\log k)\) checks.
Finding \(s\), evaluating each check, and answering
each level-ancestor query take constant time.
\end{proof}

\section{Further Applications}
\label{sec:further-app}

In this section, we present two further applications of our
\(k\)-strong tree cover: sourcewise and prioritized structures.
We first construct smaller structures for a prescribed set
of sources, and then combine these structures to obtain
better stretch for vertices of higher priority.

\subsection{Sourcewise Structures}

Source-restricted distance oracles for undirected graphs
were studied by Roditty, Thorup, and Zwick~\cite{RodittyTZ05}.
In directed graphs, sourcewise roundtrip spanners were
introduced by Zhu and Lam~\cite{ZhuLam17}.
Here, we adapt our \(k\)-strong tree-cover construction
to obtain sourcewise structures with strong-stretch guarantees.

We are given a directed graph \(G=(V,E)\) with nonnegative
edge weights and a prescribed nonempty set \(S\subseteq V\).
Write \(n=|V|\) and \(m=|E|\).
We require approximation guarantees only for ordered pairs
in \(S\times V\), while paths may use all vertices of \(G\).
Our goal is to obtain smaller structures when \(|S|\ll n\).

A \emph{sourcewise \(k\)-strong tree cover} is a collection
\(\mathcal T\) of tree-pairs such that, for every \(u\in S\)
and \(v\in V\) with finite roundtrip distance, some
\(T\in\mathcal T\) satisfies
\[
\dist_T(u\to v)
\le
\dist_G(u\to v)+(k-1)\dist_G(u\lra v).
\]
A sourcewise \(k\)-strong spanner satisfies the same inequality
with its own distances, and a sourcewise path-reporting
distance oracle returns the length of a path satisfying
this bound and supports reporting that path.

\paragraph{The basic construction.}
We apply the fast-cluster construction on \(G\), starting
with \(A_0=S\). Each vertex of \(A_{i-1}\) is sampled into
\(A_i\) independently with probability \(|S|^{-1/k}\),
for \(1\le i<k\), and \(A_k=\emptyset\).
The radius functions and clusters are defined on all of \(V\).

The stretch argument of Lemma~\ref{lem:k-strong-stretch}
applies to every pair in \(S\times V\), since its source
belongs to \(A_0\).
Lemma~\ref{lem:fast-bunch-size}, with the modified sampling
probability, gives expected overlap \(O(k|S|^{1/k})\).
There are \(|S|\) centers, each requiring \(O(k)\) pairs
of shortest-path searches, so the preprocessing time is
\(\widetilde O(k(m+n)|S|)\).
For \(k=1\), this simply constructs incoming and outgoing
shortest-path trees rooted at every source.

\paragraph{Faster preprocessing.}
Sampling only from \(S\) does not give the sparsification
bound used in the all-pairs construction: an out-neighborhood
may contain many vertices but no sources.
We therefore add a uniform sample from \(V\), whose vertices
are forced to appear at sufficiently high levels.
These additional vertices serve both as tree-pair centers
and as eliminators.

If \(|S|=1\), a single pair of shortest-path trees suffices.
Otherwise, assume \(k\ge3\), and set
\[
q=|S|^{-1/k},
\qquad
h=\lfloor k/2\rfloor+1.
\]
Independently include each vertex of \(V\) in a set \(B\)
with probability
\(
\frac{|S|^{1-h/k}}{n}.
\)
Construct the hierarchy as follows.
Set \(A_0=S\cup B\).
For \(i=1,\ldots,h\), retain every vertex of \(B\), and
include each vertex of \(A_{i-1}\setminus B\) independently
with probability \(q\).
For \(i=h+1,\ldots,k-1\), include each vertex of
\(A_{i-1}\) independently with probability \(q\).
Finally, set \(A_k=\emptyset\).

Thus \(B\subseteq A_h\), and every center outside \(S\)
appears only at a level at least \(h\).
We run the fast tree-cover construction with this hierarchy,
using the same fast clusters, radius updates, and
sparsification rule as before.
In particular, we sparsify after processing each level
\(i\ge h\).
For \(k\ge4\), we use the truncated level-\(0\) procedure
of Algorithm~\ref{alg:level-zero}, but run it only for
roots in \(S\setminus A_1\).
For \(k=3\), we compute the level-\(0\) fast clusters
directly on \(G_1\), without truncation.

As in the preceding running-time analysis, we may assume
\(k=O(\log n)\).
Larger parameters can be capped at
\(\max\{4,\lceil\log_2 n\rceil\}\).
This only improves the stretch guarantees below and changes
the stated size and query bounds by at most a constant factor.

\begin{lemma}
\label{lem:sourcewise-fast-tree-cover}
For every integer \(k\ge3\), the accelerated construction
produces a sourcewise \(k\)-strong tree cover for \(S\times V\)
in which every vertex belongs to \(O(k|S|^{1/k})\) tree-pairs
in expectation.
With high probability, its preprocessing time is
\(\widetilde O(n^2|S|^{2/3})\) for \(k=3\), and
\(\widetilde O(n^2\sqrt{|S|})\) for \(k\ge4\).
\end{lemma}

\begin{proof}
\leavevmode

\paragraph{Stretch and overlap.}
Every source still belongs to \(A_0\), and the deterministic
stretch argument is unchanged.
If the original roundtrip survives all sparsification steps,
Lemma~\ref{lem:k-strong-stretch} applies.
Otherwise, its first deleted edge is eliminated at a level
\(i\ge h>k/2\), and the replacement argument of
Lemma~\ref{lem:fast-k-strong-stretch} gives the same
\(k\)-strong bound.

To bound the overlap, condition as in
Lemma~\ref{lem:fast-bunch-size}.
For each eligible vertex in \(A_i\setminus A_{i+2}\),
the conditional probability of belonging to \(A_{i+1}\)
is at least \(q/(1+q)\).
Below level \(h\), vertices of \(B\) are either excluded
from this ordering because they also belong to \(A_{i+2}\),
or are forced into \(A_{i+1}\).
All other transitions use the usual sampling rule.
The membership choices remain independent across vertices,
so the same geometric argument applies.

Moreover, for every \(0\le i<k\),
\[
\mathbb E[|A_i|]\le 2|S|^{1-i/k}.
\]
In particular, the expected number of top-level centers
is at most \(2|S|^{1/k}\).
Consequently,
\[
\mathbb E[|\fb(v)|]=O(k|S|^{1/k})
\qquad\text{for every }v\in V.
\]

\paragraph{Preprocessing time.}
For every \(i\ge h\), the vertices of \(B\) that survive
to \(A_i\) form a uniform sample from \(V\), with inclusion
probability
\[
\frac{|S|^{1-h/k}}{n}\,q^{i-h}
=
\frac{|S|^{1-i/k}}{n}.
\]
Conditioned on the higher levels, every remaining
out-neighbor is a level-\(i\) center with probability at least
\((1-q)|S|^{1-i/k}/n\).
Since \(|S|\ge2\) and \(k=O(\log n)\), the factor
\((1-q)^{-1}=O(k)\) is absorbed by the logarithmic factors.
The proof of Lemma~\ref{lem:sparsification-size} therefore
gives, with high probability,
\[
\max_{v\in V}\deg^+_{G_i}(v)
=
\widetilde O\!\left(\frac{n}{|S|^{1-i/k}}\right),
\qquad
|E(G_i)|
=
\widetilde O\!\left(\frac{n^2}{|S|^{1-i/k}}\right).
\]
Since the last sparsification occurs at level \(h\),
\[
\max_{v\in V}\deg^+_{G_1}(v)
=
\widetilde O\!\left(\frac{n}{|S|^{1/2-1/k}}\right),
\qquad
|E(G_1)|
=
\widetilde O\!\left(\frac{n^2}{|S|^{1/2-1/k}}\right).
\]

At each level \(i\ge h\), there are
\(\widetilde O(|S|^{1-i/k})\) centers with high probability.
Their searches, including the distances needed for
sparsification, take
\[
\widetilde O\!\left(
|S|^{1-i/k}
\frac{n^2}{|S|^{1-(i+1)/k}}
\right)
=
\widetilde O(n^2|S|^{1/k})
\]
time.
At the remaining positive levels, there are at most
\(|A_1|=\widetilde O(|S|^{1-1/k})\) centers, and all
searches use \(G_1\).
Their total time is therefore
\(\widetilde O(n^2\sqrt{|S|})\).

For \(k\ge4\), the top level contains a uniform sample
from \(V\) with inclusion probability \(|S|^{1/k}/n\).
The argument of Lemma~\ref{lem:level-zero-cluster}
therefore bounds the number of expanded vertices in
each level-\(0\) search by
\(\widetilde O(n/|S|^{1/k})\), with high probability.
Together with the out-degree bound for \(G_1\), this gives
\[
\widetilde O\!\left(
\frac{n}{|S|^{1/k}}
\frac{n}{|S|^{1/2-1/k}}
\right)
=
\widetilde O\!\left(\frac{n^2}{\sqrt{|S|}}\right)
\]
time per root for scanning edges and computing the trees.
There are at most \(|S|\) roots.
The elimination tests cost
\(\widetilde O(n|S|^{1+1/k})\) in total, which also fits
within \(\widetilde O(n^2\sqrt{|S|})\).
Thus the entire construction takes
\(\widetilde O(n^2\sqrt{|S|})\) time with high probability.

For \(k=3\), sparsification at level \(2\) leaves
\(\widetilde O(n^2/|S|^{1/3})\) edges.
Computing the remaining fast clusters directly takes
\(\widetilde O(n^2|S|^{2/3})\) time with high probability.
\end{proof}

We will use the bound for \(k=3\) in the prioritized construction.
For \(k=1,2\), we use the basic construction.

\begin{corollary}
\label{cor:sourcewise-tree-cover}
For every integer \(k\ge1\), \(G\) admits a sourcewise
\(k\)-strong tree cover for \(S\times V\) in which every
vertex belongs to \(O(k|S|^{1/k})\) tree-pairs in expectation.
The expected total number of vertex occurrences is
\(O(kn|S|^{1/k})\).
The cover can be constructed in
\(\widetilde O(k(m+n)|S|)\) time.
For \(k=3\), it can also be constructed in
\(\widetilde O(n^2|S|^{2/3})\) time with high probability.
For \(k\ge4\), choosing between the basic and accelerated
constructions gives preprocessing time
\[
\widetilde O\!\left(
\min\{(m+n)|S|,\;n^2\sqrt{|S|}\}
\right)
\]
with high probability.
\end{corollary}

Taking the union of the tree-pairs immediately gives
the following spanner.

\begin{corollary}
\label{cor:sourcewise-spanner}
For every integer \(k\ge1\), \(G\) admits a sourcewise
\(k\)-strong spanner for \(S\times V\) with
\(O(kn|S|^{1/k})\) edges in expectation.
Its preprocessing bounds are those of
Corollary~\ref{cor:sourcewise-tree-cover}.
\end{corollary}

For the distance oracle, store the fast bunch of each
vertex together with the corresponding tree distances
and path-reporting information.
Given \((u,v)\in S\times V\), scan \(\fb(u)\) and consider
the route through every center whose tree-pair also
contains \(v\), exactly as in the all-pairs oracle.

\begin{corollary}
\label{cor:sourcewise-oracle}
For every integer \(k\ge1\), \(G\) admits a sourcewise
path-reporting distance oracle for \(S\times V\) with
\(k\)-strong stretch, expected space \(O(kn|S|^{1/k})\),
and expected query time \(O(k|S|^{1/k})\).
Its preprocessing bounds are those of
Corollary~\ref{cor:sourcewise-tree-cover}.
After a query, the selected path can be reported in
additional time linear in its number of edges.
\end{corollary}

\paragraph{An oracle with \(O(k)\) query time.}
For \(k\ge2\), we adapt the query procedure of
Lemma~\ref{lem:routing-linear-query}.
For every \(u\in S\setminus A_1\), store \(\fc(u,R_1)\)
in a dictionary.
For each \(v\) in this dictionary, query the preceding
oracle during preprocessing and store the returned route
length and the identifier of its tree-pair.
Given \((u,v)\), consider this stored route, if present,
and all available routes through the defined pivots of \(v\).
Return the shortest candidate.
There are \(O(k)\) candidates, each evaluated in constant time.
If a dictionary entry is present, its stored route has
\(k\)-strong stretch, which implies \((2k-2)\)-strong
stretch for \(k\ge2\).

If no dictionary entry is present and \(u\in S\setminus A_1\),
then \(v\notin\fc(u,R_1)\), exactly as in the missing
level-\(0\) case of Lemma~\ref{lem:routing-linear-query}.
Sources \(u\in S\cap A_1\) have no dictionary, so we must
also justify the pivot guarantee for these sources.
If the original roundtrip survives in \(G_1\), let \(i\ge1\)
be the level of \(u\).
Then \(R_i(u)=0\le(2i-2)\dist_G(u\lra v)\), and the
roundtrip is contained in \(G_{i+1}\).
These are precisely the second starting condition and
the containment condition used in the pivot argument,
with \(x=u\).
If an edge of the roundtrip is deleted, the replacement
argument of the same lemma applies unchanged.
Thus the pivot candidates give \((2k-2)\)-strong stretch
whenever no dictionary entry is present.

Every candidate uses a stored tree-pair, so path reporting
remains linear in the number of reported edges.
The dictionaries contain \(O(n|S|^{1/k})\) entries in
expectation.
The geometric tails of the bunch sizes bound the additional
preprocessing time by \(\widetilde O(n|S|^{2/k})\) with high
probability, which fits within the preceding preprocessing
bounds.

\begin{corollary}
\label{cor:sourcewise-oracle-linear-query}
For every integer \(k\ge2\), \(G\) admits a sourcewise
path-reporting distance oracle for \(S\times V\) with
\((2k-2)\)-strong stretch, expected space
\(O(kn|S|^{1/k})\), and \(O(k)\) query time.
With high probability, its preprocessing time satisfies
the bounds in Corollary~\ref{cor:sourcewise-tree-cover}.
After a query, the selected path can be reported in
additional time linear in its number of edges.
\end{corollary}

\subsection{Prioritized Structures}

Prioritized metric structures were introduced by Elkin,
Filtser, and Neiman~\cite{EFN15}.
The vertices are given a priority ordering
\(x_1,x_2,\ldots,x_n\), and the goal is to provide better
guarantees for pairs involving vertices of higher priority.
For undirected graphs, they obtained a path-reporting
distance oracle with space \(O(kn^{1+1/k})\) and
prioritized stretch
\(2\max\{1,\lceil k\log j/\log n\rceil\}-1\),
where \(j\) is the smaller of the two endpoint ranks.
We obtain an analogous strong-stretch guarantee for
directed graphs by combining the sourcewise structures
from the preceding subsection.

Let \(G=(V,E)\) be a directed graph with nonnegative
edge weights, \(n\ge2\) vertices, and a prescribed
priority ordering \(x_1,\ldots,x_n\).
We require guarantees for every ordered pair with finite
roundtrip distance, with the stretch determined by its
higher-priority endpoint.

\paragraph{Construction.}
Fix an integer \(k\ge4\).
For every \(i=1,\ldots,k\), let
\(
S_i=\{x_1,\ldots,x_{\lfloor n^{i/k}\rfloor}\},
\)
and construct a sourcewise \(i\)-strong tree cover for
\(S_i\times V\).
We also apply the same construction to the graph obtained
by reversing all edges, and reverse the resulting
tree-pairs back to their original directions.
These additional tree-pairs provide the corresponding
guarantee for \(V\times S_i\).

Consider an ordered pair whose higher-priority endpoint
is \(x_j\).
The smallest index \(i\) for which \(x_j\in S_i\) is
\[
i=\max\left\{1,\left\lceil
\frac{k\log j}{\log n}
\right\rceil\right\}.
\]
The corresponding sourcewise construction, in the
appropriate orientation, provides \(i\)-strong stretch.
In particular, pairs involving one of the first
\(\lfloor n^{1/k}\rfloor\) vertices receive exact routes.

\paragraph{Space and preprocessing time.}
For every vertex, the expected number of tree-pairs
containing it in the \(i\)-th construction is
\(O(i|S_i|^{1/i})=O(in^{1/k})\).
Summing over all indices and both orientations gives
expected overlap \(O(k^2n^{1/k})\), and expected total
size \(O(k^2n^{1+1/k})\).

For every \(i\ge4\), the construction of
Lemma~\ref{lem:sourcewise-fast-tree-cover} takes
\[
\widetilde O(n^2\sqrt{|S_i|})
\le
\widetilde O(n^{2+i/(2k)})
\le
\widetilde O(n^{5/2})
\]
time with high probability.
For \(i=1,2\), the basic construction takes
\[
\widetilde O((m+n)|S_i|)
\le
\widetilde O(n^{2+2/k})
\le
\widetilde O(n^{5/2}).
\]
For \(i=3\), the same lemma gives
\[
\widetilde O(n^2|S_3|^{2/3})
\le
\widetilde O(n^{2+2/k})
\le
\widetilde O(n^{5/2})
\]
with high probability.
The last inequalities use \(k\ge4\).
Since \(k=O(\log n)\) after capping larger parameters,
all constructions together take
\(\widetilde O(n^{5/2})\) time with high probability.

\begin{corollary}
\label{cor:prioritized-tree-cover}
For every integer \(k\ge4\), \(G\) admits a tree cover
in which each vertex belongs to \(O(k^2n^{1/k})\)
tree-pairs in expectation, and the expected total number
of vertex occurrences is \(O(k^2n^{1+1/k})\).
For every ordered pair \((u,v)\) with finite roundtrip
distance, if \(x_j\) is its higher-priority endpoint,
some tree-pair \(T\) satisfies
\begin{equation}
\label{eq:prioritized-strong-stretch}
\dist_T(u\to v)
\le
\dist_G(u\to v)
+
\left(
\max\left\{1,\left\lceil
\frac{k\log j}{\log n}
\right\rceil\right\}
-1
\right)\dist_G(u\lra v).
\end{equation}
The cover can be constructed in
\(\widetilde O(n^{5/2})\) time with high probability.
\end{corollary}

Taking the union of these tree-pairs gives the following.

\begin{corollary}
\label{cor:prioritized-spanner}
For every integer \(k\ge4\), \(G\) admits a spanner
\(H\subseteq G\) with \(O(k^2n^{1+1/k})\) edges in
expectation, satisfying
\eqref{eq:prioritized-strong-stretch} with \(H\) in place
of \(T\).
The spanner can be constructed in
\(\widetilde O(n^{5/2})\) time with high probability.
\end{corollary}

For the distance oracle, retain the sourcewise oracle
for each index and orientation.
Given a query \((u,v)\), identify its higher-priority
endpoint \(x_j\) and use the smallest index \(i\) for
which \(x_j\in S_i\).
This index is stored with each vertex.
If \(x_j=u\), query the corresponding oracle on \(G\);
otherwise, query the oracle on the reversed graph for
\((v,u)\) and reverse the reported path.
The expected query time is
\(O(i|S_i|^{1/i})=O(in^{1/k})\).

\begin{corollary}
\label{cor:prioritized-oracle}
For every integer \(k\ge4\), \(G\) admits a path-reporting
distance oracle with expected space \(O(k^2n^{1+1/k})\).
For every ordered query \((u,v)\) with finite roundtrip
distance, whose higher-priority endpoint is \(x_j\),
the oracle returns the length of a path satisfying
the bound in \eqref{eq:prioritized-strong-stretch},
in expected time \(O(kn^{1/k})\).
The oracle can be constructed in
\(\widetilde O(n^{5/2})\) time with high probability.
After a query, the selected path can be reported in
additional time linear in its number of edges.
\end{corollary}

Replacing each sourcewise oracle for \(i\ge2\) by the oracle
of Corollary~\ref{cor:sourcewise-oracle-linear-query}, and
using exact shortest-path trees for \(i=1\), gives the following. The query time can be reduced to \(O(\log\log k)\) by adapting the algorithm of Section~\ref{sec:PRDO} to the sourcewise setting.

\begin{corollary}
\label{cor:prioritized-oracle-linear-query}
For every integer \(k\ge4\), \(G\) admits a path-reporting
distance oracle with expected space \(O(k^2n^{1+1/k})\)
and \(O(\log \log k)\) query time.
For every ordered query \((u,v)\) with finite roundtrip
distance, whose higher-priority endpoint is \(x_j\),
the returned path length is at most
\[
\dist_G(u\to v)
+
\max\left\{0,\,
2\left\lceil\frac{k\log j}{\log n}\right\rceil-3
\right\}\dist_G(u\lra v).
\]
The oracle can be constructed in
\(\widetilde O(n^{5/2})\) time with high probability.
After a query, the corresponding path can be reported in
additional time linear in its number of edges.
\end{corollary}

Applying the tree-pair routing of Section~\ref{sec:applications} to the prioritized cover, with each label containing entries for all tree-pairs containing its vertex, gives the following.

\begin{corollary}
\label{cor:prioritized-routing}
For every integer \(k\ge4\), \(G\) admits a labeled routing
scheme with expected routing-table size and label length
\(\widetilde O(k^2n^{1/k})\) words at every vertex.
For every ordered pair \((u,v)\) with finite roundtrip distance,
whose higher-priority endpoint is \(x_j\), the route length
is at most
\[
\dist_G(u\to v)
+
\max\left\{0,\,
\left\lceil\frac{k\log j}{\log n}\right\rceil-1
\right\}\dist_G(u\lra v).
\]
The initial routing decision takes
\(\widetilde O(k^2n^{1/k})\) expected time,
and each subsequent decision takes \(O(1)\) time.
Packet headers use \(O(\log^2 n/\log\log n)\) bits.
The scheme can be constructed in
\(\widetilde O(n^{5/2})\) time with high probability.
\end{corollary}

\paragraph{Use of AI.}
ChatGPT was used to assist with drafting and revising the entire paper, checking mathematical arguments and verifying references and novelty. 
The authors reviewed the suggestions and take full responsibility for the final content, including all results, proofs, and citations.

\bibliographystyle{unsrt}
\bibliography{mybib}

\appendix




\end{document}